\documentclass[twocolumn,superscriptaddress,secnumarabic, nobibnotes, aps, prl]{revtex4-2}
\pdfoutput=1
\usepackage[usenames]{color}
\usepackage{graphicx}
\usepackage{multirow}
\usepackage{amsmath}
\usepackage{amsfonts}
\usepackage{amssymb}
\usepackage{mathrsfs}
\usepackage{url}
\usepackage{color}      % use if color is used in text
\usepackage{braket}
\usepackage[export]{adjustbox}
\usepackage{subfig}
\usepackage{setspace}
\usepackage{cleveref}
\newcommand{\bs}{\boldsymbol}

\usepackage{xr}

\makeatletter
\newcommand*{\addFileDependency}[1]{% argument=file name and extension
	\typeout{(#1)}
	\@addtofilelist{#1}
	\IfFileExists{#1}{}{\typeout{No file #1.}}
}
\makeatother

\begin{document}
\title{Engineering the Electronic Structure of Two-Dimensional Materials with Near-Field Electrostatic Effects of Self-Assembled Organic Layers}

\author{Qunfei Zhou}
\email{qunfei.zhou@northwestern.edu}
\affiliation{Materials Research Science and Engineering Center, Northwestern University, Evanston, IL 60208, USA}
\affiliation{Center for Nanoscale Materials, Argonne National Laboratory, Argonne, IL 60439, USA}
	
\author{Bukuru Anaclet}
%Pierre_Reminder: Need to check affiliation mechanism for REU+ 
\affiliation{Materials Research Science and Engineering Center, Northwestern University, Evanston, IL 60208, USA}
\affiliation{Department of Chemistry, Pomona College, 645 North College Avenue, Claremont, CA}

\author{Trevor Steiner}
%Pierre_Reminder: Need to check affiliation mechanism for REU 
\affiliation{Materials Research Science and Engineering Center, Northwestern University, Evanston, IL 60208, USA}
\affiliation{Department of Materials Science and Engineering, University of Minnesota-Twin Cities, Minneapolis, MN 55455, USA}
\affiliation{Materials Department, University of California, Santa Barbara, California 93106, USA}
%Current address: UCSB
	
\author{Michele Kotiuga}
\email{michele.kotiuga@epfl.ch}
\affiliation{Theory and Simulation of Materials (THEOS) and National Centre for Computational Design and Discovery of Novel Materials (MARVEL), \'{E}cole Polytechnique F\'{e}d\'{e}rale de Lausanne, CH-1015 Lausanne, Switzerland}

\author{Pierre Darancet}
\email{pdarancet@anl.gov}
\affiliation{Center for Nanoscale Materials, Argonne National Laboratory, Argonne, IL 60439, USA}
\affiliation{Northwestern Argonne Institute of Science and Engineering, Evanston, IL 60208, USA}
	
\date{\today}

\begin{abstract}
We compute the electronic structure of two-dimensional (2D) materials decorated with self-assembled organic monolayers using density functional theory.
We find that 2D materials are strongly impacted by near-field electrostatic effects resulting from high multipoles of the organic layer electronic density. 
We show that this effect can lead to significant ($\simeq$0.5V) modulation of the in-plane potential experienced by electrons in 2D materials within $\simeq$ 4 \AA\ from the molecular layer, with a transition between near- and far-field depending on the lateral extent of the molecules.
We develop a theory of this effect, showing that the electrostatic potential of the molecular layer can be approximated by a discretized planar charge density derived from the molecular structure and multipoles. 
Solving this model computationally and analytically, we propose implementations of this effect to generate novel electronic properties for electrons in 2D materials, such as band gap opening and anisotropic group velocity modulation for monolayer graphene from experimentally achievable molecular assemblies. 
\end{abstract}
	
\maketitle
Molecular assemblies on surfaces are known to modulate the optical and electronic properties of materials through their electrostatic properties~\cite{Kronik2006PRBDipole,Kronik2007Dipole,Ahn2006RMP,Liljeroth2017SAM2Drev,Samori2018SAM2Drev,Samori20182SAM2DAdvMat,Samori2021SAM}. In the case of bulk 3D materials, electrostatic modification occurs through far-field effects resulting from the monopole and areal dipole of the material/molecule interface~\cite{Kronik2006PRBDipole,Kronik2007Dipole,Ahn2006RMP,Kotiuga2015NL,Venka2011NL,Monti2012MolDipole,Wang2008WF,Venka2015NatNanotech,Egbert2009SAMwf}, with consequences on observables such as work functions~\cite{Kahn2012DipWF,Kahn2003SAM,De2005PRLWF}, band offsets~\cite{Goronzy2018SAMrev} and superconducting transitions~\cite{Gobbi2018SAM}. While also impacted by far-field effects~\cite{Samori2018SAM2Drev,Samori20182SAM2DAdvMat,Zhou2019NPL}, 2D electronic states such as surface states~\cite{Crommie1993Corral,Barth2007NatNano,Kawai2021SAMsurf} and electronic states in 2D materials~\cite{Samori2017SAM,Gobbi2018SAM,Samori2016Gr1D,2014TMAGr,Zboril2016SurfGrRev,shayeganfar2014EgTMA,shayeganfar2015EgF2,xiu2014Gr,Kuo2012GrPh} can also be impacted by near-field effects resulting from higher moments of the surface/interface electronic density, typically observed within $\approx 10$ \AA~\cite{Natan2007SAMrev,Martinez2015gating,Li2021Boroph} of the interface. In particular, for these systems, near-field modulation of the electrostatic potential has been proposed as a means to engineer band structures, wave functions, and topological properties ~\cite{Crommie1993Corral,Barth2007NatNano,Choi2008BandsAtom,Park2008NatPhys,Park2008PRL,gomes2012designer,Liljeroth2017topological,Liljeroth2019review,Guisinger2021Topology}. Yet, to the best of our knowledge, a theory of these near-field effects and their relationship with materials descriptors is currently lacking, hindering efficient discovery and design of these emergent electronic phases through molecular assembly. 

In this work, we derive a theory of the near-field electrostatic effects at 2D/Organic-Layer interfaces. Using density functional theory (DFT), we show that the near-field potential induced by a molecular layer on a 2D material can fluctuate by up to $\simeq$0.5V on a nanometer scale for experimentally assembled molecular layers. We show that the effect of the molecular layer is well-approximated by a discretized planar charge density, which can be derived from the structure and the multipole moments of the molecule. We solve this model numerically and analytically for three classes of organic molecules, all previously experimentally assembled on graphene and other 2D materials: (i) 2,7-dioctyl[1]-benzothieno[3,2-b][1]benzothiophene (C8-BTBT)~\cite{Hersam2019C8btbt,Hersam2020C8btbtInSe,He2014C8btbtGr},  (ii) phthalocyanine (H$_2$Pc) and perfluoro-phthalocyanine (H$_2$PcF$_{16}$)~\cite{Zhang2011FePcGr,Samori2020Pc2D,Liljeroth2014PcGr,Berndt2015PcGr}, and (iii) benzene (C$_6$H$_6$) and hexafluorobenzene (C$_6$F$_6$)~\cite{Wang2014PhGr,Smith2018PhGr,Hassan2014PhGr}. Using our results, we elucidate the molecular descriptors controlling the magnitude of the potential fluctuations and the lengthscale of the near-field to far-field transition. Finally, we show how near-field effects of phthalocyanines and benzene monolayers deposited on graphene lead to band gap opening. 

We consider monolayers of C8-BTBT, H$_2$Pc, H$_2$PcF$_{16}$, C$_6$H$_6$ and C$_6$F$_6$ on monolayer graphene and indium selenide (InSe) approximated by 3D periodic systems where the periodic images are separated by $\sim$22~\AA~of vacuum in the normal direction. All of the molecules considered in this work have inversion symmetry, leading to zero (in the gas phase) or negligible (when absorbed) dipole moments, but finite quadrupole and higher moments of their electronic densities. Beyond symmetry considerations, all molecules considered here have minimal charge transfer with the 2D materials, as shown in Supplementary Materials. All geometries of the self-assembled monolayers (SAM) are adopted from experiments~\cite{Hersam2019C8btbt,Hersam2020C8btbtInSe,He2014C8btbtGr,Zhang2011FePcGr,Samori2020Pc2D,Liljeroth2014PcGr,Berndt2015PcGr,Wang2014PhGr,Smith2018PhGr,Hassan2014PhGr}, and optimized using DFT with the exchange-correlation potential of Perdew-Burke-Ernzerhof (PBE), and Tkatchenko-Scheffler~\cite{Scheffler2009TSvdw} corrections for the van der Waals interactions. Computational details, computed geometries and binding energies of the molecules on 2D materials, and quadrupole moments are reported in the Supplementary Materials.

\begin{figure}[h]
	\centering
    \includegraphics[width=0.95\linewidth]{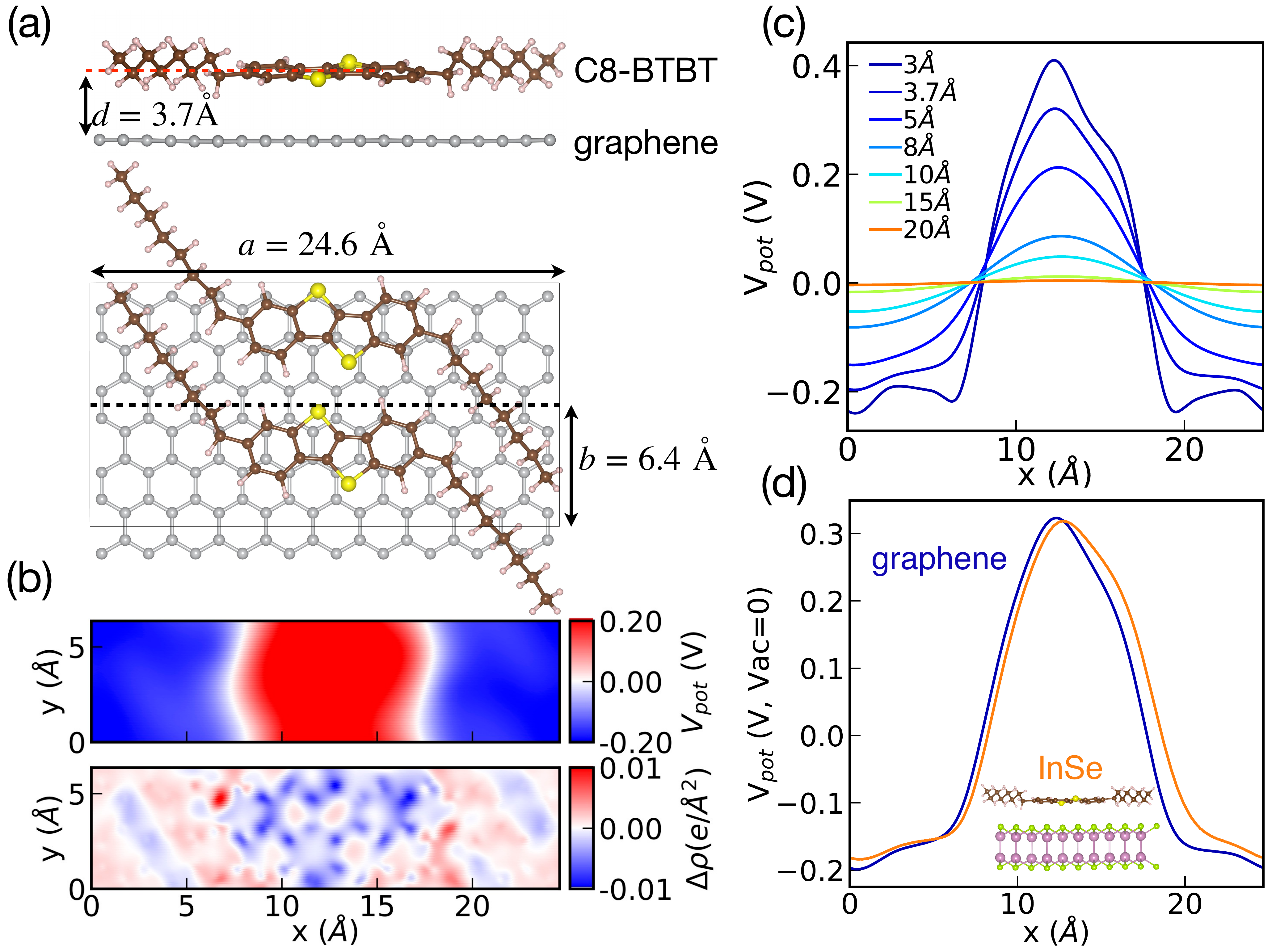}
    \caption{(a) Side and top view of C8-BTBT monolayer on graphene, with in-plane lattice parameters. (b) Electrostatic potential $V_{\textrm{pot}}(x;d)$ generated by C8-BTBT averaged along the $y$ direction, as a function of vertical distance $d$ computed with DFT. (c) V$_{\textrm{pot}} (x,y,d)$  in the graphene plane at $d=3.7$ \AA , and electronic density redistribution near the graphene plane $\int_{d-\delta z}^{{d+\delta z}} dz \Delta \rho (x,y,z)$ with $\delta z=1.0$~\AA, and $\Delta \rho = \rho$[C8-BTBT/2D] - $\rho$[C8-BTBT] - $\rho$[2D].  d) $V_{\textrm{pot}} (x,d)$ of C8-BTBT molecular layer averaged along $y$ for the configurations on graphene and InSe monolayers at $d\approx 3.7$\AA.}
    \label{fig:f1}
\end{figure}

In Fig.~\ref{fig:f1} (a), we show the optimized structure of the C8-BTBT molecular layer on graphene, and its electrostatic potential as a function of molecule-2D materials distance (Fig. \ref{fig:f1} (b)). As expected from the negligible dipole (arising from inversion symmetry in the gas phase), the electrostatic potential created by the freestanding layer decays to $0$V in the far-field, with in-plane potential fluctuations negligible at distances $\approx 20$ Å from the molecular layer. At smaller vertical distances, the features of the molecule become apparent in the potential, creating a repulsive (attractive) potential for electrons in 2D materials below the molecular center (arms).  Importantly, the magnitude of the change in potential along the molecular long axis reaches $0.57$V at the equilibrium molecule-2D material distance ($3.7$Å), with an inhomogeneous in-plane electric field exceeding 1V/nm near $x=8$Å and $x=17$Å. The overall quasi-one-dimensional periodic potential  profile added by the C8-BTBT layer is reminiscent of a Kronig–Penney model, as shown in Fig.~\ref{fig:f1}(c). Correspondingly, we observe a modulation  of the electronic density in graphene along the long axis of the molecule, indicating that graphene electrons polarize in response to this large perturbation. 

While the details of the molecular absorption and periodicity of the self-assembled molecules (SAM) depend on the 2D material, we note that, in absence of major conformational changes, the magnitude and profile of the potential is dominated by the intrinsic properties of the molecular layer, as shown by the minimal differences between the potential induced by C8-BTBT on graphene and InSe in Fig.~\ref{fig:f1}(d). This suggests that near-field effects can be derived from materials descriptors of a freestanding molecular layer. We now focus on developing such an electrostatic model.

\begin{figure*}%[H]
%\begin{figure}[H]
\centering
\includegraphics[width=0.85\linewidth]{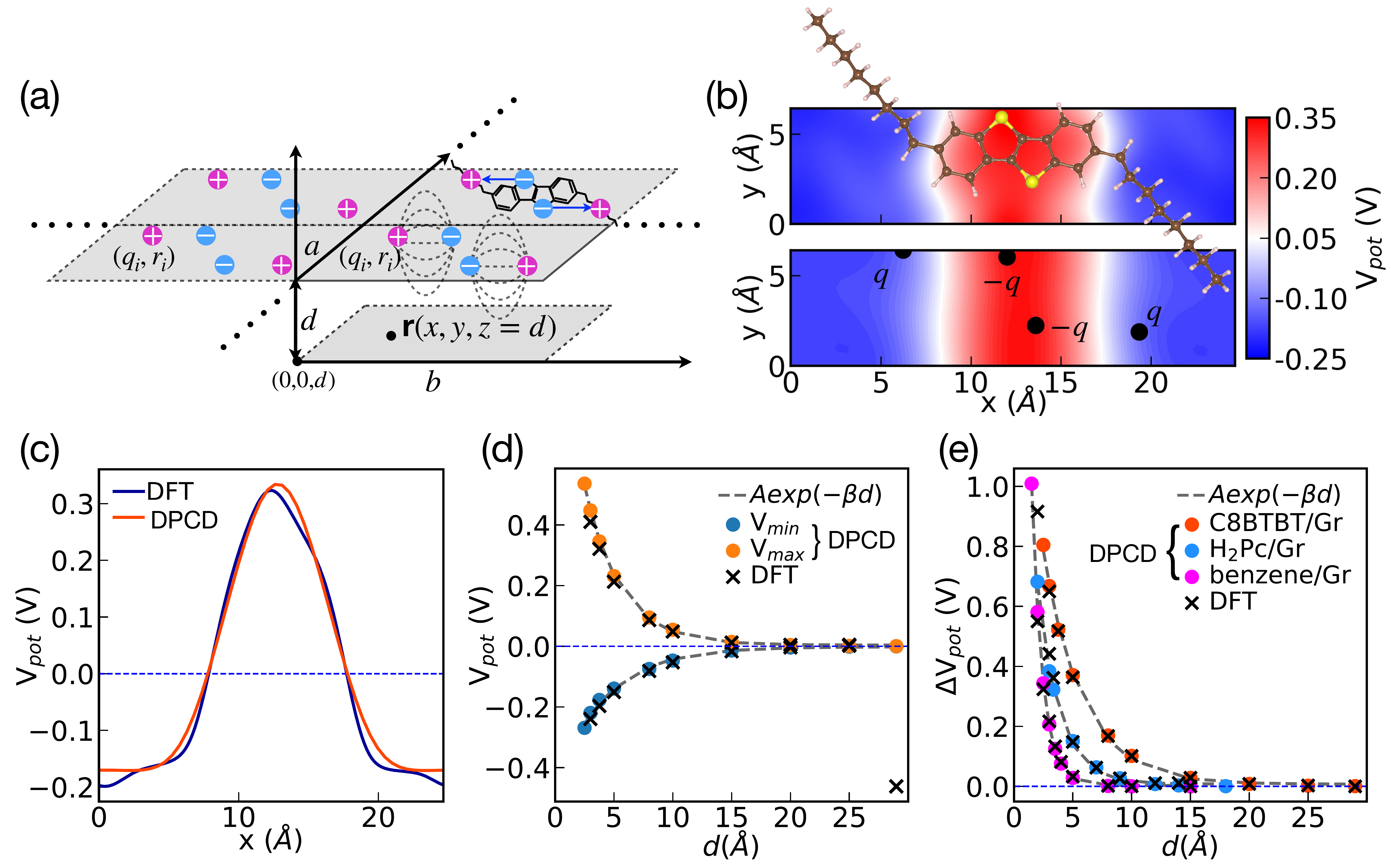}
\caption{(a) Schematics of the discretized planar charge density (DPCD) model of a molecular monolayer and its electrostatic effect on an underlying material (at a vertical distance $d$). (b) The electrostatic potential of a C8-BTBT layer in the $xy$-plane at a distance $d=3.7$ \AA\ as predicted from DFT (top) and a 4-point-charge DPCD model (bottom). The dots indicate the $x,y$ positions of the 4 point charges ($q=0.07e^{-}$) in the z=0 plane. (c) In-plane potential at $d=3.7$ \AA\ from DFT and DPCD averaged along the $y$-axis. (d) Minimum (V$_{min}$) and maximum (V$_{max}$) values in the $xy$-plane the electrostatic potential as a function of vertical distance $d$ from DPCD and DFT. (e) In-plane potential modulation $\Delta V_{pot}=V_{max}-V_{min}$ as a function of $d$ for C8-BTBT/graphene, H$_2$Pc/graphene and benzene/graphene from DPCD and DFT. Dashed lines in (d) and (e) corresponds to an exponential fit, with decay constants $\beta$ of 0.28, 0.53, 0.89 (\AA$^{-1}$) for the 3 molecules, respectively (decay constants for all molecules are summarized in Table S2).
} 
\label{fig:f2}
\end{figure*}

 As shown in Fig.~\ref{fig:f1}, Fig.~S8 and Fig.~S10 for C8-BTBT, H$_2$Pc, and C$_6$H$_6$, respectively, the periodicity and pattern of the in-plane potential generated by the molecular layers are dependent on the planar distribution of electron-withdrawing and donating groups of the molecules. Based on this observation,  we propose to represent the continuous, 3D electronic density of the SAM by a sum of point charges located in the (overall neutral) molecular plane (as shown in Fig.~\ref{fig:f2}(a)): 
 \begin{eqnarray}
 \label{Eq:density}
 \rho_\textrm{SAM}(\mathbf{r}) & \simeq & \sum_{i,n_1,n_2} q_i \delta \left( \mathbf{r} - \mathbf{r}_{i,n_1,n_2}   \right),   \nonumber \\ 
\mathbf{r}_{i,n_1,n_2} &=& (x_i,y_i,0)+n_1\mathbf{a}+n_2\mathbf{b} \quad \& \quad \sum_i q_i = 0,   
 \end{eqnarray}
where $\mathbf{a}$,$\mathbf{b}$ are the in-plane lattice parameters of the monolayer, $i$ indexes the charges $q_i$ in a single unit cell at positions $(x_i,y_i,0)$, and $n_1$ and $n_2$ are integers.  We refer to this approximation as discretized planar charge density (DPCD) model in the next. Unlike a standard multipole expansion, this model is capable of describing cases in which the lateral dimensions of the molecular features are comparable to the molecule-2D distance, typical of molecular assemblies. For simplicity, we restrict our studies to DPCD with a quadrupole but no higher moments. The coordinates and number of point charges ($\bs r_i, q_i$) are determined by the structure, the quadrupole moment of the molecule, and the out-of-plane decay of the near-field potential (see details in Section I of the Supplemental Materials). 

%\end{figure}

Results of this approximation for C8-BTBT are shown in Fig.~\ref{fig:f2}(b-d) for C8-BTBT and in Supplemental Materials for the other molecular layers. As shown in Fig.~\ref{fig:f2}(b-d), the DPCD model is able to fully capture the behavior of the electrostatic potential at typical molecule-2D distances ($d>3$ \AA). Remarkably, this is accomplished with a minimal number of point charges: 4, 5, and 7 point charges per unit cell are necessary for describing the electrostatic potential of C8-BTBT, phthalocyanine and benzene SAMs, respectively. Specifically, the model captures the in-plane profile and local electric field Fig.~\ref{fig:f2}(b-c) as well as the out-of-plane decay to the far-field. Importantly, this decay is found to be (i) exponential and (ii) strongly dependent on the SAM.  As originally explained by Natan and Kronik~\cite{Kronik2006PRBDipole,Kronik2007Dipole} for arrays of out-of-plane dipoles, this exponential decay is a result of the in-plane periodicity~\cite{Haldar2020NL} and occurs on a distance dependent upon the lattice vectors of the unit cell.  

Using the representation of the density in Eq.\ref{Eq:density}, we now analyze the physical origin of --and material parameters controlling-- the lengthscale of the near- to far-field transition. As shown in Fig. S29 of the Supplementary Materials, a direct sum in real space of the potential generated by the point charges and their periodic images typically converges on the scale of 10$^3 \sim 10^4$ periodic images for a 99\% accuracy, precluding an analytical solution. Instead, the potential quickly converges in reciprocal space. Using a Mellin transform and Poisson resummation to recast the Fourier transform of the potential (see Section VIII of Supplementary Information for full derivation), we obtain the following expression for the Coulomb potential of a neutral planar array of point charges:
\begin{align}\label{Eq:q}
V(\bs r (x,y,z=d)) = \frac{1}{4\pi \varepsilon}\sum_i \frac{q_i}{a} \sum_{\vec{k}}{}' f_{\vec{k}}(d)
\end{align}
with
\begin{align}\label{Eq:f}
	 f_{\vec{k}}(d)=\frac{\cos\left[2\pi \left(\frac{k_1(x-x_i)}{a}+\frac{k_2(y-y_i)}{b}\right)\right] e^{-2\pi d\sqrt{k_1^2/a^2+k_2^2/b^2}}}{\sqrt{k_1^2b^2/a^2+k_2^2}}
\end{align}
where $a, b$ are the lattice constants of the unit cell, and $\vec{k}=[k_1, k_2]$, which sums over the reciprocal lattice points excluding the origin. Importantly, this expression retains explicit sum over each point charge in the unit cell $i$, and converges with a minimal number of $\vec{k}$, as shown in Fig.~S29 of the Supplementary Materials. 

Using the first terms of Eq.~\ref{Eq:q}–\ref{Eq:f} to approximate the the electrostatic potential modulation $\Delta V$ resulting from the DPCD model, we obtain the following expression for the C8-BTBT monolayer:  
\begin{align} \label{Eq:4pt}
	 \Delta V  =\frac{16}{4\pi \varepsilon} \frac{Q_{xx} \sin^2(\pi x_0) }{ 2(ax_0)^2 b }e^{-2\pi d/a}  
\end{align}
where $x_0$ describes the (dimensionless) relative positions of the point charges within the unit cell, and $Q_{xx}$ is the $xx$ component of the quadrupolar tensor (see detailed derivation in Supplementary Materials). 

Eq. \ref{Eq:4pt} has several implications: First, the term $e^{-2\pi d/a} $, also found in previous works~\cite{Natan2007SAMrev,Thygesen2017NatComm,Haldar2020NL}, implies that the modulation of the near-field potential decays on a lengthscale at most comparable to the long axis of the unit cell. Second, the effect is unsurprisingly proportional to the molecular quadrupole $Q_{xx}$ (the symmetries of the point charges preventing any other non-zero multipole). Finally, the potential modulation is also impacted by the position of the point charges, an effect not captured by a multipole expansion of the potential. The latter term implies that the near-field electrostatic potentials from molecules of equal net charge quadrupole will vary with the spatial separation between charges, showing the importance of controlling the spatial extent and separation of the donor and acceptor moieties.

A direct outcome of Eqs.~\ref{Eq:q}–\ref{Eq:4pt} is that the transition between near-field and far-field can occur at a vertical distance shorter than the unit cell. This lengthscale is independent of the magnitude of the multipoles of the SAMs, and is instead related to the in-plane spacing of the donor and acceptor features. This is in agreement with the trend observed in Fig.~\ref{fig:f2} (e), in which the near-field potential of a phthalocyanine SAM decays faster than the one of C8-BTBT. %despite comparable inter-molecular distances. 

As shown above and in Supplementary Materials, the DPCD model quantitatively captures the physics of near-field electrostatic effects of molecular layers and can be used for materials screening using descriptors accounting for the multipoles and structure of molecules. We note that, unlike atomistic simulations and shown in Supplementary Materials, this model can also be used to investigate the effects of domain size and polymorphism on the in-plane electrostatic modulation, and can be combined with models of 2D dielectric screening~\cite{Quek2019screen2D} to predict the resulting induced densities and displacement fields. 

\begin{figure}[h]
	\centering
	\includegraphics[width=0.9\linewidth]{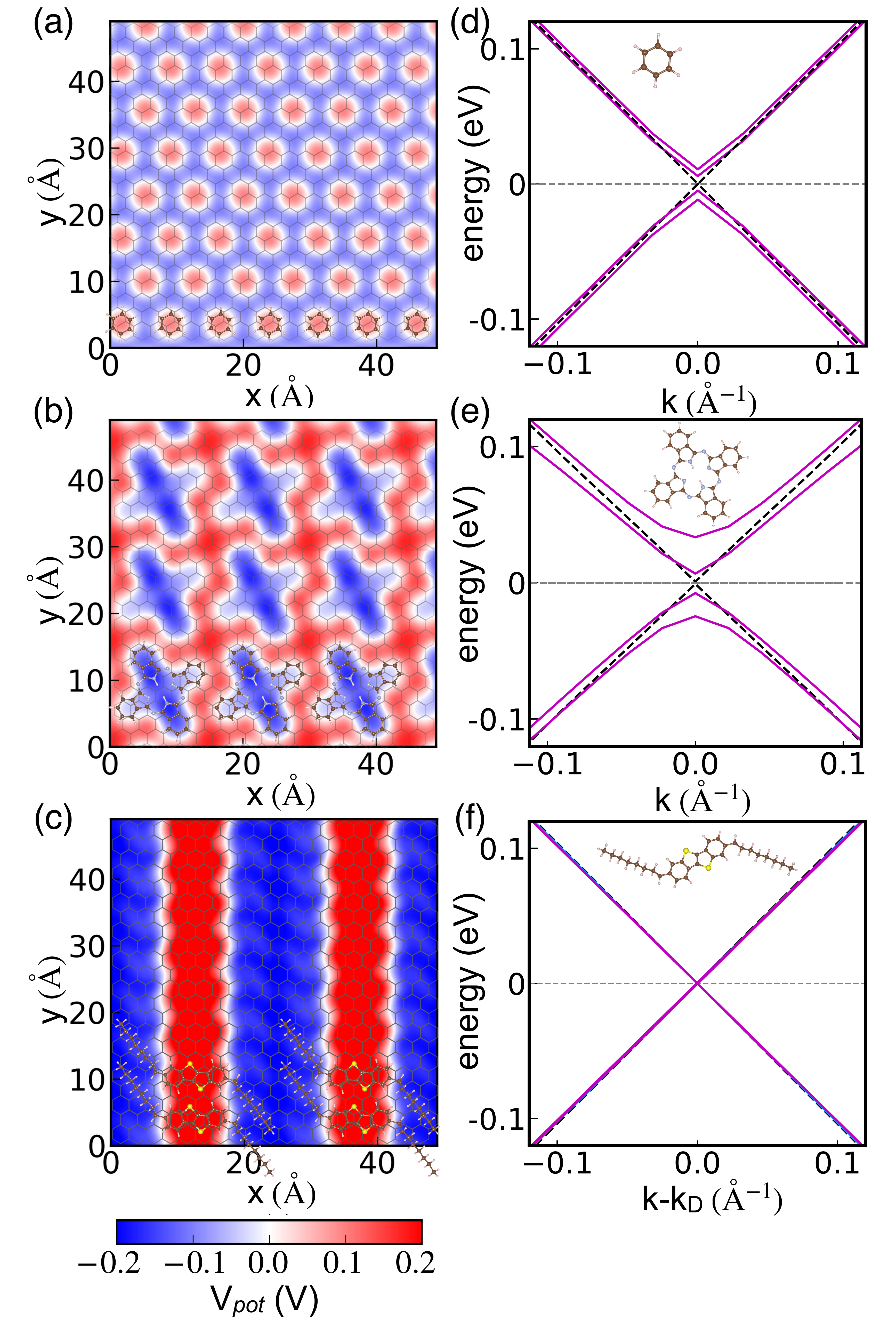}
	\caption{ In-plane electrostatic potentials in the graphene plane from (a) benzene, (b) H$_2$PcF$_{16}$, (c) C8-BTBT monolayers. (d-f) Corresponding band structures near the Dirac point (k$_D$). The dashed black lines are for pristine graphene.} 
	\label{fig:f3}
\end{figure}

Having isolated the magnitude of near-field electrostatic effects of SAMs, we now propose their use in engineering the electronic structure of 2D materials. In Fig.~\ref{fig:f3}(a-c), we show the electrostatic potential experienced by electrons in graphene decorated by SAMs of benzene, H$_2$PcF$_{16}$ and C8-BTBT, respectively (results for C$_6$F$_6$ and H$_2$Pc are included in Supplementary Materials).
All SAMs modulate the potential on the scale of $100meV$ with patterns decided by the size and shape of molecules: Muffin-tin type for benzene/graphene, Kronig–Penney type for C8-BTBT/graphene.  We note that the use of such periodic potentials on comparable lengthscales has previously been proposed as a means to tune the electronic properties of graphene and other 2D materials~\cite{Park2008NatPhys,Park2008PRL,Stroud2009PRB,Brey2009GrV,xiu2014Gr,Kuo2012GrPh,Geim2009GrRev}.  

In Fig.~\ref{fig:f3}(d-f), we show DFT band structures near the Dirac point $k_D$ for those three molecular layers. We note that, following previous works not based on DFT~\cite{Park2008NatPhys,Park2008PRL,Stroud2009PRB,Brey2009GrV}, band structures in quantitative agreement with DFT can be obtained using a tight binding model with on-site energies modulated by the external potential in Fig.~\ref{fig:f3}(d-f) (as shown in Supplementary Materials). As shown in Fig.~\ref{fig:f3}(d-e), the electrostatic potential of benzene and H$_2$Pc$_{16}$ SAMs opens a band gap at the Dirac point. We note that  band gap opening resulting from self-assembly of molecules on graphene  has been experimentally reported for different molecules~\cite{Zboril2016SurfGrRev,shayeganfar2014EgTMA,shayeganfar2015EgF2}. In contrast, there is no band gap opening for C8BTBT/graphene superlattice with a different Dirac point position. However, the group velocity is reduced by $\sim$1.2\% and $\sim$1.8\% along $y$ and $x$ directions, respectively, see Fig.~\ref{fig:f3}(f), similar to previous work with external 1D potential for graphene~\cite{Park2008NatPhys}. 

In summary, we have demonstrated in this work that self-assembled molecular layers can significantly impact the electronic structure of 2D materials and surface states through their near-field electrostatic properties. We have developed the first theory of such effects, showing that the potential generated by SAMs at a few angstroms can be quantitatively approximated by the one created by a discretized planar charge density, derived from molecular structure and multipoles. Applying this model to 5 experimentally synthesized SAMs on graphene, we elucidated the molecular descriptors involved in generating large near-field effects, showing that the spatial distribution of the donor and acceptor components of the SAM matters beyond the magnitude of its density multipoles. In particular, we showed that the lengthscale of the near- to far-field transition can be shorter than the periodicity of unit cell of the SAMs, depending on  molecular feature sizes. Finally, we showed that near-field electrostatic effects of SAMs on graphene can lead to band-structure engineering, such as band gap opening and anisotropic behavior. 

\section*{Supplementary Material}
Details of the computational methods, including the details for DFT calculations, the discretized planar charge density(DPCD) model, as well as the dielectric screening effect of the 2D material; Electrostatic potential induced by benzene, C$_6$F$_6$, H$_2$Pc, and H$_2$PcF$_{16}$ on graphene obtained from DFT and DPCD model, as well as their energy dispersion near the Dirac points for monolayer graphene; Tight-binding calculations of the potential influence on the band structures for monolayer graphene; analytical derivation of the DPCD model for the multipole effects of the molecular assemblies, numerical convergence of the DPCD model.  
Structures of the molecule/2D heterojunctions in crystallographic information file (cif) format. \\

\section*{Acknowledgments}
This work was supported by the Northwestern University MRSEC under National Science Foundation grant No.~DMR-1720139 (Q.Z., B. A., T. S., P.D.). Use of the Center for Nanoscale Materials (CNM), an Office of Science user facility, was supported by the U.S. Department of Energy, Office of Science, Office of Basic Energy Sciences, under Contract No. DE-AC02-06CH11357. We gratefully acknowledge use of the Bebop cluster in the Laboratory Computing Resource Center at Argonne National Laboratory. P.D. would like to acknowledge fruitful discussions with Jeffrey Guest, Constance Pfeiffer, Nathan Guisinger, Lincoln Lauhon, and Mark Hersam.

\bibliography{MultipoleDoping}

\begin{thebibliography}{57}%
\makeatletter
\providecommand \@ifxundefined [1]{%
 \@ifx{#1\undefined}
}%
\providecommand \@ifnum [1]{%
 \ifnum #1\expandafter \@firstoftwo
 \else \expandafter \@secondoftwo
 \fi
}%
\providecommand \@ifx [1]{%
 \ifx #1\expandafter \@firstoftwo
 \else \expandafter \@secondoftwo
 \fi
}%
\providecommand \natexlab [1]{#1}%
\providecommand \enquote  [1]{``#1''}%
\providecommand \bibnamefont  [1]{#1}%
\providecommand \bibfnamefont [1]{#1}%
\providecommand \citenamefont [1]{#1}%
\providecommand \href@noop [0]{\@secondoftwo}%
\providecommand \href [0]{\begingroup \@sanitize@url \@href}%
\providecommand \@href[1]{\@@startlink{#1}\@@href}%
\providecommand \@@href[1]{\endgroup#1\@@endlink}%
\providecommand \@sanitize@url [0]{\catcode `\\12\catcode `\$12\catcode
  `\&12\catcode `\#12\catcode `\^12\catcode `\_12\catcode `\%12\relax}%
\providecommand \@@startlink[1]{}%
\providecommand \@@endlink[0]{}%
\providecommand \url  [0]{\begingroup\@sanitize@url \@url }%
\providecommand \@url [1]{\endgroup\@href {#1}{\urlprefix }}%
\providecommand \urlprefix  [0]{URL }%
\providecommand \Eprint [0]{\href }%
\providecommand \doibase [0]{https://doi.org/}%
\providecommand \selectlanguage [0]{\@gobble}%
\providecommand \bibinfo  [0]{\@secondoftwo}%
\providecommand \bibfield  [0]{\@secondoftwo}%
\providecommand \translation [1]{[#1]}%
\providecommand \BibitemOpen [0]{}%
\providecommand \bibitemStop [0]{}%
\providecommand \bibitemNoStop [0]{.\EOS\space}%
\providecommand \EOS [0]{\spacefactor3000\relax}%
\providecommand \BibitemShut  [1]{\csname bibitem#1\endcsname}%
\let\auto@bib@innerbib\@empty
%</preamble>
\bibitem [{\citenamefont {Natan}\ \emph {et~al.}(2006)\citenamefont {Natan},
  \citenamefont {Zidon}, \citenamefont {Shapira},\ and\ \citenamefont
  {Kronik}}]{Kronik2006PRBDipole}%
  \BibitemOpen
  \bibfield  {author} {\bibinfo {author} {\bibfnamefont {A.}~\bibnamefont
  {Natan}}, \bibinfo {author} {\bibfnamefont {Y.}~\bibnamefont {Zidon}},
  \bibinfo {author} {\bibfnamefont {Y.}~\bibnamefont {Shapira}},\ and\ \bibinfo
  {author} {\bibfnamefont {L.}~\bibnamefont {Kronik}},\ }\bibfield  {title}
  {\bibinfo {title} {Cooperative effects and dipole formation at semiconductor
  and self-assembled-monolayer interfaces},\ }\href@noop {} {\bibfield
  {journal} {\bibinfo  {journal} {Phys. Rev. B}\ }\textbf {\bibinfo {volume}
  {73}},\ \bibinfo {pages} {193310} (\bibinfo {year} {2006})}\BibitemShut
  {NoStop}%
\bibitem [{\citenamefont {Deutsch}\ \emph {et~al.}(2007)\citenamefont
  {Deutsch}, \citenamefont {Natan}, \citenamefont {Shapira},\ and\
  \citenamefont {Kronik}}]{Kronik2007Dipole}%
  \BibitemOpen
  \bibfield  {author} {\bibinfo {author} {\bibfnamefont {D.}~\bibnamefont
  {Deutsch}}, \bibinfo {author} {\bibfnamefont {A.}~\bibnamefont {Natan}},
  \bibinfo {author} {\bibfnamefont {Y.}~\bibnamefont {Shapira}},\ and\ \bibinfo
  {author} {\bibfnamefont {L.}~\bibnamefont {Kronik}},\ }\bibfield  {title}
  {\bibinfo {title} {Electrostatic properties of adsorbed polar molecules:
  Opposite behavior of a single molecule and a molecular monolayer},\ }\href
  {https://doi.org/10.1021/ja068417d} {\bibfield  {journal} {\bibinfo
  {journal} {J. Am. Chem. Soc.}\ }\textbf {\bibinfo {volume} {129}},\ \bibinfo
  {pages} {2989} (\bibinfo {year} {2007})}\BibitemShut {NoStop}%
\bibitem [{\citenamefont {Ahn}\ \emph {et~al.}(2006)\citenamefont {Ahn},
  \citenamefont {Bhattacharya}, \citenamefont {Di~Ventra}, \citenamefont
  {Eckstein}, \citenamefont {Frisbie}, \citenamefont {Gershenson},
  \citenamefont {Goldman}, \citenamefont {Inoue}, \citenamefont {Mannhart},
  \citenamefont {Millis} \emph {et~al.}}]{Ahn2006RMP}%
  \BibitemOpen
  \bibfield  {author} {\bibinfo {author} {\bibfnamefont {C.}~\bibnamefont
  {Ahn}}, \bibinfo {author} {\bibfnamefont {A.}~\bibnamefont {Bhattacharya}},
  \bibinfo {author} {\bibfnamefont {M.}~\bibnamefont {Di~Ventra}}, \bibinfo
  {author} {\bibfnamefont {J.}~\bibnamefont {Eckstein}}, \bibinfo {author}
  {\bibfnamefont {C.~D.}\ \bibnamefont {Frisbie}}, \bibinfo {author}
  {\bibfnamefont {M.}~\bibnamefont {Gershenson}}, \bibinfo {author}
  {\bibfnamefont {A.}~\bibnamefont {Goldman}}, \bibinfo {author} {\bibfnamefont
  {I.}~\bibnamefont {Inoue}}, \bibinfo {author} {\bibfnamefont
  {J.}~\bibnamefont {Mannhart}}, \bibinfo {author} {\bibfnamefont {A.~J.}\
  \bibnamefont {Millis}}, \emph {et~al.},\ }\bibfield  {title} {\bibinfo
  {title} {Electrostatic modification of novel materials},\ }\href@noop {}
  {\bibfield  {journal} {\bibinfo  {journal} {Rev. Mod. Phys.}\ }\textbf
  {\bibinfo {volume} {78}},\ \bibinfo {pages} {1185} (\bibinfo {year}
  {2006})}\BibitemShut {NoStop}%
\bibitem [{\citenamefont {Kumar}\ \emph {et~al.}(2017)\citenamefont {Kumar},
  \citenamefont {Banerjee},\ and\ \citenamefont
  {Liljeroth}}]{Liljeroth2017SAM2Drev}%
  \BibitemOpen
  \bibfield  {author} {\bibinfo {author} {\bibfnamefont {A.}~\bibnamefont
  {Kumar}}, \bibinfo {author} {\bibfnamefont {K.}~\bibnamefont {Banerjee}},\
  and\ \bibinfo {author} {\bibfnamefont {P.}~\bibnamefont {Liljeroth}},\
  }\bibfield  {title} {\bibinfo {title} {Molecular assembly on two-dimensional
  materials},\ }\href@noop {} {\bibfield  {journal} {\bibinfo  {journal}
  {Nanotechnology}\ }\textbf {\bibinfo {volume} {28}},\ \bibinfo {pages}
  {082001} (\bibinfo {year} {2017})}\BibitemShut {NoStop}%
\bibitem [{\citenamefont {Bertolazzi}\ \emph {et~al.}(2018)\citenamefont
  {Bertolazzi}, \citenamefont {Gobbi}, \citenamefont {Zhao}, \citenamefont
  {Backes},\ and\ \citenamefont {Samor{\`\i}}}]{Samori2018SAM2Drev}%
  \BibitemOpen
  \bibfield  {author} {\bibinfo {author} {\bibfnamefont {S.}~\bibnamefont
  {Bertolazzi}}, \bibinfo {author} {\bibfnamefont {M.}~\bibnamefont {Gobbi}},
  \bibinfo {author} {\bibfnamefont {Y.}~\bibnamefont {Zhao}}, \bibinfo {author}
  {\bibfnamefont {C.}~\bibnamefont {Backes}},\ and\ \bibinfo {author}
  {\bibfnamefont {P.}~\bibnamefont {Samor{\`\i}}},\ }\bibfield  {title}
  {\bibinfo {title} {Molecular chemistry approaches for tuning the properties
  of two-dimensional transition metal dichalcogenides},\ }\href@noop {}
  {\bibfield  {journal} {\bibinfo  {journal} {Chem. Soc. Rev.}\ }\textbf
  {\bibinfo {volume} {47}},\ \bibinfo {pages} {6845} (\bibinfo {year}
  {2018})}\BibitemShut {NoStop}%
\bibitem [{\citenamefont {Gobbi}\ \emph
  {et~al.}(2018{\natexlab{a}})\citenamefont {Gobbi}, \citenamefont {Orgiu},\
  and\ \citenamefont {Samor{\`\i}}}]{Samori20182SAM2DAdvMat}%
  \BibitemOpen
  \bibfield  {author} {\bibinfo {author} {\bibfnamefont {M.}~\bibnamefont
  {Gobbi}}, \bibinfo {author} {\bibfnamefont {E.}~\bibnamefont {Orgiu}},\ and\
  \bibinfo {author} {\bibfnamefont {P.}~\bibnamefont {Samor{\`\i}}},\
  }\bibfield  {title} {\bibinfo {title} {When 2d materials meet molecules:
  opportunities and challenges of hybrid organic/inorganic van der waals
  heterostructures},\ }\href@noop {} {\bibfield  {journal} {\bibinfo  {journal}
  {Adv. Mater.}\ }\textbf {\bibinfo {volume} {30}},\ \bibinfo {pages} {1706103}
  (\bibinfo {year} {2018}{\natexlab{a}})}\BibitemShut {NoStop}%
\bibitem [{\citenamefont {Wang}\ \emph {et~al.}(2021)\citenamefont {Wang},
  \citenamefont {Iglesias}, \citenamefont {Gali}, \citenamefont {Beljonne},\
  and\ \citenamefont {Samor{\`\i}}}]{Samori2021SAM}%
  \BibitemOpen
  \bibfield  {author} {\bibinfo {author} {\bibfnamefont {Y.}~\bibnamefont
  {Wang}}, \bibinfo {author} {\bibfnamefont {D.}~\bibnamefont {Iglesias}},
  \bibinfo {author} {\bibfnamefont {S.~M.}\ \bibnamefont {Gali}}, \bibinfo
  {author} {\bibfnamefont {D.}~\bibnamefont {Beljonne}},\ and\ \bibinfo
  {author} {\bibfnamefont {P.}~\bibnamefont {Samor{\`\i}}},\ }\bibfield
  {title} {\bibinfo {title} {Light-programmable logic-in-memory in 2d
  semiconductors enabled by supramolecular functionalization: Photoresponsive
  collective effect of aligned molecular dipoles},\ }\href@noop {} {\bibfield
  {journal} {\bibinfo  {journal} {ACS Nano}\ } (\bibinfo {year}
  {2021})}\BibitemShut {NoStop}%
\bibitem [{\citenamefont {Kotiuga}\ \emph {et~al.}(2015)\citenamefont
  {Kotiuga}, \citenamefont {Darancet}, \citenamefont {Arroyo}, \citenamefont
  {Venkataraman},\ and\ \citenamefont {Neaton}}]{Kotiuga2015NL}%
  \BibitemOpen
  \bibfield  {author} {\bibinfo {author} {\bibfnamefont {M.}~\bibnamefont
  {Kotiuga}}, \bibinfo {author} {\bibfnamefont {P.}~\bibnamefont {Darancet}},
  \bibinfo {author} {\bibfnamefont {C.~R.}\ \bibnamefont {Arroyo}}, \bibinfo
  {author} {\bibfnamefont {L.}~\bibnamefont {Venkataraman}},\ and\ \bibinfo
  {author} {\bibfnamefont {J.~B.}\ \bibnamefont {Neaton}},\ }\bibfield  {title}
  {\bibinfo {title} {Adsorption-induced solvent-based electrostatic gating of
  charge transport through molecular junctions},\ }\href@noop {} {\bibfield
  {journal} {\bibinfo  {journal} {Nano Lett.}\ }\textbf {\bibinfo {volume}
  {15}},\ \bibinfo {pages} {4498} (\bibinfo {year} {2015})}\BibitemShut
  {NoStop}%
\bibitem [{\citenamefont {Fatemi}\ \emph {et~al.}(2011)\citenamefont {Fatemi},
  \citenamefont {Kamenetska}, \citenamefont {Neaton},\ and\ \citenamefont
  {Venkataraman}}]{Venka2011NL}%
  \BibitemOpen
  \bibfield  {author} {\bibinfo {author} {\bibfnamefont {V.}~\bibnamefont
  {Fatemi}}, \bibinfo {author} {\bibfnamefont {M.}~\bibnamefont {Kamenetska}},
  \bibinfo {author} {\bibfnamefont {J.}~\bibnamefont {Neaton}},\ and\ \bibinfo
  {author} {\bibfnamefont {L.}~\bibnamefont {Venkataraman}},\ }\bibfield
  {title} {\bibinfo {title} {Environmental control of single-molecule junction
  transport},\ }\href@noop {} {\bibfield  {journal} {\bibinfo  {journal} {Nano
  Lett.}\ }\textbf {\bibinfo {volume} {11}},\ \bibinfo {pages} {1988} (\bibinfo
  {year} {2011})}\BibitemShut {NoStop}%
\bibitem [{\citenamefont {Monti}(2012)}]{Monti2012MolDipole}%
  \BibitemOpen
  \bibfield  {author} {\bibinfo {author} {\bibfnamefont {O.~L.}\ \bibnamefont
  {Monti}},\ }\bibfield  {title} {\bibinfo {title} {Understanding interfacial
  electronic structure and charge transfer: an electrostatic perspective},\
  }\href@noop {} {\bibfield  {journal} {\bibinfo  {journal} {J. Phys. Chem.
  Lett.}\ }\textbf {\bibinfo {volume} {3}},\ \bibinfo {pages} {2342} (\bibinfo
  {year} {2012})}\BibitemShut {NoStop}%
\bibitem [{\citenamefont {Wang}\ \emph {et~al.}(2008)\citenamefont {Wang},
  \citenamefont {Prodan}, \citenamefont {Car},\ and\ \citenamefont
  {Selloni}}]{Wang2008WF}%
  \BibitemOpen
  \bibfield  {author} {\bibinfo {author} {\bibfnamefont {J.-g.}\ \bibnamefont
  {Wang}}, \bibinfo {author} {\bibfnamefont {E.}~\bibnamefont {Prodan}},
  \bibinfo {author} {\bibfnamefont {R.}~\bibnamefont {Car}},\ and\ \bibinfo
  {author} {\bibfnamefont {A.}~\bibnamefont {Selloni}},\ }\bibfield  {title}
  {\bibinfo {title} {Band alignment in molecular devices: Influence of
  anchoring group and metal work function},\ }\href@noop {} {\bibfield
  {journal} {\bibinfo  {journal} {Phys. Rev. B}\ }\textbf {\bibinfo {volume}
  {77}},\ \bibinfo {pages} {245443} (\bibinfo {year} {2008})}\BibitemShut
  {NoStop}%
\bibitem [{\citenamefont {Capozzi}\ \emph {et~al.}(2015)\citenamefont
  {Capozzi}, \citenamefont {Xia}, \citenamefont {Adak}, \citenamefont {Dell},
  \citenamefont {Liu}, \citenamefont {Taylor}, \citenamefont {Neaton},
  \citenamefont {Campos},\ and\ \citenamefont
  {Venkataraman}}]{Venka2015NatNanotech}%
  \BibitemOpen
  \bibfield  {author} {\bibinfo {author} {\bibfnamefont {B.}~\bibnamefont
  {Capozzi}}, \bibinfo {author} {\bibfnamefont {J.}~\bibnamefont {Xia}},
  \bibinfo {author} {\bibfnamefont {O.}~\bibnamefont {Adak}}, \bibinfo {author}
  {\bibfnamefont {E.~J.}\ \bibnamefont {Dell}}, \bibinfo {author}
  {\bibfnamefont {Z.-F.}\ \bibnamefont {Liu}}, \bibinfo {author} {\bibfnamefont
  {J.~C.}\ \bibnamefont {Taylor}}, \bibinfo {author} {\bibfnamefont {J.~B.}\
  \bibnamefont {Neaton}}, \bibinfo {author} {\bibfnamefont {L.~M.}\
  \bibnamefont {Campos}},\ and\ \bibinfo {author} {\bibfnamefont
  {L.}~\bibnamefont {Venkataraman}},\ }\bibfield  {title} {\bibinfo {title}
  {Single-molecule diodes with high rectification ratios through environmental
  control},\ }\href@noop {} {\bibfield  {journal} {\bibinfo  {journal} {Nat.
  Nanotechnol.}\ }\textbf {\bibinfo {volume} {10}},\ \bibinfo {pages} {522}
  (\bibinfo {year} {2015})}\BibitemShut {NoStop}%
\bibitem [{\citenamefont {Rissner}\ \emph {et~al.}(2009)\citenamefont
  {Rissner}, \citenamefont {Rangger}, \citenamefont {Hofmann}, \citenamefont
  {Track}, \citenamefont {Heimel},\ and\ \citenamefont
  {Zojer}}]{Egbert2009SAMwf}%
  \BibitemOpen
  \bibfield  {author} {\bibinfo {author} {\bibfnamefont {F.}~\bibnamefont
  {Rissner}}, \bibinfo {author} {\bibfnamefont {G.~M.}\ \bibnamefont
  {Rangger}}, \bibinfo {author} {\bibfnamefont {O.~T.}\ \bibnamefont
  {Hofmann}}, \bibinfo {author} {\bibfnamefont {A.~M.}\ \bibnamefont {Track}},
  \bibinfo {author} {\bibfnamefont {G.}~\bibnamefont {Heimel}},\ and\ \bibinfo
  {author} {\bibfnamefont {E.}~\bibnamefont {Zojer}},\ }\bibfield  {title}
  {\bibinfo {title} {Understanding the electronic structure of
  metal/sam/organic- semiconductor heterojunctions},\ }\href@noop {} {\bibfield
   {journal} {\bibinfo  {journal} {ACS Nano}\ }\textbf {\bibinfo {volume}
  {3}},\ \bibinfo {pages} {3513} (\bibinfo {year} {2009})}\BibitemShut
  {NoStop}%
\bibitem [{\citenamefont {Zhou}\ \emph {et~al.}(2012)\citenamefont {Zhou},
  \citenamefont {Fuentes-Hernandez}, \citenamefont {Shim}, \citenamefont
  {Meyer}, \citenamefont {Giordano}, \citenamefont {Li}, \citenamefont
  {Winget}, \citenamefont {Papadopoulos}, \citenamefont {Cheun}, \citenamefont
  {Kim} \emph {et~al.}}]{Kahn2012DipWF}%
  \BibitemOpen
  \bibfield  {author} {\bibinfo {author} {\bibfnamefont {Y.}~\bibnamefont
  {Zhou}}, \bibinfo {author} {\bibfnamefont {C.}~\bibnamefont
  {Fuentes-Hernandez}}, \bibinfo {author} {\bibfnamefont {J.}~\bibnamefont
  {Shim}}, \bibinfo {author} {\bibfnamefont {J.}~\bibnamefont {Meyer}},
  \bibinfo {author} {\bibfnamefont {A.~J.}\ \bibnamefont {Giordano}}, \bibinfo
  {author} {\bibfnamefont {H.}~\bibnamefont {Li}}, \bibinfo {author}
  {\bibfnamefont {P.}~\bibnamefont {Winget}}, \bibinfo {author} {\bibfnamefont
  {T.}~\bibnamefont {Papadopoulos}}, \bibinfo {author} {\bibfnamefont
  {H.}~\bibnamefont {Cheun}}, \bibinfo {author} {\bibfnamefont
  {J.}~\bibnamefont {Kim}}, \emph {et~al.},\ }\bibfield  {title} {\bibinfo
  {title} {A universal method to produce low--work function electrodes for
  organic electronics},\ }\href@noop {} {\bibfield  {journal} {\bibinfo
  {journal} {Science}\ }\textbf {\bibinfo {volume} {336}},\ \bibinfo {pages}
  {327} (\bibinfo {year} {2012})}\BibitemShut {NoStop}%
\bibitem [{\citenamefont {Kahn}\ \emph {et~al.}(2003)\citenamefont {Kahn},
  \citenamefont {Koch},\ and\ \citenamefont {Gao}}]{Kahn2003SAM}%
  \BibitemOpen
  \bibfield  {author} {\bibinfo {author} {\bibfnamefont {A.}~\bibnamefont
  {Kahn}}, \bibinfo {author} {\bibfnamefont {N.}~\bibnamefont {Koch}},\ and\
  \bibinfo {author} {\bibfnamefont {W.}~\bibnamefont {Gao}},\ }\bibfield
  {title} {\bibinfo {title} {Electronic structure and electrical properties of
  interfaces between metals and $\pi$-conjugated molecular films},\ }\href@noop
  {} {\bibfield  {journal} {\bibinfo  {journal} {J. Polym. Sci. Pol. Phys.}\
  }\textbf {\bibinfo {volume} {41}},\ \bibinfo {pages} {2529} (\bibinfo {year}
  {2003})}\BibitemShut {NoStop}%
\bibitem [{\citenamefont {De~Renzi}\ \emph {et~al.}(2005)\citenamefont
  {De~Renzi}, \citenamefont {Rousseau}, \citenamefont {Marchetto},
  \citenamefont {Biagi}, \citenamefont {Scandolo},\ and\ \citenamefont
  {Del~Pennino}}]{De2005PRLWF}%
  \BibitemOpen
  \bibfield  {author} {\bibinfo {author} {\bibfnamefont {V.}~\bibnamefont
  {De~Renzi}}, \bibinfo {author} {\bibfnamefont {R.}~\bibnamefont {Rousseau}},
  \bibinfo {author} {\bibfnamefont {D.}~\bibnamefont {Marchetto}}, \bibinfo
  {author} {\bibfnamefont {R.}~\bibnamefont {Biagi}}, \bibinfo {author}
  {\bibfnamefont {S.}~\bibnamefont {Scandolo}},\ and\ \bibinfo {author}
  {\bibfnamefont {U.}~\bibnamefont {Del~Pennino}},\ }\bibfield  {title}
  {\bibinfo {title} {Metal work-function changes induced by organic adsorbates:
  A combined experimental and theoretical study},\ }\href@noop {} {\bibfield
  {journal} {\bibinfo  {journal} {Phys. Rev. Lett.}\ }\textbf {\bibinfo
  {volume} {95}},\ \bibinfo {pages} {046804} (\bibinfo {year}
  {2005})}\BibitemShut {NoStop}%
\bibitem [{\citenamefont {Goronzy}\ \emph {et~al.}(2018)\citenamefont
  {Goronzy}, \citenamefont {Ebrahimi}, \citenamefont {Rosei}, \citenamefont
  {Arramel}, \citenamefont {Fang}, \citenamefont {De~Feyter}, \citenamefont
  {Tait}, \citenamefont {Wang}, \citenamefont {Beton}, \citenamefont {Wee}
  \emph {et~al.}}]{Goronzy2018SAMrev}%
  \BibitemOpen
  \bibfield  {author} {\bibinfo {author} {\bibfnamefont {D.~P.}\ \bibnamefont
  {Goronzy}}, \bibinfo {author} {\bibfnamefont {M.}~\bibnamefont {Ebrahimi}},
  \bibinfo {author} {\bibfnamefont {F.}~\bibnamefont {Rosei}}, \bibinfo
  {author} {\bibnamefont {Arramel}}, \bibinfo {author} {\bibfnamefont
  {Y.}~\bibnamefont {Fang}}, \bibinfo {author} {\bibfnamefont {S.}~\bibnamefont
  {De~Feyter}}, \bibinfo {author} {\bibfnamefont {S.~L.}\ \bibnamefont {Tait}},
  \bibinfo {author} {\bibfnamefont {C.}~\bibnamefont {Wang}}, \bibinfo {author}
  {\bibfnamefont {P.~H.}\ \bibnamefont {Beton}}, \bibinfo {author}
  {\bibfnamefont {A.~T.}\ \bibnamefont {Wee}}, \emph {et~al.},\ }\bibfield
  {title} {\bibinfo {title} {Supramolecular assemblies on surfaces:
  nanopatterning, functionality, and reactivity},\ }\href@noop {} {\bibfield
  {journal} {\bibinfo  {journal} {ACS Nano}\ }\textbf {\bibinfo {volume}
  {12}},\ \bibinfo {pages} {7445} (\bibinfo {year} {2018})}\BibitemShut
  {NoStop}%
\bibitem [{\citenamefont {Gobbi}\ \emph
  {et~al.}(2018{\natexlab{b}})\citenamefont {Gobbi}, \citenamefont {Bonacchi},
  \citenamefont {Lian}, \citenamefont {Vercouter}, \citenamefont {Bertolazzi},
  \citenamefont {Zyska}, \citenamefont {Timpel}, \citenamefont {Tatti},
  \citenamefont {Olivier}, \citenamefont {Hecht} \emph
  {et~al.}}]{Gobbi2018SAM}%
  \BibitemOpen
  \bibfield  {author} {\bibinfo {author} {\bibfnamefont {M.}~\bibnamefont
  {Gobbi}}, \bibinfo {author} {\bibfnamefont {S.}~\bibnamefont {Bonacchi}},
  \bibinfo {author} {\bibfnamefont {J.~X.}\ \bibnamefont {Lian}}, \bibinfo
  {author} {\bibfnamefont {A.}~\bibnamefont {Vercouter}}, \bibinfo {author}
  {\bibfnamefont {S.}~\bibnamefont {Bertolazzi}}, \bibinfo {author}
  {\bibfnamefont {B.}~\bibnamefont {Zyska}}, \bibinfo {author} {\bibfnamefont
  {M.}~\bibnamefont {Timpel}}, \bibinfo {author} {\bibfnamefont
  {R.}~\bibnamefont {Tatti}}, \bibinfo {author} {\bibfnamefont
  {Y.}~\bibnamefont {Olivier}}, \bibinfo {author} {\bibfnamefont
  {S.}~\bibnamefont {Hecht}}, \emph {et~al.},\ }\bibfield  {title} {\bibinfo
  {title} {Collective molecular switching in hybrid superlattices for
  light-modulated two-dimensional electronics},\ }\href@noop {} {\bibfield
  {journal} {\bibinfo  {journal} {Nat. Commun.}\ }\textbf {\bibinfo {volume}
  {9}},\ \bibinfo {pages} {1} (\bibinfo {year}
  {2018}{\natexlab{b}})}\BibitemShut {NoStop}%
\bibitem [{\citenamefont {Zhou}\ \emph {et~al.}(2019)\citenamefont {Zhou},
  \citenamefont {Cho}, \citenamefont {Yang}, \citenamefont {Weiss},
  \citenamefont {Berkelbach},\ and\ \citenamefont {Darancet}}]{Zhou2019NPL}%
  \BibitemOpen
  \bibfield  {author} {\bibinfo {author} {\bibfnamefont {Q.}~\bibnamefont
  {Zhou}}, \bibinfo {author} {\bibfnamefont {Y.}~\bibnamefont {Cho}}, \bibinfo
  {author} {\bibfnamefont {S.}~\bibnamefont {Yang}}, \bibinfo {author}
  {\bibfnamefont {E.~A.}\ \bibnamefont {Weiss}}, \bibinfo {author}
  {\bibfnamefont {T.~C.}\ \bibnamefont {Berkelbach}},\ and\ \bibinfo {author}
  {\bibfnamefont {P.}~\bibnamefont {Darancet}},\ }\bibfield  {title} {\bibinfo
  {title} {Large band edge tunability in colloidal nanoplatelets},\ }\href@noop
  {} {\bibfield  {journal} {\bibinfo  {journal} {Nano Lett.}\ }\textbf
  {\bibinfo {volume} {19}},\ \bibinfo {pages} {7124} (\bibinfo {year}
  {2019})}\BibitemShut {NoStop}%
\bibitem [{\citenamefont {Crommie}\ \emph {et~al.}(1993)\citenamefont
  {Crommie}, \citenamefont {Lutz},\ and\ \citenamefont
  {Eigler}}]{Crommie1993Corral}%
  \BibitemOpen
  \bibfield  {author} {\bibinfo {author} {\bibfnamefont {M.~F.}\ \bibnamefont
  {Crommie}}, \bibinfo {author} {\bibfnamefont {C.~P.}\ \bibnamefont {Lutz}},\
  and\ \bibinfo {author} {\bibfnamefont {D.~M.}\ \bibnamefont {Eigler}},\
  }\bibfield  {title} {\bibinfo {title} {Confinement of electrons to quantum
  corrals on a metal surface},\ }\href@noop {} {\bibfield  {journal} {\bibinfo
  {journal} {Science}\ }\textbf {\bibinfo {volume} {262}},\ \bibinfo {pages}
  {218} (\bibinfo {year} {1993})}\BibitemShut {NoStop}%
\bibitem [{\citenamefont {Pennec}\ \emph {et~al.}(2007)\citenamefont {Pennec},
  \citenamefont {Auw{\"a}rter}, \citenamefont {Schiffrin}, \citenamefont
  {Weber-Bargioni}, \citenamefont {Riemann},\ and\ \citenamefont
  {Barth}}]{Barth2007NatNano}%
  \BibitemOpen
  \bibfield  {author} {\bibinfo {author} {\bibfnamefont {Y.}~\bibnamefont
  {Pennec}}, \bibinfo {author} {\bibfnamefont {W.}~\bibnamefont
  {Auw{\"a}rter}}, \bibinfo {author} {\bibfnamefont {A.}~\bibnamefont
  {Schiffrin}}, \bibinfo {author} {\bibfnamefont {A.}~\bibnamefont
  {Weber-Bargioni}}, \bibinfo {author} {\bibfnamefont {A.}~\bibnamefont
  {Riemann}},\ and\ \bibinfo {author} {\bibfnamefont {J.}~\bibnamefont
  {Barth}},\ }\bibfield  {title} {\bibinfo {title} {Supramolecular gratings for
  tuneable confinement of electrons on metal surfaces},\ }\href@noop {}
  {\bibfield  {journal} {\bibinfo  {journal} {Nat. Nanotechnol.}\ }\textbf
  {\bibinfo {volume} {2}},\ \bibinfo {pages} {99} (\bibinfo {year}
  {2007})}\BibitemShut {NoStop}%
\bibitem [{\citenamefont {Kawai}\ \emph {et~al.}(2021)\citenamefont {Kawai},
  \citenamefont {Kher-Elden}, \citenamefont {Sadeghi}, \citenamefont {Abd
  El-Fattah}, \citenamefont {Sun}, \citenamefont {Izumi}, \citenamefont
  {Minakata}, \citenamefont {Takeda},\ and\ \citenamefont
  {Lobo-Checa}}]{Kawai2021SAMsurf}%
  \BibitemOpen
  \bibfield  {author} {\bibinfo {author} {\bibfnamefont {S.}~\bibnamefont
  {Kawai}}, \bibinfo {author} {\bibfnamefont {M.~A.}\ \bibnamefont
  {Kher-Elden}}, \bibinfo {author} {\bibfnamefont {A.}~\bibnamefont {Sadeghi}},
  \bibinfo {author} {\bibfnamefont {Z.~M.}\ \bibnamefont {Abd El-Fattah}},
  \bibinfo {author} {\bibfnamefont {K.}~\bibnamefont {Sun}}, \bibinfo {author}
  {\bibfnamefont {S.}~\bibnamefont {Izumi}}, \bibinfo {author} {\bibfnamefont
  {S.}~\bibnamefont {Minakata}}, \bibinfo {author} {\bibfnamefont
  {Y.}~\bibnamefont {Takeda}},\ and\ \bibinfo {author} {\bibfnamefont
  {J.}~\bibnamefont {Lobo-Checa}},\ }\bibfield  {title} {\bibinfo {title} {Near
  fermi superatom state stabilized by surface state resonances in a multiporous
  molecular network},\ }\href@noop {} {\bibfield  {journal} {\bibinfo
  {journal} {Nano Lett.}\ } (\bibinfo {year} {2021})}\BibitemShut {NoStop}%
\bibitem [{\citenamefont {Gobbi}\ \emph {et~al.}(2017)\citenamefont {Gobbi},
  \citenamefont {Bonacchi}, \citenamefont {Lian}, \citenamefont {Liu},
  \citenamefont {Wang}, \citenamefont {Stoeckel}, \citenamefont {Squillaci},
  \citenamefont {D’avino}, \citenamefont {Narita}, \citenamefont {M{\"u}llen}
  \emph {et~al.}}]{Samori2017SAM}%
  \BibitemOpen
  \bibfield  {author} {\bibinfo {author} {\bibfnamefont {M.}~\bibnamefont
  {Gobbi}}, \bibinfo {author} {\bibfnamefont {S.}~\bibnamefont {Bonacchi}},
  \bibinfo {author} {\bibfnamefont {J.~X.}\ \bibnamefont {Lian}}, \bibinfo
  {author} {\bibfnamefont {Y.}~\bibnamefont {Liu}}, \bibinfo {author}
  {\bibfnamefont {X.-Y.}\ \bibnamefont {Wang}}, \bibinfo {author}
  {\bibfnamefont {M.-A.}\ \bibnamefont {Stoeckel}}, \bibinfo {author}
  {\bibfnamefont {M.~A.}\ \bibnamefont {Squillaci}}, \bibinfo {author}
  {\bibfnamefont {G.}~\bibnamefont {D’avino}}, \bibinfo {author}
  {\bibfnamefont {A.}~\bibnamefont {Narita}}, \bibinfo {author} {\bibfnamefont
  {K.}~\bibnamefont {M{\"u}llen}}, \emph {et~al.},\ }\bibfield  {title}
  {\bibinfo {title} {Periodic potentials in hybrid van der waals
  heterostructures formed by supramolecular lattices on graphene},\ }\href@noop
  {} {\bibfield  {journal} {\bibinfo  {journal} {Nat. Commun.}\ }\textbf
  {\bibinfo {volume} {8}},\ \bibinfo {pages} {1} (\bibinfo {year}
  {2017})}\BibitemShut {NoStop}%
\bibitem [{\citenamefont {Xia}\ \emph {et~al.}(2016)\citenamefont {Xia},
  \citenamefont {Leonardi}, \citenamefont {Gobbi}, \citenamefont {Liu},
  \citenamefont {Bellani}, \citenamefont {Liscio}, \citenamefont {Kovtun},
  \citenamefont {Li}, \citenamefont {Feng}, \citenamefont {Orgiu} \emph
  {et~al.}}]{Samori2016Gr1D}%
  \BibitemOpen
  \bibfield  {author} {\bibinfo {author} {\bibfnamefont {Z.}~\bibnamefont
  {Xia}}, \bibinfo {author} {\bibfnamefont {F.}~\bibnamefont {Leonardi}},
  \bibinfo {author} {\bibfnamefont {M.}~\bibnamefont {Gobbi}}, \bibinfo
  {author} {\bibfnamefont {Y.}~\bibnamefont {Liu}}, \bibinfo {author}
  {\bibfnamefont {V.}~\bibnamefont {Bellani}}, \bibinfo {author} {\bibfnamefont
  {A.}~\bibnamefont {Liscio}}, \bibinfo {author} {\bibfnamefont
  {A.}~\bibnamefont {Kovtun}}, \bibinfo {author} {\bibfnamefont
  {R.}~\bibnamefont {Li}}, \bibinfo {author} {\bibfnamefont {X.}~\bibnamefont
  {Feng}}, \bibinfo {author} {\bibfnamefont {E.}~\bibnamefont {Orgiu}}, \emph
  {et~al.},\ }\bibfield  {title} {\bibinfo {title} {Electrochemical
  functionalization of graphene at the nanoscale with self-assembling diazonium
  salts},\ }\href@noop {} {\bibfield  {journal} {\bibinfo  {journal} {ACS
  Nano}\ }\textbf {\bibinfo {volume} {10}},\ \bibinfo {pages} {7125} (\bibinfo
  {year} {2016})}\BibitemShut {NoStop}%
\bibitem [{\citenamefont {Shayeganfar}\ and\ \citenamefont
  {Rochefort}(2014{\natexlab{a}})}]{2014TMAGr}%
  \BibitemOpen
  \bibfield  {author} {\bibinfo {author} {\bibfnamefont {F.}~\bibnamefont
  {Shayeganfar}}\ and\ \bibinfo {author} {\bibfnamefont {A.}~\bibnamefont
  {Rochefort}},\ }\bibfield  {title} {\bibinfo {title} {Electronic properties
  of self-assembled trimesic acid monolayer on graphene},\ }\href@noop {}
  {\bibfield  {journal} {\bibinfo  {journal} {Langmuir}\ }\textbf {\bibinfo
  {volume} {30}},\ \bibinfo {pages} {9707} (\bibinfo {year}
  {2014}{\natexlab{a}})}\BibitemShut {NoStop}%
\bibitem [{\citenamefont {Georgakilas}\ \emph {et~al.}(2016)\citenamefont
  {Georgakilas}, \citenamefont {Tiwari}, \citenamefont {Kemp}, \citenamefont
  {Perman}, \citenamefont {Bourlinos}, \citenamefont {Kim},\ and\ \citenamefont
  {Zboril}}]{Zboril2016SurfGrRev}%
  \BibitemOpen
  \bibfield  {author} {\bibinfo {author} {\bibfnamefont {V.}~\bibnamefont
  {Georgakilas}}, \bibinfo {author} {\bibfnamefont {J.~N.}\ \bibnamefont
  {Tiwari}}, \bibinfo {author} {\bibfnamefont {K.~C.}\ \bibnamefont {Kemp}},
  \bibinfo {author} {\bibfnamefont {J.~A.}\ \bibnamefont {Perman}}, \bibinfo
  {author} {\bibfnamefont {A.~B.}\ \bibnamefont {Bourlinos}}, \bibinfo {author}
  {\bibfnamefont {K.~S.}\ \bibnamefont {Kim}},\ and\ \bibinfo {author}
  {\bibfnamefont {R.}~\bibnamefont {Zboril}},\ }\bibfield  {title} {\bibinfo
  {title} {Noncovalent functionalization of graphene and graphene oxide for
  energy materials, biosensing, catalytic, and biomedical applications},\
  }\href@noop {} {\bibfield  {journal} {\bibinfo  {journal} {Chem. Rev.}\
  }\textbf {\bibinfo {volume} {116}},\ \bibinfo {pages} {5464} (\bibinfo {year}
  {2016})}\BibitemShut {NoStop}%
\bibitem [{\citenamefont {Shayeganfar}\ and\ \citenamefont
  {Rochefort}(2014{\natexlab{b}})}]{shayeganfar2014EgTMA}%
  \BibitemOpen
  \bibfield  {author} {\bibinfo {author} {\bibfnamefont {F.}~\bibnamefont
  {Shayeganfar}}\ and\ \bibinfo {author} {\bibfnamefont {A.}~\bibnamefont
  {Rochefort}},\ }\bibfield  {title} {\bibinfo {title} {Electronic properties
  of self-assembled trimesic acid monolayer on graphene},\ }\href@noop {}
  {\bibfield  {journal} {\bibinfo  {journal} {Langmuir}\ }\textbf {\bibinfo
  {volume} {30}},\ \bibinfo {pages} {9707} (\bibinfo {year}
  {2014}{\natexlab{b}})}\BibitemShut {NoStop}%
\bibitem [{\citenamefont {Shayeganfar}(2015)}]{shayeganfar2015EgF2}%
  \BibitemOpen
  \bibfield  {author} {\bibinfo {author} {\bibfnamefont {F.}~\bibnamefont
  {Shayeganfar}},\ }\bibfield  {title} {\bibinfo {title} {Energy gap tuning of
  graphene layers with single molecular f2 adsorption},\ }\href@noop {}
  {\bibfield  {journal} {\bibinfo  {journal} {J. Phys. Chem. C}\ }\textbf
  {\bibinfo {volume} {119}},\ \bibinfo {pages} {12681} (\bibinfo {year}
  {2015})}\BibitemShut {NoStop}%
\bibitem [{\citenamefont {Xiu}\ \emph {et~al.}(2014)\citenamefont {Xiu},
  \citenamefont {Gong}, \citenamefont {Wang}, \citenamefont {Liang},
  \citenamefont {Chen},\ and\ \citenamefont {Kawazoe}}]{xiu2014Gr}%
  \BibitemOpen
  \bibfield  {author} {\bibinfo {author} {\bibfnamefont {S.}~\bibnamefont
  {Xiu}}, \bibinfo {author} {\bibfnamefont {L.}~\bibnamefont {Gong}}, \bibinfo
  {author} {\bibfnamefont {V.}~\bibnamefont {Wang}}, \bibinfo {author}
  {\bibfnamefont {Y.}~\bibnamefont {Liang}}, \bibinfo {author} {\bibfnamefont
  {G.}~\bibnamefont {Chen}},\ and\ \bibinfo {author} {\bibfnamefont
  {Y.}~\bibnamefont {Kawazoe}},\ }\bibfield  {title} {\bibinfo {title}
  {Degenerate perturbation in band-gap opening of graphene superlattice},\
  }\href@noop {} {\bibfield  {journal} {\bibinfo  {journal} {J. Phys. Chem. C}\
  }\textbf {\bibinfo {volume} {118}},\ \bibinfo {pages} {8174} (\bibinfo {year}
  {2014})}\BibitemShut {NoStop}%
\bibitem [{\citenamefont {Chang}\ \emph {et~al.}(2012)\citenamefont {Chang},
  \citenamefont {Fan}, \citenamefont {Li},\ and\ \citenamefont
  {Kuo}}]{Kuo2012GrPh}%
  \BibitemOpen
  \bibfield  {author} {\bibinfo {author} {\bibfnamefont {C.-H.}\ \bibnamefont
  {Chang}}, \bibinfo {author} {\bibfnamefont {X.}~\bibnamefont {Fan}}, \bibinfo
  {author} {\bibfnamefont {L.-J.}\ \bibnamefont {Li}},\ and\ \bibinfo {author}
  {\bibfnamefont {J.-L.}\ \bibnamefont {Kuo}},\ }\bibfield  {title} {\bibinfo
  {title} {Band gap tuning of graphene by adsorption of aromatic molecules},\
  }\href@noop {} {\bibfield  {journal} {\bibinfo  {journal} {J. Phys. Chem. C}\
  }\textbf {\bibinfo {volume} {116}},\ \bibinfo {pages} {13788} (\bibinfo
  {year} {2012})}\BibitemShut {NoStop}%
\bibitem [{\citenamefont {Natan}\ \emph {et~al.}(2007)\citenamefont {Natan},
  \citenamefont {Kronik}, \citenamefont {Haick},\ and\ \citenamefont
  {Tung}}]{Natan2007SAMrev}%
  \BibitemOpen
  \bibfield  {author} {\bibinfo {author} {\bibfnamefont {A.}~\bibnamefont
  {Natan}}, \bibinfo {author} {\bibfnamefont {L.}~\bibnamefont {Kronik}},
  \bibinfo {author} {\bibfnamefont {H.}~\bibnamefont {Haick}},\ and\ \bibinfo
  {author} {\bibfnamefont {R.~T.}\ \bibnamefont {Tung}},\ }\bibfield  {title}
  {\bibinfo {title} {Electrostatic properties of ideal and non-ideal polar
  organic monolayers: Implications for electronic devices},\ }\href@noop {}
  {\bibfield  {journal} {\bibinfo  {journal} {Adv. Mater.}\ }\textbf {\bibinfo
  {volume} {19}},\ \bibinfo {pages} {4103} (\bibinfo {year}
  {2007})}\BibitemShut {NoStop}%
\bibitem [{\citenamefont {Mart{\'\i}nez-Blanco}\ \emph
  {et~al.}(2015)\citenamefont {Mart{\'\i}nez-Blanco}, \citenamefont {Nacci},
  \citenamefont {Erwin}, \citenamefont {Kanisawa}, \citenamefont {Locane},
  \citenamefont {Thomas}, \citenamefont {Von~Oppen}, \citenamefont {Brouwer},\
  and\ \citenamefont {F{\"o}lsch}}]{Martinez2015gating}%
  \BibitemOpen
  \bibfield  {author} {\bibinfo {author} {\bibfnamefont {J.}~\bibnamefont
  {Mart{\'\i}nez-Blanco}}, \bibinfo {author} {\bibfnamefont {C.}~\bibnamefont
  {Nacci}}, \bibinfo {author} {\bibfnamefont {S.~C.}\ \bibnamefont {Erwin}},
  \bibinfo {author} {\bibfnamefont {K.}~\bibnamefont {Kanisawa}}, \bibinfo
  {author} {\bibfnamefont {E.}~\bibnamefont {Locane}}, \bibinfo {author}
  {\bibfnamefont {M.}~\bibnamefont {Thomas}}, \bibinfo {author} {\bibfnamefont
  {F.}~\bibnamefont {Von~Oppen}}, \bibinfo {author} {\bibfnamefont {P.~W.}\
  \bibnamefont {Brouwer}},\ and\ \bibinfo {author} {\bibfnamefont
  {S.}~\bibnamefont {F{\"o}lsch}},\ }\bibfield  {title} {\bibinfo {title}
  {Gating a single-molecule transistor with individual atoms},\ }\href@noop {}
  {\bibfield  {journal} {\bibinfo  {journal} {Nat. Phys.}\ }\textbf {\bibinfo
  {volume} {11}},\ \bibinfo {pages} {640} (\bibinfo {year} {2015})}\BibitemShut
  {NoStop}%
\bibitem [{\citenamefont {Li}\ \emph {et~al.}(2021)\citenamefont {Li},
  \citenamefont {Kolluru}, \citenamefont {Rahn}, \citenamefont {Schwenker},
  \citenamefont {Li}, \citenamefont {Hennig}, \citenamefont {Darancet},
  \citenamefont {Chan},\ and\ \citenamefont {Hersam}}]{Li2021Boroph}%
  \BibitemOpen
  \bibfield  {author} {\bibinfo {author} {\bibfnamefont {Q.}~\bibnamefont
  {Li}}, \bibinfo {author} {\bibfnamefont {V.~S.~C.}\ \bibnamefont {Kolluru}},
  \bibinfo {author} {\bibfnamefont {M.~S.}\ \bibnamefont {Rahn}}, \bibinfo
  {author} {\bibfnamefont {E.}~\bibnamefont {Schwenker}}, \bibinfo {author}
  {\bibfnamefont {S.}~\bibnamefont {Li}}, \bibinfo {author} {\bibfnamefont
  {R.~G.}\ \bibnamefont {Hennig}}, \bibinfo {author} {\bibfnamefont
  {P.}~\bibnamefont {Darancet}}, \bibinfo {author} {\bibfnamefont {M.~K.}\
  \bibnamefont {Chan}},\ and\ \bibinfo {author} {\bibfnamefont {M.~C.}\
  \bibnamefont {Hersam}},\ }\bibfield  {title} {\bibinfo {title} {Synthesis of
  borophane polymorphs through hydrogenation of borophene},\ }\href@noop {}
  {\bibfield  {journal} {\bibinfo  {journal} {Science}\ }\textbf {\bibinfo
  {volume} {371}},\ \bibinfo {pages} {1143} (\bibinfo {year}
  {2021})}\BibitemShut {NoStop}%
\bibitem [{\citenamefont {Choi}\ \emph {et~al.}(2008)\citenamefont {Choi},
  \citenamefont {Kang}, \citenamefont {Ryang},\ and\ \citenamefont
  {Yeom}}]{Choi2008BandsAtom}%
  \BibitemOpen
  \bibfield  {author} {\bibinfo {author} {\bibfnamefont {W.~H.}\ \bibnamefont
  {Choi}}, \bibinfo {author} {\bibfnamefont {P.~G.}\ \bibnamefont {Kang}},
  \bibinfo {author} {\bibfnamefont {K.~D.}\ \bibnamefont {Ryang}},\ and\
  \bibinfo {author} {\bibfnamefont {H.~W.}\ \bibnamefont {Yeom}},\ }\bibfield
  {title} {\bibinfo {title} {Band-structure engineering of gold atomic wires on
  silicon by controlled doping},\ }\href@noop {} {\bibfield  {journal}
  {\bibinfo  {journal} {Phys. Rev. Lett.}\ }\textbf {\bibinfo {volume} {100}},\
  \bibinfo {pages} {126801} (\bibinfo {year} {2008})}\BibitemShut {NoStop}%
\bibitem [{\citenamefont {Park}\ \emph
  {et~al.}(2008{\natexlab{a}})\citenamefont {Park}, \citenamefont {Yang},
  \citenamefont {Son}, \citenamefont {Cohen},\ and\ \citenamefont
  {Louie}}]{Park2008NatPhys}%
  \BibitemOpen
  \bibfield  {author} {\bibinfo {author} {\bibfnamefont {C.-H.}\ \bibnamefont
  {Park}}, \bibinfo {author} {\bibfnamefont {L.}~\bibnamefont {Yang}}, \bibinfo
  {author} {\bibfnamefont {Y.-W.}\ \bibnamefont {Son}}, \bibinfo {author}
  {\bibfnamefont {M.~L.}\ \bibnamefont {Cohen}},\ and\ \bibinfo {author}
  {\bibfnamefont {S.~G.}\ \bibnamefont {Louie}},\ }\bibfield  {title} {\bibinfo
  {title} {Anisotropic behaviours of massless dirac fermions in graphene under
  periodic potentials},\ }\href@noop {} {\bibfield  {journal} {\bibinfo
  {journal} {Nat. Phys.}\ }\textbf {\bibinfo {volume} {4}},\ \bibinfo {pages}
  {213} (\bibinfo {year} {2008}{\natexlab{a}})}\BibitemShut {NoStop}%
\bibitem [{\citenamefont {Park}\ \emph
  {et~al.}(2008{\natexlab{b}})\citenamefont {Park}, \citenamefont {Yang},
  \citenamefont {Son}, \citenamefont {Cohen},\ and\ \citenamefont
  {Louie}}]{Park2008PRL}%
  \BibitemOpen
  \bibfield  {author} {\bibinfo {author} {\bibfnamefont {C.-H.}\ \bibnamefont
  {Park}}, \bibinfo {author} {\bibfnamefont {L.}~\bibnamefont {Yang}}, \bibinfo
  {author} {\bibfnamefont {Y.-W.}\ \bibnamefont {Son}}, \bibinfo {author}
  {\bibfnamefont {M.~L.}\ \bibnamefont {Cohen}},\ and\ \bibinfo {author}
  {\bibfnamefont {S.~G.}\ \bibnamefont {Louie}},\ }\bibfield  {title} {\bibinfo
  {title} {New generation of massless dirac fermions in graphene under external
  periodic potentials},\ }\href@noop {} {\bibfield  {journal} {\bibinfo
  {journal} {Phys. Rev. Lett.}\ }\textbf {\bibinfo {volume} {101}},\ \bibinfo
  {pages} {126804} (\bibinfo {year} {2008}{\natexlab{b}})}\BibitemShut
  {NoStop}%
\bibitem [{\citenamefont {Gomes}\ \emph {et~al.}(2012)\citenamefont {Gomes},
  \citenamefont {Mar}, \citenamefont {Ko}, \citenamefont {Guinea},\ and\
  \citenamefont {Manoharan}}]{gomes2012designer}%
  \BibitemOpen
  \bibfield  {author} {\bibinfo {author} {\bibfnamefont {K.~K.}\ \bibnamefont
  {Gomes}}, \bibinfo {author} {\bibfnamefont {W.}~\bibnamefont {Mar}}, \bibinfo
  {author} {\bibfnamefont {W.}~\bibnamefont {Ko}}, \bibinfo {author}
  {\bibfnamefont {F.}~\bibnamefont {Guinea}},\ and\ \bibinfo {author}
  {\bibfnamefont {H.~C.}\ \bibnamefont {Manoharan}},\ }\bibfield  {title}
  {\bibinfo {title} {Designer dirac fermions and topological phases in
  molecular graphene},\ }\href@noop {} {\bibfield  {journal} {\bibinfo
  {journal} {Nature}\ }\textbf {\bibinfo {volume} {483}},\ \bibinfo {pages}
  {306} (\bibinfo {year} {2012})}\BibitemShut {NoStop}%
\bibitem [{\citenamefont {Drost}\ \emph {et~al.}(2017)\citenamefont {Drost},
  \citenamefont {Ojanen}, \citenamefont {Harju},\ and\ \citenamefont
  {Liljeroth}}]{Liljeroth2017topological}%
  \BibitemOpen
  \bibfield  {author} {\bibinfo {author} {\bibfnamefont {R.}~\bibnamefont
  {Drost}}, \bibinfo {author} {\bibfnamefont {T.}~\bibnamefont {Ojanen}},
  \bibinfo {author} {\bibfnamefont {A.}~\bibnamefont {Harju}},\ and\ \bibinfo
  {author} {\bibfnamefont {P.}~\bibnamefont {Liljeroth}},\ }\bibfield  {title}
  {\bibinfo {title} {Topological states in engineered atomic lattices},\
  }\href@noop {} {\bibfield  {journal} {\bibinfo  {journal} {Nat. Phys.}\
  }\textbf {\bibinfo {volume} {13}},\ \bibinfo {pages} {668} (\bibinfo {year}
  {2017})}\BibitemShut {NoStop}%
\bibitem [{\citenamefont {Yan}\ and\ \citenamefont
  {Liljeroth}(2019)}]{Liljeroth2019review}%
  \BibitemOpen
  \bibfield  {author} {\bibinfo {author} {\bibfnamefont {L.}~\bibnamefont
  {Yan}}\ and\ \bibinfo {author} {\bibfnamefont {P.}~\bibnamefont
  {Liljeroth}},\ }\bibfield  {title} {\bibinfo {title} {Engineered electronic
  states in atomically precise artificial lattices and graphene nanoribbons},\
  }\href@noop {} {\bibfield  {journal} {\bibinfo  {journal} {Adv. Phys. X}\
  }\textbf {\bibinfo {volume} {4}},\ \bibinfo {pages} {1651672} (\bibinfo
  {year} {2019})}\BibitemShut {NoStop}%
\bibitem [{\citenamefont {Trainer}\ \emph {et~al.}(2021)\citenamefont
  {Trainer}, \citenamefont {Srinivasan}, \citenamefont {Fisher}, \citenamefont
  {Zhang}, \citenamefont {Pfeiffer}, \citenamefont {Hla}, \citenamefont
  {Darancet},\ and\ \citenamefont {Guisinger}}]{Guisinger2021Topology}%
  \BibitemOpen
  \bibfield  {author} {\bibinfo {author} {\bibfnamefont {D.~J.}\ \bibnamefont
  {Trainer}}, \bibinfo {author} {\bibfnamefont {S.}~\bibnamefont {Srinivasan}},
  \bibinfo {author} {\bibfnamefont {B.~L.}\ \bibnamefont {Fisher}}, \bibinfo
  {author} {\bibfnamefont {Y.}~\bibnamefont {Zhang}}, \bibinfo {author}
  {\bibfnamefont {C.~R.}\ \bibnamefont {Pfeiffer}}, \bibinfo {author}
  {\bibfnamefont {S.-W.}\ \bibnamefont {Hla}}, \bibinfo {author} {\bibfnamefont
  {P.}~\bibnamefont {Darancet}},\ and\ \bibinfo {author} {\bibfnamefont
  {N.~P.}\ \bibnamefont {Guisinger}},\ }\bibfield  {title} {\bibinfo {title}
  {Manipulating topology in tailored artificial graphene nanoribbons},\
  }\href@noop {} {\bibfield  {journal} {\bibinfo  {journal} {arXiv preprint
  arXiv:2104.11334}\ } (\bibinfo {year} {2021})}\BibitemShut {NoStop}%
\bibitem [{\citenamefont {Liu}\ \emph {et~al.}(2019)\citenamefont {Liu},
  \citenamefont {Balla}, \citenamefont {Sangwan}, \citenamefont {Usta},
  \citenamefont {Facchetti}, \citenamefont {Marks},\ and\ \citenamefont
  {Hersam}}]{Hersam2019C8btbt}%
  \BibitemOpen
  \bibfield  {author} {\bibinfo {author} {\bibfnamefont {X.}~\bibnamefont
  {Liu}}, \bibinfo {author} {\bibfnamefont {I.}~\bibnamefont {Balla}}, \bibinfo
  {author} {\bibfnamefont {V.~K.}\ \bibnamefont {Sangwan}}, \bibinfo {author}
  {\bibfnamefont {H.}~\bibnamefont {Usta}}, \bibinfo {author} {\bibfnamefont
  {A.}~\bibnamefont {Facchetti}}, \bibinfo {author} {\bibfnamefont {T.~J.}\
  \bibnamefont {Marks}},\ and\ \bibinfo {author} {\bibfnamefont {M.~C.}\
  \bibnamefont {Hersam}},\ }\bibfield  {title} {\bibinfo {title} {Ultrahigh
  vacuum self-assembly of rotationally commensurate c8-btbt/mos2/graphene
  mixed-dimensional heterostructures},\ }\href@noop {} {\bibfield  {journal}
  {\bibinfo  {journal} {Chem. Mater.}\ }\textbf {\bibinfo {volume} {31}},\
  \bibinfo {pages} {1761} (\bibinfo {year} {2019})}\BibitemShut {NoStop}%
\bibitem [{\citenamefont {Li}\ \emph {et~al.}(2020)\citenamefont {Li},
  \citenamefont {Zhong}, \citenamefont {Henning}, \citenamefont {Sangwan},
  \citenamefont {Zhou}, \citenamefont {Liu}, \citenamefont {Rahn},
  \citenamefont {Wells}, \citenamefont {Park}, \citenamefont {Luxa} \emph
  {et~al.}}]{Hersam2020C8btbtInSe}%
  \BibitemOpen
  \bibfield  {author} {\bibinfo {author} {\bibfnamefont {S.}~\bibnamefont
  {Li}}, \bibinfo {author} {\bibfnamefont {C.}~\bibnamefont {Zhong}}, \bibinfo
  {author} {\bibfnamefont {A.}~\bibnamefont {Henning}}, \bibinfo {author}
  {\bibfnamefont {V.~K.}\ \bibnamefont {Sangwan}}, \bibinfo {author}
  {\bibfnamefont {Q.}~\bibnamefont {Zhou}}, \bibinfo {author} {\bibfnamefont
  {X.}~\bibnamefont {Liu}}, \bibinfo {author} {\bibfnamefont {M.~S.}\
  \bibnamefont {Rahn}}, \bibinfo {author} {\bibfnamefont {S.~A.}\ \bibnamefont
  {Wells}}, \bibinfo {author} {\bibfnamefont {H.~Y.}\ \bibnamefont {Park}},
  \bibinfo {author} {\bibfnamefont {J.}~\bibnamefont {Luxa}}, \emph {et~al.},\
  }\bibfield  {title} {\bibinfo {title} {Molecular-scale characterization of
  photoinduced charge separation in mixed-dimensional inse--organic van der
  waals heterostructures},\ }\href@noop {} {\bibfield  {journal} {\bibinfo
  {journal} {ACS Nano}\ }\textbf {\bibinfo {volume} {14}},\ \bibinfo {pages}
  {3509} (\bibinfo {year} {2020})}\BibitemShut {NoStop}%
\bibitem [{\citenamefont {He}\ \emph {et~al.}(2014)\citenamefont {He},
  \citenamefont {Zhang}, \citenamefont {Wu}, \citenamefont {Xu}, \citenamefont
  {Nan}, \citenamefont {Liu}, \citenamefont {Yao}, \citenamefont {Wang},
  \citenamefont {Yuan}, \citenamefont {Li} \emph {et~al.}}]{He2014C8btbtGr}%
  \BibitemOpen
  \bibfield  {author} {\bibinfo {author} {\bibfnamefont {D.}~\bibnamefont
  {He}}, \bibinfo {author} {\bibfnamefont {Y.}~\bibnamefont {Zhang}}, \bibinfo
  {author} {\bibfnamefont {Q.}~\bibnamefont {Wu}}, \bibinfo {author}
  {\bibfnamefont {R.}~\bibnamefont {Xu}}, \bibinfo {author} {\bibfnamefont
  {H.}~\bibnamefont {Nan}}, \bibinfo {author} {\bibfnamefont {J.}~\bibnamefont
  {Liu}}, \bibinfo {author} {\bibfnamefont {J.}~\bibnamefont {Yao}}, \bibinfo
  {author} {\bibfnamefont {Z.}~\bibnamefont {Wang}}, \bibinfo {author}
  {\bibfnamefont {S.}~\bibnamefont {Yuan}}, \bibinfo {author} {\bibfnamefont
  {Y.}~\bibnamefont {Li}}, \emph {et~al.},\ }\bibfield  {title} {\bibinfo
  {title} {Two-dimensional quasi-freestanding molecular crystals for
  high-performance organic field-effect transistors},\ }\href@noop {}
  {\bibfield  {journal} {\bibinfo  {journal} {Nat. Commun.}\ }\textbf {\bibinfo
  {volume} {5}},\ \bibinfo {pages} {1} (\bibinfo {year} {2014})}\BibitemShut
  {NoStop}%
\bibitem [{\citenamefont {Zhang}\ \emph {et~al.}(2011)\citenamefont {Zhang},
  \citenamefont {Sun}, \citenamefont {Low}, \citenamefont {Zhang},
  \citenamefont {Pan}, \citenamefont {Liu}, \citenamefont {Mao}, \citenamefont
  {Zhou}, \citenamefont {Guo}, \citenamefont {Du} \emph
  {et~al.}}]{Zhang2011FePcGr}%
  \BibitemOpen
  \bibfield  {author} {\bibinfo {author} {\bibfnamefont {H.}~\bibnamefont
  {Zhang}}, \bibinfo {author} {\bibfnamefont {J.}~\bibnamefont {Sun}}, \bibinfo
  {author} {\bibfnamefont {T.}~\bibnamefont {Low}}, \bibinfo {author}
  {\bibfnamefont {L.}~\bibnamefont {Zhang}}, \bibinfo {author} {\bibfnamefont
  {Y.}~\bibnamefont {Pan}}, \bibinfo {author} {\bibfnamefont {Q.}~\bibnamefont
  {Liu}}, \bibinfo {author} {\bibfnamefont {J.}~\bibnamefont {Mao}}, \bibinfo
  {author} {\bibfnamefont {H.}~\bibnamefont {Zhou}}, \bibinfo {author}
  {\bibfnamefont {H.}~\bibnamefont {Guo}}, \bibinfo {author} {\bibfnamefont
  {S.}~\bibnamefont {Du}}, \emph {et~al.},\ }\bibfield  {title} {\bibinfo
  {title} {Assembly of iron phthalocyanine and pentacene molecules on a
  graphene monolayer grown on ru (0001)},\ }\href@noop {} {\bibfield  {journal}
  {\bibinfo  {journal} {Phys. Rev. B}\ }\textbf {\bibinfo {volume} {84}},\
  \bibinfo {pages} {245436} (\bibinfo {year} {2011})}\BibitemShut {NoStop}%
\bibitem [{\citenamefont {Wang}\ \emph {et~al.}(2020)\citenamefont {Wang},
  \citenamefont {Gali}, \citenamefont {Slassi}, \citenamefont {Beljonne},\ and\
  \citenamefont {Samor{\`\i}}}]{Samori2020Pc2D}%
  \BibitemOpen
  \bibfield  {author} {\bibinfo {author} {\bibfnamefont {Y.}~\bibnamefont
  {Wang}}, \bibinfo {author} {\bibfnamefont {S.~M.}\ \bibnamefont {Gali}},
  \bibinfo {author} {\bibfnamefont {A.}~\bibnamefont {Slassi}}, \bibinfo
  {author} {\bibfnamefont {D.}~\bibnamefont {Beljonne}},\ and\ \bibinfo
  {author} {\bibfnamefont {P.}~\bibnamefont {Samor{\`\i}}},\ }\bibfield
  {title} {\bibinfo {title} {Collective dipole-dominated doping of monolayer
  mos2: Orientation and magnitude control via the supramolecular approach},\
  }\href@noop {} {\bibfield  {journal} {\bibinfo  {journal} {Adv. Funct.
  Mater.}\ }\textbf {\bibinfo {volume} {30}},\ \bibinfo {pages} {2002846}
  (\bibinfo {year} {2020})}\BibitemShut {NoStop}%
\bibitem [{\citenamefont {J{\"a}rvinen}\ \emph {et~al.}(2014)\citenamefont
  {J{\"a}rvinen}, \citenamefont {H{\"a}m{\"a}l{\"a}inen}, \citenamefont
  {Ij{\"a}s}, \citenamefont {Harju},\ and\ \citenamefont
  {Liljeroth}}]{Liljeroth2014PcGr}%
  \BibitemOpen
  \bibfield  {author} {\bibinfo {author} {\bibfnamefont {P.}~\bibnamefont
  {J{\"a}rvinen}}, \bibinfo {author} {\bibfnamefont {S.~K.}\ \bibnamefont
  {H{\"a}m{\"a}l{\"a}inen}}, \bibinfo {author} {\bibfnamefont {M.}~\bibnamefont
  {Ij{\"a}s}}, \bibinfo {author} {\bibfnamefont {A.}~\bibnamefont {Harju}},\
  and\ \bibinfo {author} {\bibfnamefont {P.}~\bibnamefont {Liljeroth}},\
  }\bibfield  {title} {\bibinfo {title} {Self-assembly and orbital imaging of
  metal phthalocyanines on a graphene model surface},\ }\href@noop {}
  {\bibfield  {journal} {\bibinfo  {journal} {J. Phys. Chem. C}\ }\textbf
  {\bibinfo {volume} {118}},\ \bibinfo {pages} {13320} (\bibinfo {year}
  {2014})}\BibitemShut {NoStop}%
\bibitem [{\citenamefont {Altenburg}\ \emph {et~al.}(2015)\citenamefont
  {Altenburg}, \citenamefont {Lattelais}, \citenamefont {Wang}, \citenamefont
  {Bocquet},\ and\ \citenamefont {Berndt}}]{Berndt2015PcGr}%
  \BibitemOpen
  \bibfield  {author} {\bibinfo {author} {\bibfnamefont {S.~J.}\ \bibnamefont
  {Altenburg}}, \bibinfo {author} {\bibfnamefont {M.}~\bibnamefont
  {Lattelais}}, \bibinfo {author} {\bibfnamefont {B.}~\bibnamefont {Wang}},
  \bibinfo {author} {\bibfnamefont {M.-L.}\ \bibnamefont {Bocquet}},\ and\
  \bibinfo {author} {\bibfnamefont {R.}~\bibnamefont {Berndt}},\ }\bibfield
  {title} {\bibinfo {title} {Reaction of phthalocyanines with graphene on ir
  (111)},\ }\href@noop {} {\bibfield  {journal} {\bibinfo  {journal} {J. Am.
  Chem. Soc.}\ }\textbf {\bibinfo {volume} {137}},\ \bibinfo {pages} {9452}
  (\bibinfo {year} {2015})}\BibitemShut {NoStop}%
\bibitem [{\citenamefont {Wang}\ \emph {et~al.}(2014)\citenamefont {Wang},
  \citenamefont {Zhang},\ and\ \citenamefont {Wang}}]{Wang2014PhGr}%
  \BibitemOpen
  \bibfield  {author} {\bibinfo {author} {\bibfnamefont {W.}~\bibnamefont
  {Wang}}, \bibinfo {author} {\bibfnamefont {Y.}~\bibnamefont {Zhang}},\ and\
  \bibinfo {author} {\bibfnamefont {Y.-B.}\ \bibnamefont {Wang}},\ }\bibfield
  {title} {\bibinfo {title} {Noncovalent $\pi$-$\pi$ interaction between
  graphene and aromatic molecule: Structure, energy, and nature},\ }\href@noop
  {} {\bibfield  {journal} {\bibinfo  {journal} {J. Chem. Phys.}\ }\textbf
  {\bibinfo {volume} {140}},\ \bibinfo {pages} {094302} (\bibinfo {year}
  {2014})}\BibitemShut {NoStop}%
\bibitem [{\citenamefont {Smith}\ and\ \citenamefont
  {Kay}(2018)}]{Smith2018PhGr}%
  \BibitemOpen
  \bibfield  {author} {\bibinfo {author} {\bibfnamefont {R.~S.}\ \bibnamefont
  {Smith}}\ and\ \bibinfo {author} {\bibfnamefont {B.~D.}\ \bibnamefont
  {Kay}},\ }\bibfield  {title} {\bibinfo {title} {Desorption of benzene, 1, 3,
  5-trifluorobenzene, and hexafluorobenzene from a graphene surface: The effect
  of lateral interactions on the desorption kinetics},\ }\href@noop {}
  {\bibfield  {journal} {\bibinfo  {journal} {J. Phys. Chem. Lett.}\ }\textbf
  {\bibinfo {volume} {9}},\ \bibinfo {pages} {2632} (\bibinfo {year}
  {2018})}\BibitemShut {NoStop}%
\bibitem [{\citenamefont {Hassan}\ \emph {et~al.}(2014)\citenamefont {Hassan},
  \citenamefont {Walter},\ and\ \citenamefont {Moseler}}]{Hassan2014PhGr}%
  \BibitemOpen
  \bibfield  {author} {\bibinfo {author} {\bibfnamefont {M.}~\bibnamefont
  {Hassan}}, \bibinfo {author} {\bibfnamefont {M.}~\bibnamefont {Walter}},\
  and\ \bibinfo {author} {\bibfnamefont {M.}~\bibnamefont {Moseler}},\
  }\bibfield  {title} {\bibinfo {title} {Interactions of polymers with reduced
  graphene oxide: van der waals binding energies of benzene on graphene with
  defects},\ }\href@noop {} {\bibfield  {journal} {\bibinfo  {journal} {Phys.
  Chem. Chem. Phys.}\ }\textbf {\bibinfo {volume} {16}},\ \bibinfo {pages} {33}
  (\bibinfo {year} {2014})}\BibitemShut {NoStop}%
\bibitem [{\citenamefont {Tkatchenko}\ and\ \citenamefont
  {Scheffler}(2009)}]{Scheffler2009TSvdw}%
  \BibitemOpen
  \bibfield  {author} {\bibinfo {author} {\bibfnamefont {A.}~\bibnamefont
  {Tkatchenko}}\ and\ \bibinfo {author} {\bibfnamefont {M.}~\bibnamefont
  {Scheffler}},\ }\bibfield  {title} {\bibinfo {title} {Accurate molecular van
  der waals interactions from ground-state electron density and free-atom
  reference data},\ }\href@noop {} {\bibfield  {journal} {\bibinfo  {journal}
  {Phys. Rev. Lett.}\ }\textbf {\bibinfo {volume} {102}},\ \bibinfo {pages}
  {073005} (\bibinfo {year} {2009})}\BibitemShut {NoStop}%
\bibitem [{\citenamefont {Haldar}\ \emph {et~al.}(2020)\citenamefont {Haldar},
  \citenamefont {Cortes}, \citenamefont {Darancet},\ and\ \citenamefont
  {Sharifzadeh}}]{Haldar2020NL}%
  \BibitemOpen
  \bibfield  {author} {\bibinfo {author} {\bibfnamefont {A.}~\bibnamefont
  {Haldar}}, \bibinfo {author} {\bibfnamefont {C.~L.}\ \bibnamefont {Cortes}},
  \bibinfo {author} {\bibfnamefont {P.}~\bibnamefont {Darancet}},\ and\
  \bibinfo {author} {\bibfnamefont {S.}~\bibnamefont {Sharifzadeh}},\
  }\bibfield  {title} {\bibinfo {title} {Microscopic theory of plasmons in
  substrate-supported borophene},\ }\href@noop {} {\bibfield  {journal}
  {\bibinfo  {journal} {Nano Lett.}\ }\textbf {\bibinfo {volume} {20}},\
  \bibinfo {pages} {2986} (\bibinfo {year} {2020})}\BibitemShut {NoStop}%
\bibitem [{\citenamefont {Gjerding}\ \emph {et~al.}(2017)\citenamefont
  {Gjerding}, \citenamefont {Petersen}, \citenamefont {Pedersen}, \citenamefont
  {Mortensen},\ and\ \citenamefont {Thygesen}}]{Thygesen2017NatComm}%
  \BibitemOpen
  \bibfield  {author} {\bibinfo {author} {\bibfnamefont {M.~N.}\ \bibnamefont
  {Gjerding}}, \bibinfo {author} {\bibfnamefont {R.}~\bibnamefont {Petersen}},
  \bibinfo {author} {\bibfnamefont {T.~G.}\ \bibnamefont {Pedersen}}, \bibinfo
  {author} {\bibfnamefont {N.~A.}\ \bibnamefont {Mortensen}},\ and\ \bibinfo
  {author} {\bibfnamefont {K.~S.}\ \bibnamefont {Thygesen}},\ }\bibfield
  {title} {\bibinfo {title} {Layered van der waals crystals with hyperbolic
  light dispersion},\ }\href@noop {} {\bibfield  {journal} {\bibinfo  {journal}
  {Nat. Commun.}\ }\textbf {\bibinfo {volume} {8}},\ \bibinfo {pages} {1}
  (\bibinfo {year} {2017})}\BibitemShut {NoStop}%
\bibitem [{\citenamefont {Noori}\ \emph {et~al.}(2019)\citenamefont {Noori},
  \citenamefont {Cheng}, \citenamefont {Xuan},\ and\ \citenamefont
  {Quek}}]{Quek2019screen2D}%
  \BibitemOpen
  \bibfield  {author} {\bibinfo {author} {\bibfnamefont {K.}~\bibnamefont
  {Noori}}, \bibinfo {author} {\bibfnamefont {N.~L.~Q.}\ \bibnamefont {Cheng}},
  \bibinfo {author} {\bibfnamefont {F.}~\bibnamefont {Xuan}},\ and\ \bibinfo
  {author} {\bibfnamefont {S.~Y.}\ \bibnamefont {Quek}},\ }\bibfield  {title}
  {\bibinfo {title} {Dielectric screening by 2d substrates},\ }\href@noop {}
  {\bibfield  {journal} {\bibinfo  {journal} {2D Mater.}\ }\textbf {\bibinfo
  {volume} {6}},\ \bibinfo {pages} {035036} (\bibinfo {year}
  {2019})}\BibitemShut {NoStop}%
\bibitem [{\citenamefont {Tiwari}\ and\ \citenamefont
  {Stroud}(2009)}]{Stroud2009PRB}%
  \BibitemOpen
  \bibfield  {author} {\bibinfo {author} {\bibfnamefont {R.~P.}\ \bibnamefont
  {Tiwari}}\ and\ \bibinfo {author} {\bibfnamefont {D.}~\bibnamefont
  {Stroud}},\ }\bibfield  {title} {\bibinfo {title} {Tunable band gap in
  graphene with a noncentrosymmetric superlattice potential},\ }\href@noop {}
  {\bibfield  {journal} {\bibinfo  {journal} {Phys. Rev. B}\ }\textbf {\bibinfo
  {volume} {79}},\ \bibinfo {pages} {205435} (\bibinfo {year}
  {2009})}\BibitemShut {NoStop}%
\bibitem [{\citenamefont {Brey}\ and\ \citenamefont
  {Fertig}(2009)}]{Brey2009GrV}%
  \BibitemOpen
  \bibfield  {author} {\bibinfo {author} {\bibfnamefont {L.}~\bibnamefont
  {Brey}}\ and\ \bibinfo {author} {\bibfnamefont {H.}~\bibnamefont {Fertig}},\
  }\bibfield  {title} {\bibinfo {title} {Emerging zero modes for graphene in a
  periodic potential},\ }\href@noop {} {\bibfield  {journal} {\bibinfo
  {journal} {Phys. Rev. Lett.}\ }\textbf {\bibinfo {volume} {103}},\ \bibinfo
  {pages} {046809} (\bibinfo {year} {2009})}\BibitemShut {NoStop}%
\bibitem [{\citenamefont {Castro~Neto}\ \emph {et~al.}(2009)\citenamefont
  {Castro~Neto}, \citenamefont {Guinea}, \citenamefont {Peres}, \citenamefont
  {Novoselov},\ and\ \citenamefont {Geim}}]{Geim2009GrRev}%
  \BibitemOpen
  \bibfield  {author} {\bibinfo {author} {\bibfnamefont {A.~H.}\ \bibnamefont
  {Castro~Neto}}, \bibinfo {author} {\bibfnamefont {F.}~\bibnamefont {Guinea}},
  \bibinfo {author} {\bibfnamefont {N.~M.~R.}\ \bibnamefont {Peres}}, \bibinfo
  {author} {\bibfnamefont {K.~S.}\ \bibnamefont {Novoselov}},\ and\ \bibinfo
  {author} {\bibfnamefont {A.~K.}\ \bibnamefont {Geim}},\ }\bibfield  {title}
  {\bibinfo {title} {The electronic properties of graphene},\ }\href@noop {}
  {\bibfield  {journal} {\bibinfo  {journal} {Rev. Mod. Phys.}\ }\textbf
  {\bibinfo {volume} {81}},\ \bibinfo {pages} {109} (\bibinfo {year}
  {2009})}\BibitemShut {NoStop}%
\end{thebibliography}%


\begin{thebibliography}{14}%
\makeatletter
\providecommand \@ifxundefined [1]{%
 \@ifx{#1\undefined}
}%
\providecommand \@ifnum [1]{%
 \ifnum #1\expandafter \@firstoftwo
 \else \expandafter \@secondoftwo
 \fi
}%
\providecommand \@ifx [1]{%
 \ifx #1\expandafter \@firstoftwo
 \else \expandafter \@secondoftwo
 \fi
}%
\providecommand \natexlab [1]{#1}%
\providecommand \enquote  [1]{``#1''}%
\providecommand \bibnamefont  [1]{#1}%
\providecommand \bibfnamefont [1]{#1}%
\providecommand \citenamefont [1]{#1}%
\providecommand \href@noop [0]{\@secondoftwo}%
\providecommand \href [0]{\begingroup \@sanitize@url \@href}%
\providecommand \@href[1]{\@@startlink{#1}\@@href}%
\providecommand \@@href[1]{\endgroup#1\@@endlink}%
\providecommand \@sanitize@url [0]{\catcode `\\12\catcode `\$12\catcode
  `\&12\catcode `\#12\catcode `\^12\catcode `\_12\catcode `\%12\relax}%
\providecommand \@@startlink[1]{}%
\providecommand \@@endlink[0]{}%
\providecommand \url  [0]{\begingroup\@sanitize@url \@url }%
\providecommand \@url [1]{\endgroup\@href {#1}{\urlprefix }}%
\providecommand \urlprefix  [0]{URL }%
\providecommand \Eprint [0]{\href }%
\providecommand \doibase [0]{https://doi.org/}%
\providecommand \selectlanguage [0]{\@gobble}%
\providecommand \bibinfo  [0]{\@secondoftwo}%
\providecommand \bibfield  [0]{\@secondoftwo}%
\providecommand \translation [1]{[#1]}%
\providecommand \BibitemOpen [0]{}%
\providecommand \bibitemStop [0]{}%
\providecommand \bibitemNoStop [0]{.\EOS\space}%
\providecommand \EOS [0]{\spacefactor3000\relax}%
\providecommand \BibitemShut  [1]{\csname bibitem#1\endcsname}%
\let\auto@bib@innerbib\@empty
%</preamble>
\bibitem [{\citenamefont {Giannozzi}\ \emph {et~al.}(2009)\citenamefont
  {Giannozzi}, \citenamefont {Baroni}, \citenamefont {Bonini}, \citenamefont
  {Calandra}, \citenamefont {Car}, \citenamefont {Cavazzoni}, \citenamefont
  {Ceresoli}, \citenamefont {Chiarotti}, \citenamefont {Cococcioni},
  \citenamefont {Dabo}, \citenamefont {Corso}, \citenamefont {de~Gironcoli},
  \citenamefont {Fabris}, \citenamefont {Fratesi}, \citenamefont {Gebauer},
  \citenamefont {Gerstmann}, \citenamefont {Gougoussis}, \citenamefont
  {Kokalj}, \citenamefont {Lazzeri}, \citenamefont {Martin-Samos},
  \citenamefont {Marzari}, \citenamefont {Mauri}, \citenamefont {Mazzarello},
  \citenamefont {Paolini}, \citenamefont {Pasquarello}, \citenamefont
  {Paulatto}, \citenamefont {Sbraccia}, \citenamefont {Scandolo}, \citenamefont
  {Sclauzero}, \citenamefont {Seitsonen}, \citenamefont {Smogunov},
  \citenamefont {Umari},\ and\ \citenamefont {Wentzcovitch}}]{Giannozzi2009QE}%
  \BibitemOpen
  \bibfield  {author} {\bibinfo {author} {\bibfnamefont {P.}~\bibnamefont
  {Giannozzi}}, \bibinfo {author} {\bibfnamefont {S.}~\bibnamefont {Baroni}},
  \bibinfo {author} {\bibfnamefont {N.}~\bibnamefont {Bonini}}, \bibinfo
  {author} {\bibfnamefont {M.}~\bibnamefont {Calandra}}, \bibinfo {author}
  {\bibfnamefont {R.}~\bibnamefont {Car}}, \bibinfo {author} {\bibfnamefont
  {C.}~\bibnamefont {Cavazzoni}}, \bibinfo {author} {\bibfnamefont
  {D.}~\bibnamefont {Ceresoli}}, \bibinfo {author} {\bibfnamefont {G.~L.}\
  \bibnamefont {Chiarotti}}, \bibinfo {author} {\bibfnamefont {M.}~\bibnamefont
  {Cococcioni}}, \bibinfo {author} {\bibfnamefont {I.}~\bibnamefont {Dabo}},
  \bibinfo {author} {\bibfnamefont {A.~D.}\ \bibnamefont {Corso}}, \bibinfo
  {author} {\bibfnamefont {S.}~\bibnamefont {de~Gironcoli}}, \bibinfo {author}
  {\bibfnamefont {S.}~\bibnamefont {Fabris}}, \bibinfo {author} {\bibfnamefont
  {G.}~\bibnamefont {Fratesi}}, \bibinfo {author} {\bibfnamefont
  {R.}~\bibnamefont {Gebauer}}, \bibinfo {author} {\bibfnamefont
  {U.}~\bibnamefont {Gerstmann}}, \bibinfo {author} {\bibfnamefont
  {C.}~\bibnamefont {Gougoussis}}, \bibinfo {author} {\bibfnamefont
  {A.}~\bibnamefont {Kokalj}}, \bibinfo {author} {\bibfnamefont
  {M.}~\bibnamefont {Lazzeri}}, \bibinfo {author} {\bibfnamefont
  {L.}~\bibnamefont {Martin-Samos}}, \bibinfo {author} {\bibfnamefont
  {N.}~\bibnamefont {Marzari}}, \bibinfo {author} {\bibfnamefont
  {F.}~\bibnamefont {Mauri}}, \bibinfo {author} {\bibfnamefont
  {R.}~\bibnamefont {Mazzarello}}, \bibinfo {author} {\bibfnamefont
  {S.}~\bibnamefont {Paolini}}, \bibinfo {author} {\bibfnamefont
  {A.}~\bibnamefont {Pasquarello}}, \bibinfo {author} {\bibfnamefont
  {L.}~\bibnamefont {Paulatto}}, \bibinfo {author} {\bibfnamefont
  {C.}~\bibnamefont {Sbraccia}}, \bibinfo {author} {\bibfnamefont
  {S.}~\bibnamefont {Scandolo}}, \bibinfo {author} {\bibfnamefont
  {G.}~\bibnamefont {Sclauzero}}, \bibinfo {author} {\bibfnamefont {A.~P.}\
  \bibnamefont {Seitsonen}}, \bibinfo {author} {\bibfnamefont {A.}~\bibnamefont
  {Smogunov}}, \bibinfo {author} {\bibfnamefont {P.}~\bibnamefont {Umari}},\
  and\ \bibinfo {author} {\bibfnamefont {R.~M.}\ \bibnamefont {Wentzcovitch}},\
  }\bibfield  {title} {\bibinfo {title} {{QUANTUM} {ESPRESSO}: a modular and
  open-source software project for quantum simulations of materials},\
  }\href@noop {} {\bibfield  {journal} {\bibinfo  {journal} {J. Phys.: Condens.
  Matter}\ }\textbf {\bibinfo {volume} {21}},\ \bibinfo {pages} {395502}
  (\bibinfo {year} {2009})}\BibitemShut {NoStop}%
\bibitem [{\citenamefont {Giannozzi}\ \emph {et~al.}(2017)\citenamefont
  {Giannozzi}, \citenamefont {Andreussi}, \citenamefont {Brumme}, \citenamefont
  {Bunau}, \citenamefont {Nardelli}, \citenamefont {Calandra}, \citenamefont
  {Car}, \citenamefont {Cavazzoni}, \citenamefont {Ceresoli}, \citenamefont
  {Cococcioni}, \citenamefont {Colonna}, \citenamefont {Carnimeo},
  \citenamefont {Corso}, \citenamefont {de~Gironcoli}, \citenamefont {Delugas},
  \citenamefont {DiStasio}, \citenamefont {Ferretti}, \citenamefont {Floris},
  \citenamefont {Fratesi}, \citenamefont {Fugallo}, \citenamefont {Gebauer},
  \citenamefont {Gerstmann}, \citenamefont {Giustino}, \citenamefont {Gorni},
  \citenamefont {Jia}, \citenamefont {Kawamura}, \citenamefont {Ko},
  \citenamefont {Kokalj}, \citenamefont {K{\"u}{\c{c}}{\"u}kbenli},
  \citenamefont {Lazzeri}, \citenamefont {Marsili}, \citenamefont {Marzari},
  \citenamefont {Mauri}, \citenamefont {Nguyen}, \citenamefont {Nguyen},
  \citenamefont {de-la Roza}, \citenamefont {Paulatto}, \citenamefont
  {Ponc{\'{e}}}, \citenamefont {Rocca}, \citenamefont {Sabatini}, \citenamefont
  {Santra}, \citenamefont {Schlipf}, \citenamefont {Seitsonen}, \citenamefont
  {Smogunov}, \citenamefont {Timrov}, \citenamefont {Thonhauser}, \citenamefont
  {Umari}, \citenamefont {Vast}, \citenamefont {Wu},\ and\ \citenamefont
  {Baroni}}]{Giannozzi2017QE}%
  \BibitemOpen
  \bibfield  {author} {\bibinfo {author} {\bibfnamefont {P.}~\bibnamefont
  {Giannozzi}}, \bibinfo {author} {\bibfnamefont {O.}~\bibnamefont
  {Andreussi}}, \bibinfo {author} {\bibfnamefont {T.}~\bibnamefont {Brumme}},
  \bibinfo {author} {\bibfnamefont {O.}~\bibnamefont {Bunau}}, \bibinfo
  {author} {\bibfnamefont {M.~B.}\ \bibnamefont {Nardelli}}, \bibinfo {author}
  {\bibfnamefont {M.}~\bibnamefont {Calandra}}, \bibinfo {author}
  {\bibfnamefont {R.}~\bibnamefont {Car}}, \bibinfo {author} {\bibfnamefont
  {C.}~\bibnamefont {Cavazzoni}}, \bibinfo {author} {\bibfnamefont
  {D.}~\bibnamefont {Ceresoli}}, \bibinfo {author} {\bibfnamefont
  {M.}~\bibnamefont {Cococcioni}}, \bibinfo {author} {\bibfnamefont
  {N.}~\bibnamefont {Colonna}}, \bibinfo {author} {\bibfnamefont
  {I.}~\bibnamefont {Carnimeo}}, \bibinfo {author} {\bibfnamefont {A.~D.}\
  \bibnamefont {Corso}}, \bibinfo {author} {\bibfnamefont {S.}~\bibnamefont
  {de~Gironcoli}}, \bibinfo {author} {\bibfnamefont {P.}~\bibnamefont
  {Delugas}}, \bibinfo {author} {\bibfnamefont {R.~A.}\ \bibnamefont
  {DiStasio}}, \bibinfo {author} {\bibfnamefont {A.}~\bibnamefont {Ferretti}},
  \bibinfo {author} {\bibfnamefont {A.}~\bibnamefont {Floris}}, \bibinfo
  {author} {\bibfnamefont {G.}~\bibnamefont {Fratesi}}, \bibinfo {author}
  {\bibfnamefont {G.}~\bibnamefont {Fugallo}}, \bibinfo {author} {\bibfnamefont
  {R.}~\bibnamefont {Gebauer}}, \bibinfo {author} {\bibfnamefont
  {U.}~\bibnamefont {Gerstmann}}, \bibinfo {author} {\bibfnamefont
  {F.}~\bibnamefont {Giustino}}, \bibinfo {author} {\bibfnamefont
  {T.}~\bibnamefont {Gorni}}, \bibinfo {author} {\bibfnamefont
  {J.}~\bibnamefont {Jia}}, \bibinfo {author} {\bibfnamefont {M.}~\bibnamefont
  {Kawamura}}, \bibinfo {author} {\bibfnamefont {H.-Y.}\ \bibnamefont {Ko}},
  \bibinfo {author} {\bibfnamefont {A.}~\bibnamefont {Kokalj}}, \bibinfo
  {author} {\bibfnamefont {E.}~\bibnamefont {K{\"u}{\c{c}}{\"u}kbenli}},
  \bibinfo {author} {\bibfnamefont {M.}~\bibnamefont {Lazzeri}}, \bibinfo
  {author} {\bibfnamefont {M.}~\bibnamefont {Marsili}}, \bibinfo {author}
  {\bibfnamefont {N.}~\bibnamefont {Marzari}}, \bibinfo {author} {\bibfnamefont
  {F.}~\bibnamefont {Mauri}}, \bibinfo {author} {\bibfnamefont {N.~L.}\
  \bibnamefont {Nguyen}}, \bibinfo {author} {\bibfnamefont {H.-V.}\
  \bibnamefont {Nguyen}}, \bibinfo {author} {\bibfnamefont {A.~O.}\
  \bibnamefont {de-la Roza}}, \bibinfo {author} {\bibfnamefont
  {L.}~\bibnamefont {Paulatto}}, \bibinfo {author} {\bibfnamefont
  {S.}~\bibnamefont {Ponc{\'{e}}}}, \bibinfo {author} {\bibfnamefont
  {D.}~\bibnamefont {Rocca}}, \bibinfo {author} {\bibfnamefont
  {R.}~\bibnamefont {Sabatini}}, \bibinfo {author} {\bibfnamefont
  {B.}~\bibnamefont {Santra}}, \bibinfo {author} {\bibfnamefont
  {M.}~\bibnamefont {Schlipf}}, \bibinfo {author} {\bibfnamefont {A.~P.}\
  \bibnamefont {Seitsonen}}, \bibinfo {author} {\bibfnamefont {A.}~\bibnamefont
  {Smogunov}}, \bibinfo {author} {\bibfnamefont {I.}~\bibnamefont {Timrov}},
  \bibinfo {author} {\bibfnamefont {T.}~\bibnamefont {Thonhauser}}, \bibinfo
  {author} {\bibfnamefont {P.}~\bibnamefont {Umari}}, \bibinfo {author}
  {\bibfnamefont {N.}~\bibnamefont {Vast}}, \bibinfo {author} {\bibfnamefont
  {X.}~\bibnamefont {Wu}},\ and\ \bibinfo {author} {\bibfnamefont
  {S.}~\bibnamefont {Baroni}},\ }\bibfield  {title} {\bibinfo {title} {Advanced
  capabilities for materials modelling with quantum {ESPRESSO}},\ }\href@noop
  {} {\bibfield  {journal} {\bibinfo  {journal} {J. Phys.: Condens. Matter}\
  }\textbf {\bibinfo {volume} {29}},\ \bibinfo {pages} {465901} (\bibinfo
  {year} {2017})}\BibitemShut {NoStop}%
\bibitem [{\citenamefont {Hamann}(2013)}]{Hamann2013}%
  \BibitemOpen
  \bibfield  {author} {\bibinfo {author} {\bibfnamefont {D.}~\bibnamefont
  {Hamann}},\ }\bibfield  {title} {\bibinfo {title} {Optimized norm-conserving
  vanderbilt pseudopotentials},\ }\href@noop {} {\bibfield  {journal} {\bibinfo
   {journal} {Phys. Rev. B}\ }\textbf {\bibinfo {volume} {88}},\ \bibinfo
  {pages} {085117} (\bibinfo {year} {2013})}\BibitemShut {NoStop}%
\bibitem [{\citenamefont {Van~Setten}\ \emph {et~al.}(2018)\citenamefont
  {Van~Setten}, \citenamefont {Giantomassi}, \citenamefont {Bousquet},
  \citenamefont {Verstraete}, \citenamefont {Hamann}, \citenamefont {Gonze},\
  and\ \citenamefont {Rignanese}}]{Van2018pseudodojo}%
  \BibitemOpen
  \bibfield  {author} {\bibinfo {author} {\bibfnamefont {M.}~\bibnamefont
  {Van~Setten}}, \bibinfo {author} {\bibfnamefont {M.}~\bibnamefont
  {Giantomassi}}, \bibinfo {author} {\bibfnamefont {E.}~\bibnamefont
  {Bousquet}}, \bibinfo {author} {\bibfnamefont {M.~J.}\ \bibnamefont
  {Verstraete}}, \bibinfo {author} {\bibfnamefont {D.~R.}\ \bibnamefont
  {Hamann}}, \bibinfo {author} {\bibfnamefont {X.}~\bibnamefont {Gonze}},\ and\
  \bibinfo {author} {\bibfnamefont {G.-M.}\ \bibnamefont {Rignanese}},\
  }\bibfield  {title} {\bibinfo {title} {The pseudodojo: Training and grading a
  85 element optimized norm-conserving pseudopotential table},\ }\href@noop {}
  {\bibfield  {journal} {\bibinfo  {journal} {Comput. Phys. Commun.}\ }\textbf
  {\bibinfo {volume} {226}},\ \bibinfo {pages} {39} (\bibinfo {year}
  {2018})}\BibitemShut {NoStop}%
\bibitem [{\citenamefont {Perdew}\ \emph {et~al.}(1996)\citenamefont {Perdew},
  \citenamefont {Burke},\ and\ \citenamefont {Ernzerhof}}]{Perdew1996}%
  \BibitemOpen
  \bibfield  {author} {\bibinfo {author} {\bibfnamefont {J.~P.}\ \bibnamefont
  {Perdew}}, \bibinfo {author} {\bibfnamefont {K.}~\bibnamefont {Burke}},\ and\
  \bibinfo {author} {\bibfnamefont {M.}~\bibnamefont {Ernzerhof}},\ }\bibfield
  {title} {\bibinfo {title} {Generalized gradient approximation made simple},\
  }\href@noop {} {\bibfield  {journal} {\bibinfo  {journal} {Phys. Rev. Lett.}\
  }\textbf {\bibinfo {volume} {77}},\ \bibinfo {pages} {3865} (\bibinfo {year}
  {1996})}\BibitemShut {NoStop}%
\bibitem [{\citenamefont {Tkatchenko}\ and\ \citenamefont
  {Scheffler}(2009)}]{Scheffler2009TSvdw}%
  \BibitemOpen
  \bibfield  {author} {\bibinfo {author} {\bibfnamefont {A.}~\bibnamefont
  {Tkatchenko}}\ and\ \bibinfo {author} {\bibfnamefont {M.}~\bibnamefont
  {Scheffler}},\ }\bibfield  {title} {\bibinfo {title} {Accurate molecular van
  der waals interactions from ground-state electron density and free-atom
  reference data},\ }\href@noop {} {\bibfield  {journal} {\bibinfo  {journal}
  {Phys. Rev. Lett.}\ }\textbf {\bibinfo {volume} {102}},\ \bibinfo {pages}
  {073005} (\bibinfo {year} {2009})}\BibitemShut {NoStop}%
\bibitem [{\citenamefont {Noori}\ \emph {et~al.}(2019)\citenamefont {Noori},
  \citenamefont {Cheng}, \citenamefont {Xuan},\ and\ \citenamefont
  {Quek}}]{Quek2019screen2D}%
  \BibitemOpen
  \bibfield  {author} {\bibinfo {author} {\bibfnamefont {K.}~\bibnamefont
  {Noori}}, \bibinfo {author} {\bibfnamefont {N.~L.~Q.}\ \bibnamefont {Cheng}},
  \bibinfo {author} {\bibfnamefont {F.}~\bibnamefont {Xuan}},\ and\ \bibinfo
  {author} {\bibfnamefont {S.~Y.}\ \bibnamefont {Quek}},\ }\bibfield  {title}
  {\bibinfo {title} {Dielectric screening by 2d substrates},\ }\href@noop {}
  {\bibfield  {journal} {\bibinfo  {journal} {2D Mater.}\ }\textbf {\bibinfo
  {volume} {6}},\ \bibinfo {pages} {035036} (\bibinfo {year}
  {2019})}\BibitemShut {NoStop}%
\bibitem [{\citenamefont {Iversen}\ \emph {et~al.}(1998)\citenamefont
  {Iversen}, \citenamefont {Kharkats},\ and\ \citenamefont
  {Ulstrup}}]{Iversen1998ScreenSlab}%
  \BibitemOpen
  \bibfield  {author} {\bibinfo {author} {\bibfnamefont {G.}~\bibnamefont
  {Iversen}}, \bibinfo {author} {\bibfnamefont {Y.~I.}\ \bibnamefont
  {Kharkats}},\ and\ \bibinfo {author} {\bibfnamefont {J.}~\bibnamefont
  {Ulstrup}},\ }\bibfield  {title} {\bibinfo {title} {Simple dielectric image
  charge models for electrostatic interactions in metalloproteins},\
  }\href@noop {} {\bibfield  {journal} {\bibinfo  {journal} {Mol. Phys.}\
  }\textbf {\bibinfo {volume} {94}},\ \bibinfo {pages} {297} (\bibinfo {year}
  {1998})}\BibitemShut {NoStop}%
\bibitem [{\citenamefont {Bessler}\ \emph {et~al.}(2019)\citenamefont
  {Bessler}, \citenamefont {Duerig},\ and\ \citenamefont
  {Koren}}]{Gr2019dielectric}%
  \BibitemOpen
  \bibfield  {author} {\bibinfo {author} {\bibfnamefont {R.}~\bibnamefont
  {Bessler}}, \bibinfo {author} {\bibfnamefont {U.}~\bibnamefont {Duerig}},\
  and\ \bibinfo {author} {\bibfnamefont {E.}~\bibnamefont {Koren}},\ }\bibfield
   {title} {\bibinfo {title} {The dielectric constant of a bilayer graphene
  interface},\ }\href@noop {} {\bibfield  {journal} {\bibinfo  {journal}
  {Nanoscale Adv.}\ }\textbf {\bibinfo {volume} {1}},\ \bibinfo {pages} {1702}
  (\bibinfo {year} {2019})}\BibitemShut {NoStop}%
\bibitem [{\citenamefont {Errandonea}\ \emph {et~al.}(1999)\citenamefont
  {Errandonea}, \citenamefont {Segura}, \citenamefont {Munoz},\ and\
  \citenamefont {Chevy}}]{InSe1999Eps}%
  \BibitemOpen
  \bibfield  {author} {\bibinfo {author} {\bibfnamefont {D.}~\bibnamefont
  {Errandonea}}, \bibinfo {author} {\bibfnamefont {A.}~\bibnamefont {Segura}},
  \bibinfo {author} {\bibfnamefont {V.}~\bibnamefont {Munoz}},\ and\ \bibinfo
  {author} {\bibfnamefont {A.}~\bibnamefont {Chevy}},\ }\bibfield  {title}
  {\bibinfo {title} {Effects of pressure and temperature on the dielectric
  constant of gas, gase, and inse: Role of the electronic contribution},\
  }\href@noop {} {\bibfield  {journal} {\bibinfo  {journal} {Phys. Rev. B}\
  }\textbf {\bibinfo {volume} {60}},\ \bibinfo {pages} {15866} (\bibinfo {year}
  {1999})}\BibitemShut {NoStop}%
\bibitem [{\citenamefont {Natan}\ \emph {et~al.}(2007)\citenamefont {Natan},
  \citenamefont {Kronik}, \citenamefont {Haick},\ and\ \citenamefont
  {Tung}}]{Natan2007SAMrev}%
  \BibitemOpen
  \bibfield  {author} {\bibinfo {author} {\bibfnamefont {A.}~\bibnamefont
  {Natan}}, \bibinfo {author} {\bibfnamefont {L.}~\bibnamefont {Kronik}},
  \bibinfo {author} {\bibfnamefont {H.}~\bibnamefont {Haick}},\ and\ \bibinfo
  {author} {\bibfnamefont {R.~T.}\ \bibnamefont {Tung}},\ }\bibfield  {title}
  {\bibinfo {title} {Electrostatic properties of ideal and non-ideal polar
  organic monolayers: Implications for electronic devices},\ }\href@noop {}
  {\bibfield  {journal} {\bibinfo  {journal} {Adv. Mater.}\ }\textbf {\bibinfo
  {volume} {19}},\ \bibinfo {pages} {4103} (\bibinfo {year}
  {2007})}\BibitemShut {NoStop}%
\bibitem [{\citenamefont {Chang}\ \emph {et~al.}(2012)\citenamefont {Chang},
  \citenamefont {Fan}, \citenamefont {Li},\ and\ \citenamefont
  {Kuo}}]{Kuo2012GrPh}%
  \BibitemOpen
  \bibfield  {author} {\bibinfo {author} {\bibfnamefont {C.-H.}\ \bibnamefont
  {Chang}}, \bibinfo {author} {\bibfnamefont {X.}~\bibnamefont {Fan}}, \bibinfo
  {author} {\bibfnamefont {L.-J.}\ \bibnamefont {Li}},\ and\ \bibinfo {author}
  {\bibfnamefont {J.-L.}\ \bibnamefont {Kuo}},\ }\bibfield  {title} {\bibinfo
  {title} {Band gap tuning of graphene by adsorption of aromatic molecules},\
  }\href@noop {} {\bibfield  {journal} {\bibinfo  {journal} {J. Phys. Chem. C}\
  }\textbf {\bibinfo {volume} {116}},\ \bibinfo {pages} {13788} (\bibinfo
  {year} {2012})}\BibitemShut {NoStop}%
\bibitem [{\citenamefont {Vanderbilt}(2018)}]{vanderbilt_PythTB2018}%
  \BibitemOpen
  \bibfield  {author} {\bibinfo {author} {\bibfnamefont {D.}~\bibnamefont
  {Vanderbilt}},\ }\bibinfo {title} {The pythtb package},\ in\ \href
  {https://doi.org/10.1017/9781316662205.012} {\emph {\bibinfo {booktitle}
  {Berry Phases in Electronic Structure Theory: Electric Polarization, Orbital
  Magnetization and Topological Insulators}}}\ (\bibinfo  {publisher}
  {Cambridge University Press},\ \bibinfo {year} {2018})\ p.\ \bibinfo {pages}
  {327–362}\BibitemShut {NoStop}%
\bibitem [{\citenamefont {Castro~Neto}\ \emph {et~al.}(2009)\citenamefont
  {Castro~Neto}, \citenamefont {Guinea}, \citenamefont {Peres}, \citenamefont
  {Novoselov},\ and\ \citenamefont {Geim}}]{Geim2009GrRev}%
  \BibitemOpen
  \bibfield  {author} {\bibinfo {author} {\bibfnamefont {A.~H.}\ \bibnamefont
  {Castro~Neto}}, \bibinfo {author} {\bibfnamefont {F.}~\bibnamefont {Guinea}},
  \bibinfo {author} {\bibfnamefont {N.~M.~R.}\ \bibnamefont {Peres}}, \bibinfo
  {author} {\bibfnamefont {K.~S.}\ \bibnamefont {Novoselov}},\ and\ \bibinfo
  {author} {\bibfnamefont {A.~K.}\ \bibnamefont {Geim}},\ }\bibfield  {title}
  {\bibinfo {title} {The electronic properties of graphene},\ }\href@noop {}
  {\bibfield  {journal} {\bibinfo  {journal} {Rev. Mod. Phys.}\ }\textbf
  {\bibinfo {volume} {81}},\ \bibinfo {pages} {109} (\bibinfo {year}
  {2009})}\BibitemShut {NoStop}%
\end{thebibliography}%

\end{document}

\title{Engineering the Electronic Structure of Two-Dimensional Materials with Near-Field Electrostatic Effects of Self-Assembled Organic Layers}
	
\author{Qunfei Zhou}
\email{qunfei.zhou@northwestern.edu}
\affiliation{Materials Research Science and Engineering Center, Northwestern University, Evanston, IL 60208, USA}
\affiliation{Center for Nanoscale Materials, Argonne National Laboratory, Argonne, IL 60439, USA}
	
\author{Bukuru Anaclet}
%Pierre_Reminder: Need to check affiliation mechanism for REU+ 
\affiliation{Materials Research Science and Engineering Center, Northwestern University, Evanston, IL 60208, USA}
\affiliation{Department of Chemistry, Pomona College, 645 North College Avenue, Claremont, CA}

\author{Trevor Steiner}
%Pierre_Reminder: Need to check affiliation mechanism for REU 
\affiliation{Materials Research Science and Engineering Center, Northwestern University, Evanston, IL 60208, USA}
\affiliation{Department of Materials Science and Engineering, University of Minnesota-Twin Cities, Minneapolis, MN 55455, USA}
\affiliation{Materials Department, University of California, Santa Barbara, California 93106, USA}

\author{Michele Kotiuga}
\email{michele.kotiuga@epfl.ch}
\affiliation{Theory and Simulation of Materials (THEOS) and National Centre for Computational Design and Discovery of Novel Materials (MARVEL), \'{E}cole Polytechnique F\'{e}d\'{e}rale de Lausanne, CH-1015 Lausanne, Switzerland}
	
\author{Pierre Darancet}
\email{pdarancet@anl.gov}
\affiliation{Center for Nanoscale Materials, Argonne National Laboratory, Argonne, IL 60439, USA}
\affiliation{Northwestern Argonne Institute of Science and Engineering, Evanston, IL 60208, USA}
	
\date{\today}

\maketitle

\tableofcontents

\section{Computational Methods}	\label{SecSI:method}
\subsection{Computational Details} \label{Sec:DFT}
All Density Functional Theory based calculations are performed using the Quantum-Espresso package~\cite{Giannozzi2009QE,Giannozzi2017QE}. Optimized norm-conserving Vanderbilt pseudopotentials~\cite{Hamann2013} are obtained from the PseudoDojo library~\cite{Van2018pseudodojo}. Exchange-correlation potentials use Perdew-Burke-Ernzerhof (PBE) parametrization of the generalized gradient approximation~\cite{Perdew1996}. The plane-wave cutoff energy is 90 Ry for all systems studied in this work.$k$-point sampling of 20$\times$20 is used for graphene with in-plane size of 2.46$\times$2.46 \AA, or equivalent for the heterojunctions with other dimensions. The vacuum regions are all around 22 \AA. Convergences of total and electronic energies are 10$^{-3}$ eV/atom, and $10^{-6}$ eV, respectively. For all the heterojunction supercells with molecules on graphene, van der Waals interactions are described using the Thatchenko-Scheffler~\cite{Scheffler2009TSvdw} dispersion corrections. The binding energy $E_b$ for molecules on 2D graphene are computed as:
\begin{equation}
	E_b=\frac{E[mol/2D] - N_{mol}E[mol] - E[2D]}{N_{mol}}
\end{equation}
where $E[mol/2D], E[mol], E[2D]$ are the total energies for the relaxed structures of the molecule/2D supercell, molecule, and 2D, respectively. $N_{mol}$ is the number of molecules for each molecule/2D supercell.

\subsection{Discretized Planar Charge Density (DPCD) Model} \label{Sec:DPCD}

\begin{figure}[H]
	\includegraphics[width=0.7\linewidth, center]{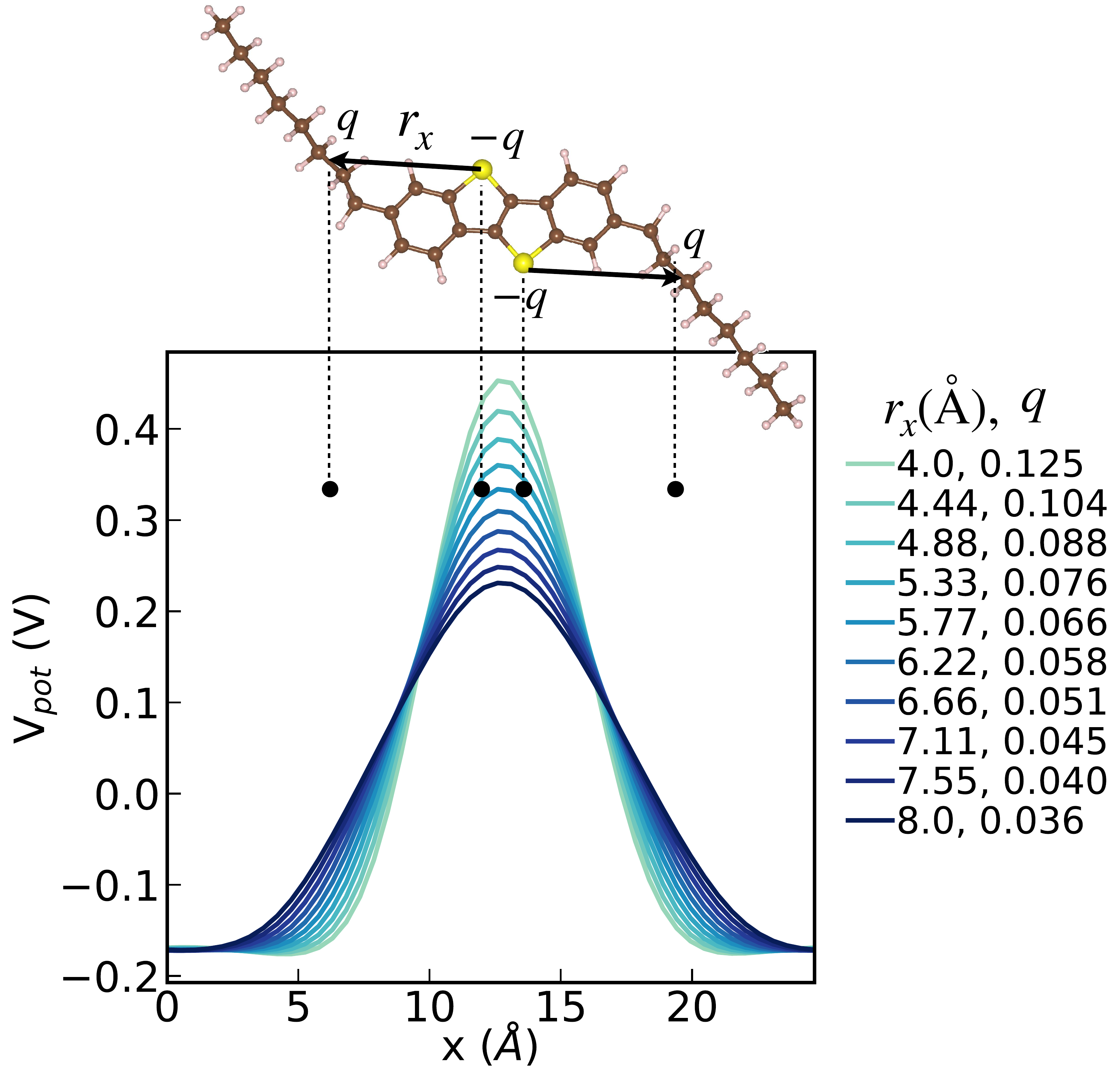}
	\caption{In-plane potential from C8-BTBT layer averaged along $x$ direction, as a function of $l_x$ where $q$ is from $\bs Q_x=\sum_i q_i \bs r_{i,x}^2$ and $r_y$ for each $q$ charge from $\bs Q_y=\sum_i q_i \bs r_{i,y}^2$. $\bs Q_x, \bs Q_y$ are the $x,y$ components the quadrupole moment, respectively. The $-q$ charges are at the positions of the two S atoms. The values of $l_x, q$ of 5.77 \AA, 0.07, respectively, lead to potential $V_{pot}$ comparable to DFT-calculated results.}
	\label{figSI:DPCD}
\end{figure}

By calculating the in-plane potential as a function of varying positions and values of discretized planar charges, and comparing it with the DFT-calculated results, we can determine the discretized planar charge density to represent the multipole effect of the molecules using only a few point charges. As shown in Fig.~\ref{figSI:DPCD} for C8-BTBT molecular layers, we can represent each molecule by using only four point charges, where the two negative charges are at the position of the S atoms, and the charge density is determined according to the quadrupole moment in the $x$ and $y$ direction, $\bs Q_x=\sum_i q_i \bs r_{i,x}^2$ and $\bs Q_y=\sum_i q_i \bs r_{i,y}^2$, where $\bs Q_x, \bs Q_y$ are the $x,y$ components the quadrupole moment, respectively. We perform similar determining procedure for benzene/C$_6$F$_6$ and H$_2$Pc(F$_{16}$) molecules, see their discretized planar charge densities shown in Fig.~\ref{figSI:Vbenzene} and Fig.~\ref{figSI:VPc}. 

\subsection{Dielectric Screening Effect of the 2D Substrate}\label{Sec:scr}

Due to the polarization of 2D graphene, the in-plane potential provided by the organic molecular layers can be electrostatically screened, an effect dependent on the molecule-2D distance and dielectric constant of the 2D substrate~\cite{Quek2019screen2D,Iversen1998ScreenSlab}. We denote the electrostatic potential of the molecular layer as $V_{pot}$, and that of the molecular layer after screening by the 2D substrate as $V^{scr}_{pot}$, which can be computed as below from DFT:
\begin{equation}
	V^{scr,DFT}_{pot} = V^{DFT}_{pot} [mol/2D] - V^{DFT}_{pot}[2D]
\end{equation}
where $V_{pot} [mol/2D], V_{pot}[2D]$ are the DFT-calculated potentials for molecule/2D heterostructure and 2D substrate, respectively.

Using the discretized planar charge density model, we determine the discrete point charges $q_i$ representing the multipole effects of each surface organic molecule at $(x_i, y_i, z+d)$ with $z$ the position of 2D material and $d$ the vertical molecule-2D distance. For each point charge $q_i$, the image charge $p_i$ is~\cite{Quek2019screen2D}:
\begin{equation}
	p_i = -\frac{\varepsilon-1}{\varepsilon+1} q_i
\end{equation}
where $\varepsilon$ is the dielectric constant for the 2D substrate, and the position of $p_i$ is $(x_i, y_i, z-d)$. Therefore we can add the electrostatic potential from $q_i$ and their image charges $p_i$, leading to the screened potential as~\cite{Quek2019screen2D}:
\begin{equation}
	V^{scr,DPCD}_{pot} = (1-\frac{\varepsilon-1}{\varepsilon+1}) V^{DPCD}_{pot}
\end{equation}
where $V^{DPCD}_{pot}$ is the potential obtained using Eq.2 from charges $q_i$. For 2D substrate graphene and InSe, we use dielectric constant of 6~\cite{Gr2019dielectric}, and 7.6~\cite{InSe1999Eps}.
%dielectric constant of monolayer graphene is $\simeq$ 3 ~\cite{Santos2013GrDielectric,Crommie2012GrDielectric}

\section{Molecules and Structures}
\begin{figure}[H]
	\includegraphics[width=0.8\linewidth, center]{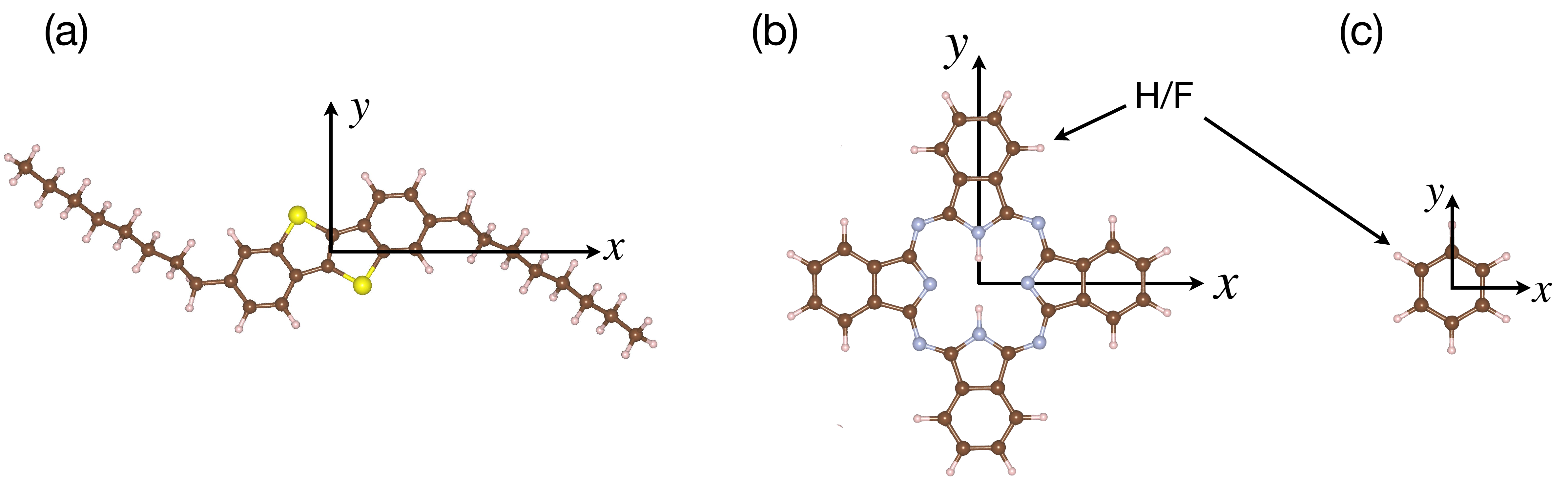}
	\caption{Molecular structures for (a) C8-BTBT, (b) phthalocyanine/perfluorophthalocyanine, H$_2$Pc/H$_2$PcF$_{16}$, (c) benzene/hexafluorobenzene, C$_6$H$_6$/C$_6$F$_6$. Their DFT-calculated quadrupole moments along $x$ and $y$ directions are included in Table~\ref{tbl:QStr}.} 
	\label{figSI:mol}
\end{figure}

\begin{table}[H]
	\small
	\captionsetup{margin={1.0cm,1.0cm}}
	\caption{\ Quadrupole moments along $x$ (Q$_x$) and $y$ (Q$_y$) directions as shown in Fig.~\ref{figSI:mol}, supercell size N$\times$M in unit of primitive 2D cell (2.464$\times$2.46 \AA for graphene, and 4.09$\times$4.09 for InSe), molecules per supercell N$_{mol}$, binding energies E$_b$, inter-molecular distance along $x$ and $y$ axes ($l_x$, $l_y$), normal distance $d$ from molecular center layer to graphene, and areal interface dipole $P/A$.} 
	\label{tbl:QStr}
	\begin{tabular*}{0.95\textwidth}{@{\extracolsep{\fill}}ccccccc}
		\hline
		molecules & Q$_x$/Q$_y$ (D.\AA) & N$\times$M & N$_{mol}$  & $d$ (\AA) & E$_b$ (eV/molecule) & $P/A$ (Debye/nm$^2$) \\ \hline %& $l_x/l_y$
		benzene 				& 3.60/3.60  & 3$\times$3 & 1 & 3.31 & 0.73 & 0.30 \\ %& 7.39/7.39
		C$_6$F$_6$			& -3.52/-3.52 & 3$\times$3 & 1 & 3.32 & 0.71 & 0.10  \\ %& 7.39/7.39
		H$_2$Pc				   & 6.89/32.26 & 6$\times$7$\sqrt{3}$ & 2  & 3.35 & 3.28 & 0.21  \\ %& 14.80/14.95
		H$_2$PcF$_{16}$  & -9.71/-35.04 & 6$\times$7$\sqrt{3}$ & 2 & 3.34 & 3.63 & 0.16 \\ %& 14.80/14.95
		C8-BTBT				   & 26.90/-0.98  & 10$\times$3$\sqrt{3}$ & 2 & 3.76 & 2.65 & 0.20 \\ \hline %& 6.41/24.66
		C8-BTBT(on InSe)  & 26.90/-0.98  & 6$\times \sqrt{3}$ & 1  & 3.72 & 1.86 & 0.52 \\ % & 7.09/24.57
		\hline
	\end{tabular*} 
\end{table}	

\begin{figure}[H]
	\includegraphics[width=0.6\linewidth, center]{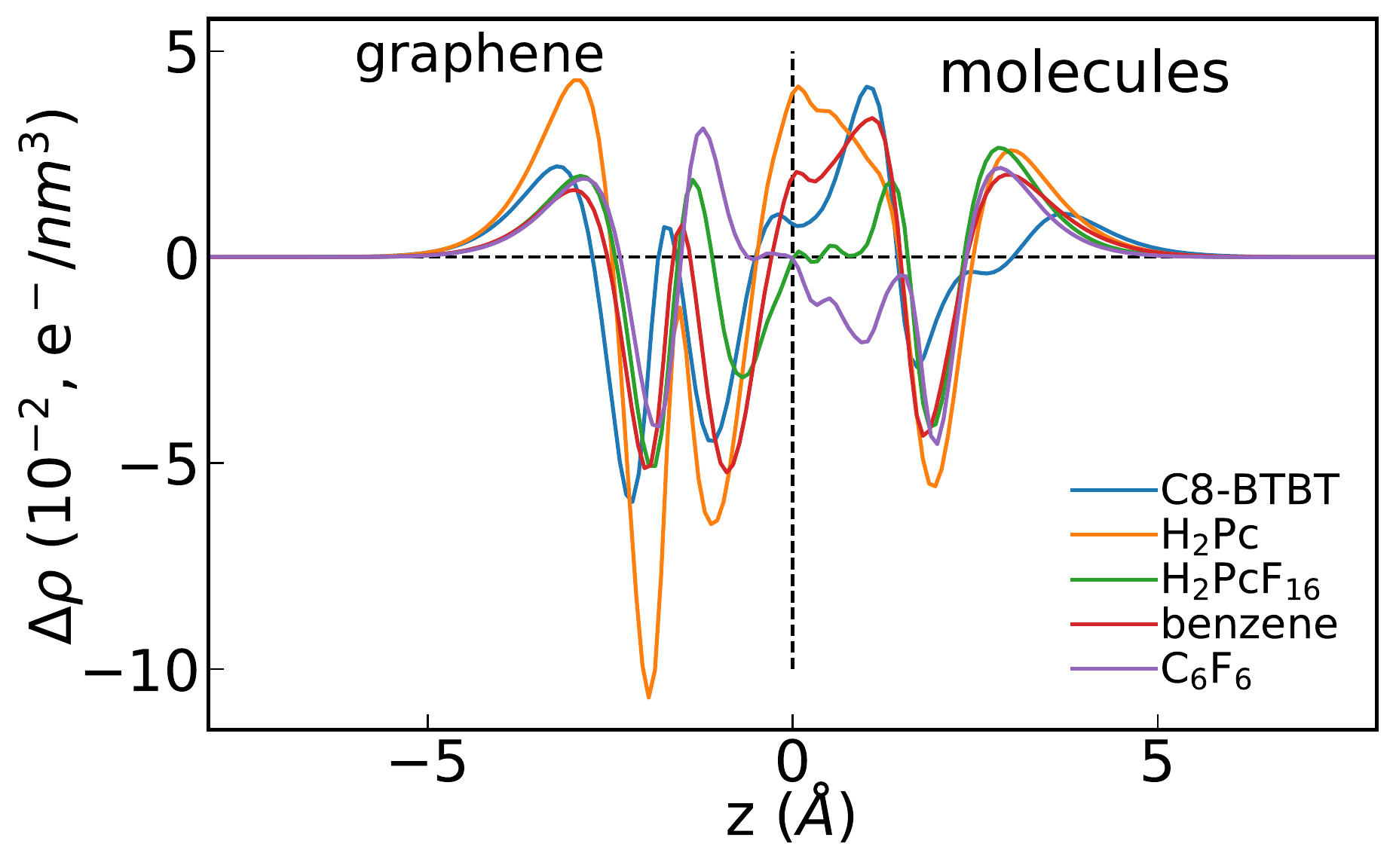}
	\caption{Change in charge density $\Delta \rho_z$ along the $z$ direction computed as $\Delta \rho = \rho$[SAM/2D] - $\rho$[SAM] - $\rho$[2D]. The vertical dashed line depicts the center of the interface region. The interface areal dipole values computed from $\Delta \rho_z$ are included in Table~\ref{tbl:QStr}.} 
	\label{figSI:chg}
\end{figure}

\section{Results for Benzene/graphene, H$_2$Pc(F$_{16}$)/graphene and Screening Effect}
\begin{figure}[H]
	\includegraphics[width=0.6\linewidth, center]{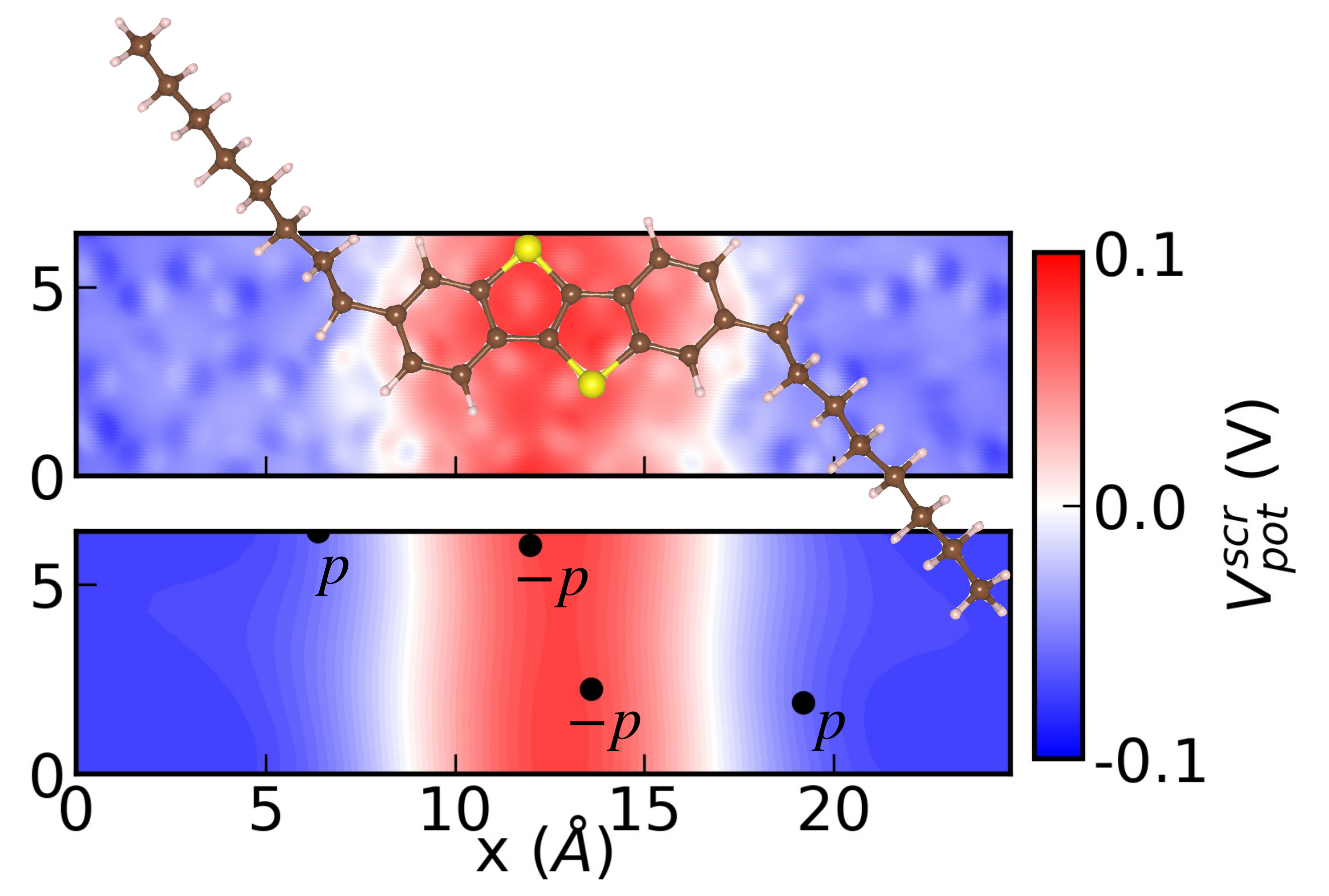}
	\caption{In-plane potentials for C8-BTBT/graphene from DFT (top) and DPCD model (bottom) after 2D screening. The effective charge \textit{p}=0.02.}
	\label{figSI:dVbtbt}
\end{figure}

\begin{figure}[H]
	\includegraphics[width=0.95\linewidth, center]{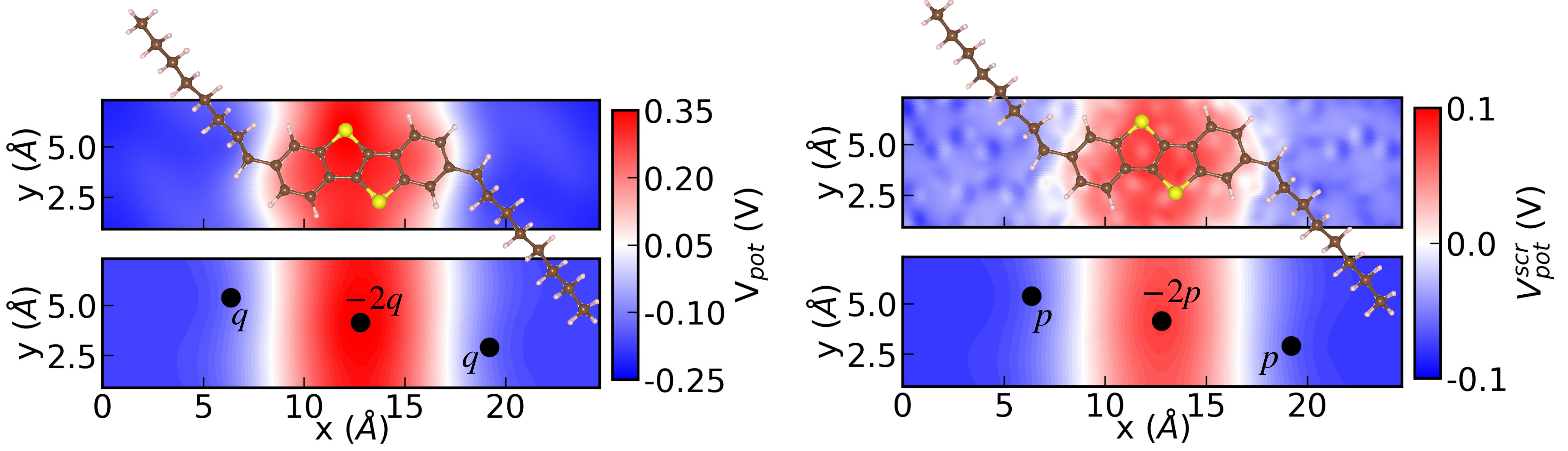}
	\caption{In-plane potentials for C8-BTBT/graphene from DFT (top) and DPCD (bottom) using three point charges representing each molecule (a) before and (b) after 2D screening. The value of the charges are $q=0.07$ and $p=0.02$. } 
	\label{figSI:dVbtbt3pt}
\end{figure}

\begin{figure}[H]
	\includegraphics[width=0.6\linewidth, center]{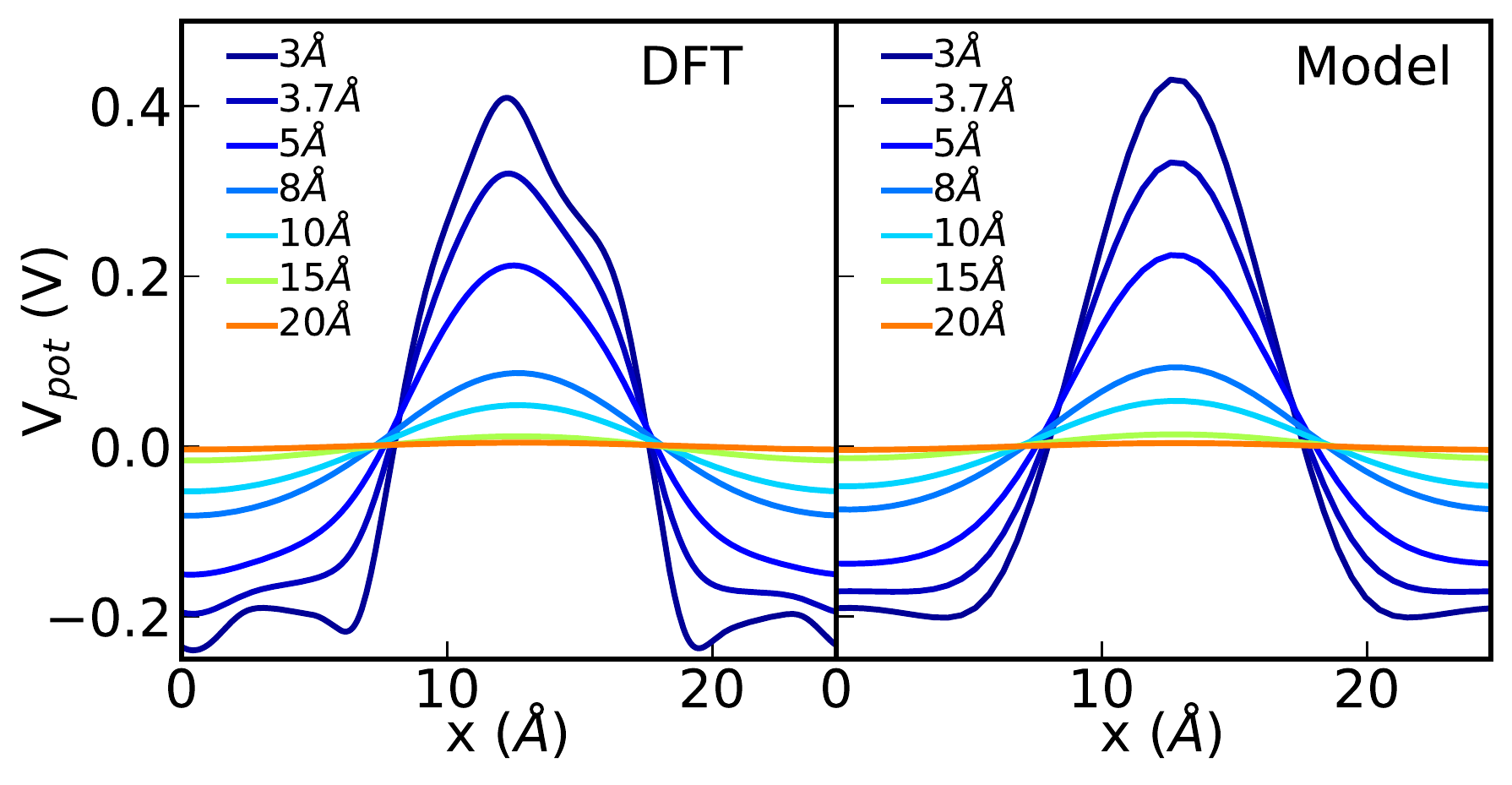}
	\caption{Electrostatic potential averaged along $y$-axis for C8-BTBT/graphene as a function of molecule-graphene vertical distance $d$, from DFT (left) and DPCD model (right).} 
	\label{figSI:Vdmodel}
\end{figure}

\begin{figure}[H]
	\includegraphics[width=0.48\linewidth, center]{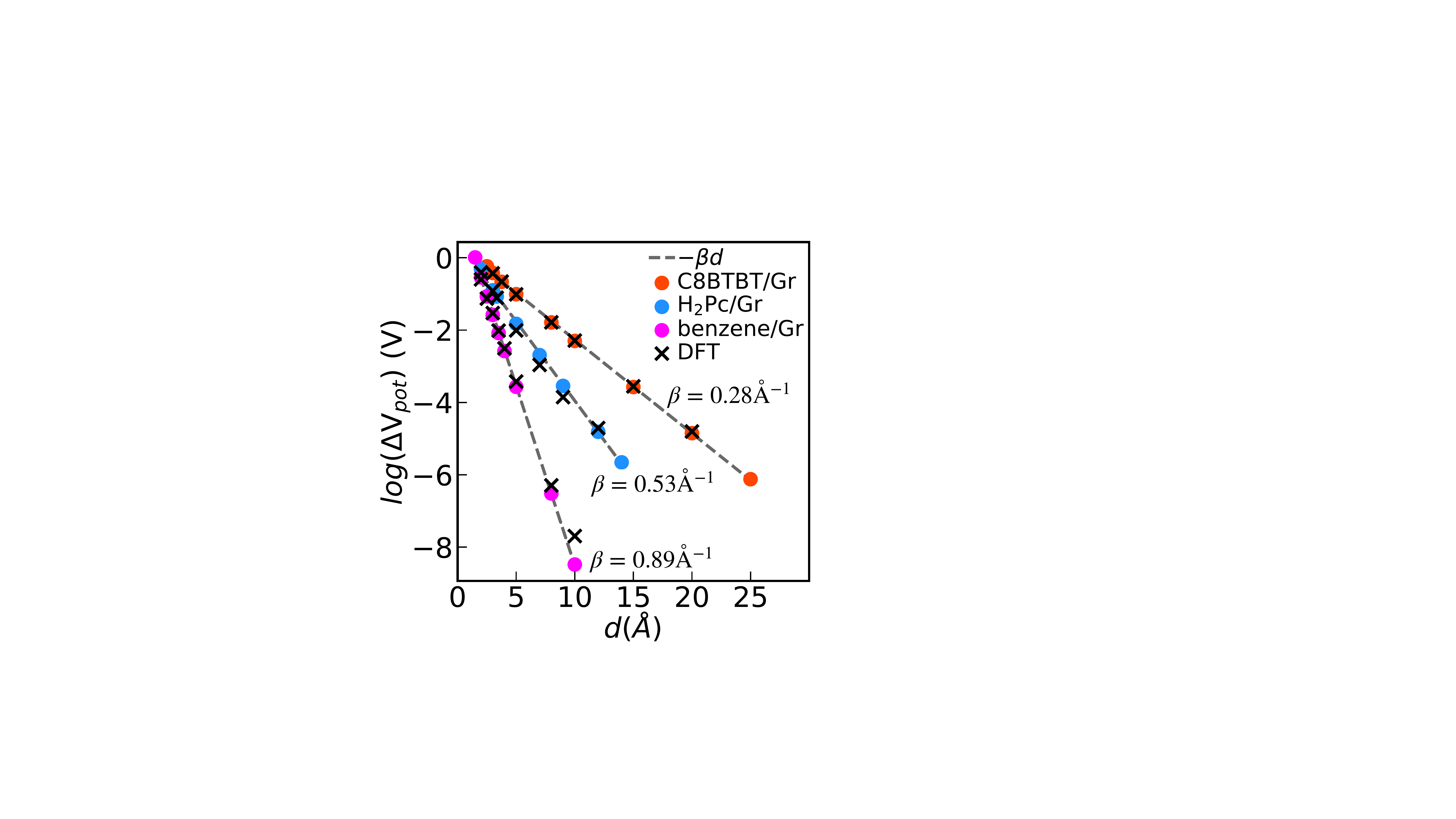}
	\caption{In-plane modulation of potential $\Delta V_{pot}$ as a function of $d$ on a logarithmic scale. The dashed lines are fitted from dotted data using function $log(\Delta V_{pot})=-\beta d + C$.} 
	\label{figSI:Vlog}
\end{figure}

\begin{figure}[H]
	\includegraphics[width=0.48\linewidth, center]{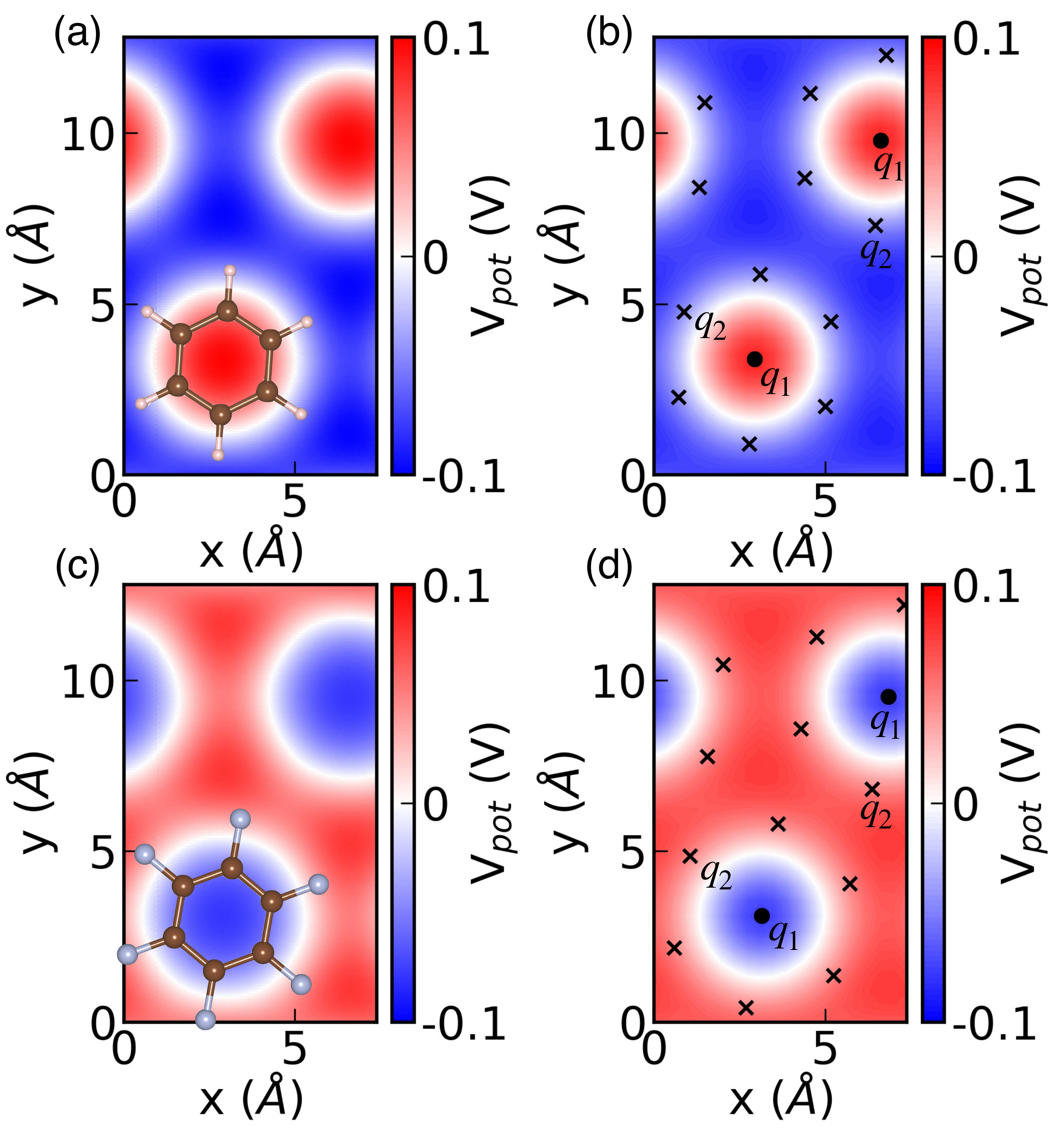}
	\caption{In-plane potentials for benzene/graphene and hexafluorobenzene/graphene from (a) DFT, and (b) DPCD with charge distributions as shown by black dots ($q_1$) and crosses ($q_2$) where $q_1=-6q_2$. $q_2$ values for benzene and hexafluorobenzene are 0.040 and -0.033, respectively. Those values are obtained from the DFT-calculated quadrupole moments, see Fig.~\ref{figSI:mol}. The potential values are shifted so that V$_{pot,min}$=-V$_{pot,max}$.} 
	\label{figSI:Vbenzene}
\end{figure}

\begin{figure}[H]
	\includegraphics[width=0.48\linewidth, center]{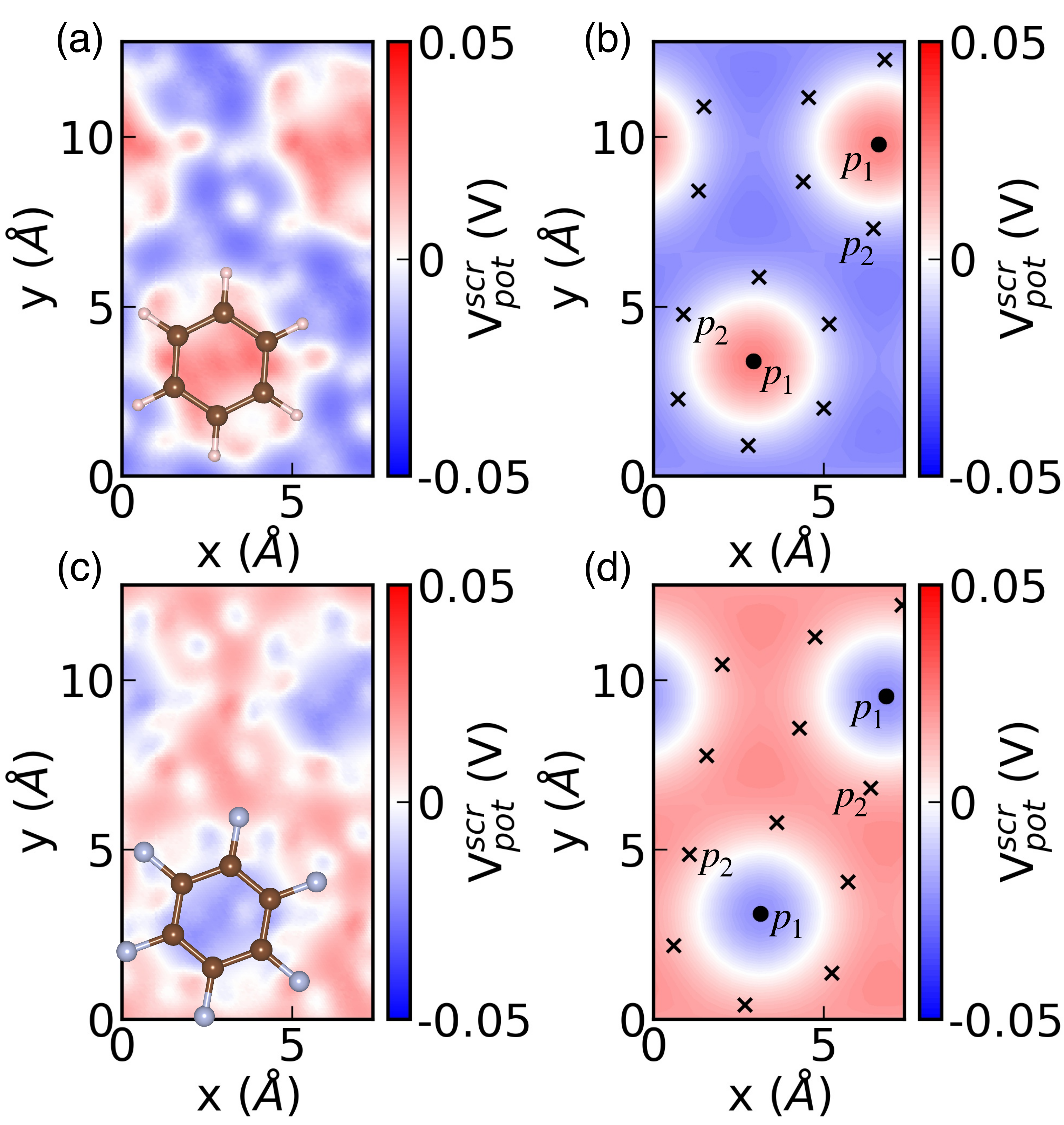}
	\caption{In-plane potential after 2D screening for benzene/graphene and hexafluorobenzene/graphene from (a) DFT, and (b) DPCD model with same point charge distributions as in Fig.~\ref{figSI:Vbenzene}. The effective charge $p_1$ for benzene and hexafluorobenzene are 0.013 and 0.011, respectively. The potential values are shifted so that V$_{pot,min}$=-V$_{pot,max}$.} 
	\label{figSI:dVbenzene}
\end{figure}

\begin{figure}[H]
	\includegraphics[width=0.9\linewidth, center]{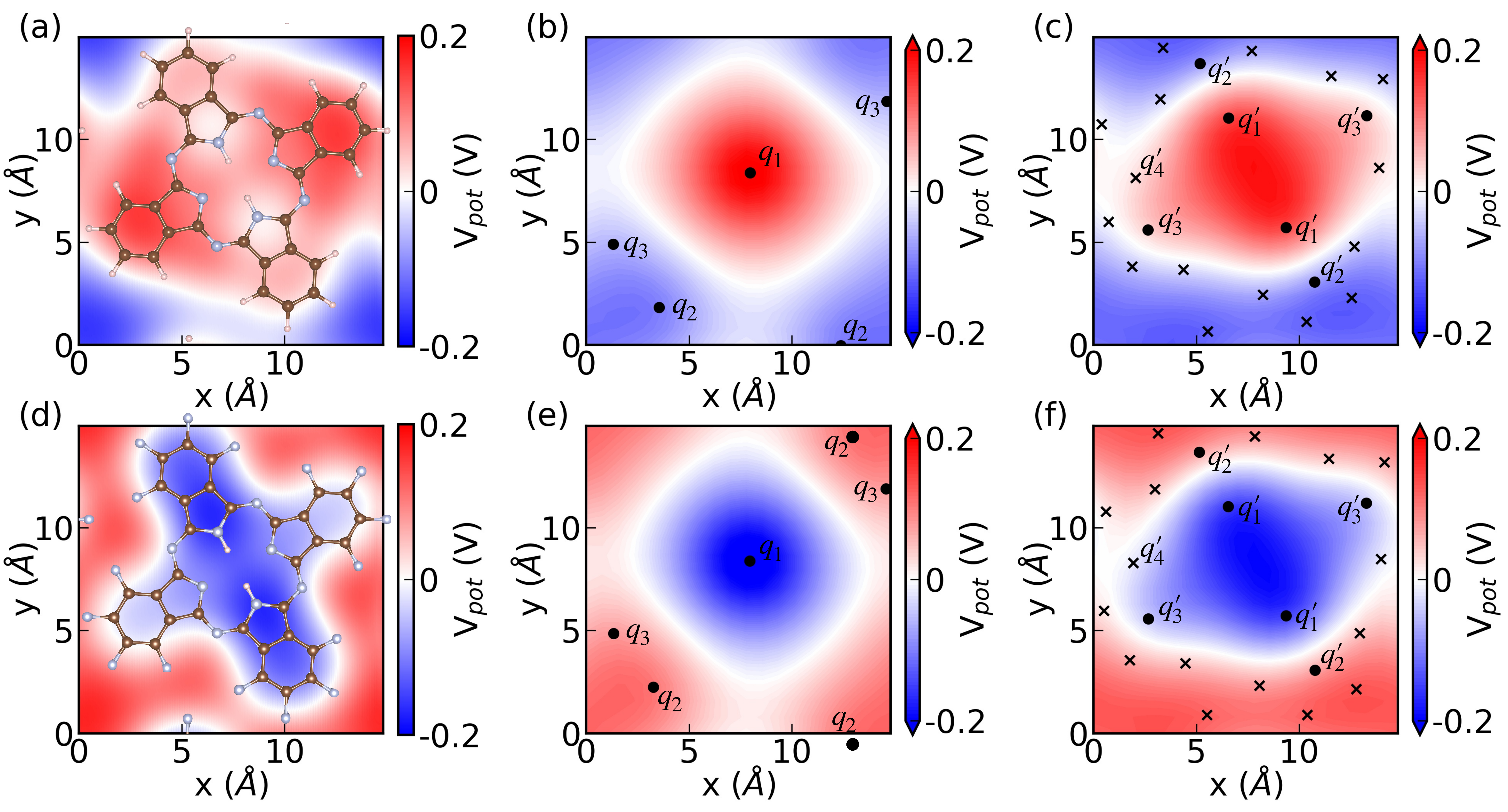}
	\caption{In-plane potentials for H$_2$Pc/graphene from (a) DFT, and (b-c) DPCD model with charge distributed as shown by black dots and crosses. Same for H$_2$PcF$_{16}$/graphene in (d-f). All the potential values are shifted so that V$_{pot,min}$=-V$_{pot,max}$. In (c) and (f), all the crosses are charges $q_4'$ at positions of the peripheral H/F atoms. The values of ($q_1, q_2, q_3$) for H$_2$Pc and H$_2$PcF$_{16}$ are (-0.100, 0.013, 0.037) and (0.109, -0.018, -0.036), 
	respectively. The values of ($q_1', q_2', q_3', q_4'$) for H$_2$Pc and H$_2$PcF$_{16}$ are (-0.098, 0.029, -0.068, 0.017) and (0.096, -0.018, 0.078, -0.020), respectively, as obtained from the DFT-calculated quadrupole moments, see Fig.~\ref{figSI:mol}.} 
	\label{figSI:VPc}
\end{figure}

\begin{figure}[H]
	\includegraphics[width=0.9\linewidth, center]{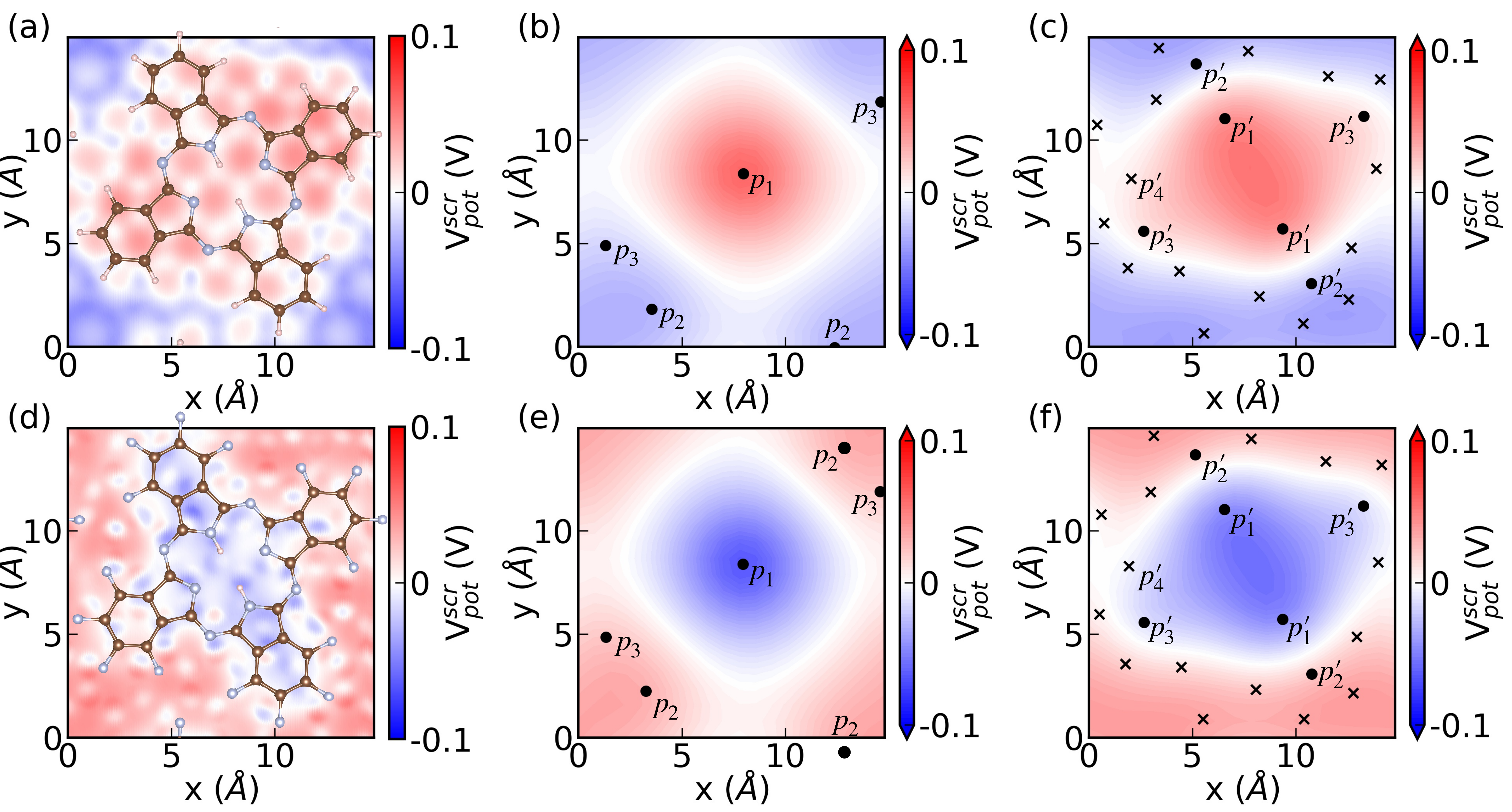}
	\caption{Same as in Fig.~\ref{figSI:VPc} but for potentials after 2D screening.}
	\label{figSI:dVPc}
\end{figure}

\begin{table}[H]
	\small
	\captionsetup{margin={1.0cm,1.0cm}}
	\caption{In-plane modulation of the electrostatic potential $\Delta V_{pot}$ and 2D screened potential modulation $\Delta V_{pot}^{scr}$ in unit of volts, by self-assembled molecules from DFT and DPCD model. The parameters in the fitted function $\Delta V=A$exp$(-\beta d)$ for the distance dependence of potential. Decay length $d_{max}$ defined as the distance at which the electrostatic potential goes to be negligible ($< 10^{-3} $ V). And the inter-molecular distance along $x$ and $y$ axes ($l_x$, $l_y$). } 
	\label{tbl:dV}
	\centering
	\begin{tabular*}{0.95\textwidth}{@{\extracolsep{\fill}}ccccccccc c}
		\hline
		\multirow{2}{*}{molecule} & \multicolumn{2}{c}{$\Delta V_{pot}$} & \multicolumn{2}{c}{$\Delta V_{pot}^{scr}$} & \multicolumn{2}{c}{$\beta$ (\AA$^{-1}$)} & \multirow{2}{*}{$A$ (V)} & \multirow{2}{*}{$d_{max}$ (\AA)} & $l_x/l_y$ (\AA) \\ \cline{2-3} \cline{4-5} \cline{6-7}
		& DFT & model & DFT & model & DFT & model  & & &  \\ \hline
		C8-BTBT 			& 0.569 & 0.524 & 0.162 & 0.150 & 0.28 & 0.30  &  1.62 & 26.0 & 24.66/6.41 \\
		H$_2$Pc 		     & 0.316 & 0.320 & 0.092 & 0.091 & 0.53 & 0.55 & 1.98 & 14.0 & 14.80/14.95 \\
		H$_2$PF$_{16}$  & 0.344 & 0.350 & 0.107 & 0.100 & 0.71 & 0.54 & 2.13 & 14.0 & 14.80/14.95 \\ 
		benzene 			 & 0.196 & 0.174 & 0.053	& 0.050 & 0.89 & 1.02 & 5.13 & 8.2 & 7.39/7.39 \\
		C$_6$F$_6$  	  & 0.157 & 0.155 & 0.042 & 0.044  &0.96 & 1.07  & 5.01 & 8.1 & 7.39/7.39 \\
		\hline
	\end{tabular*} 
\end{table}

\section{Impacts on the In-Plane Electrostatic Potential Modulation} \label{Sec:f-V}

\begin{figure}[H]
	\includegraphics[width=0.9\linewidth, center]{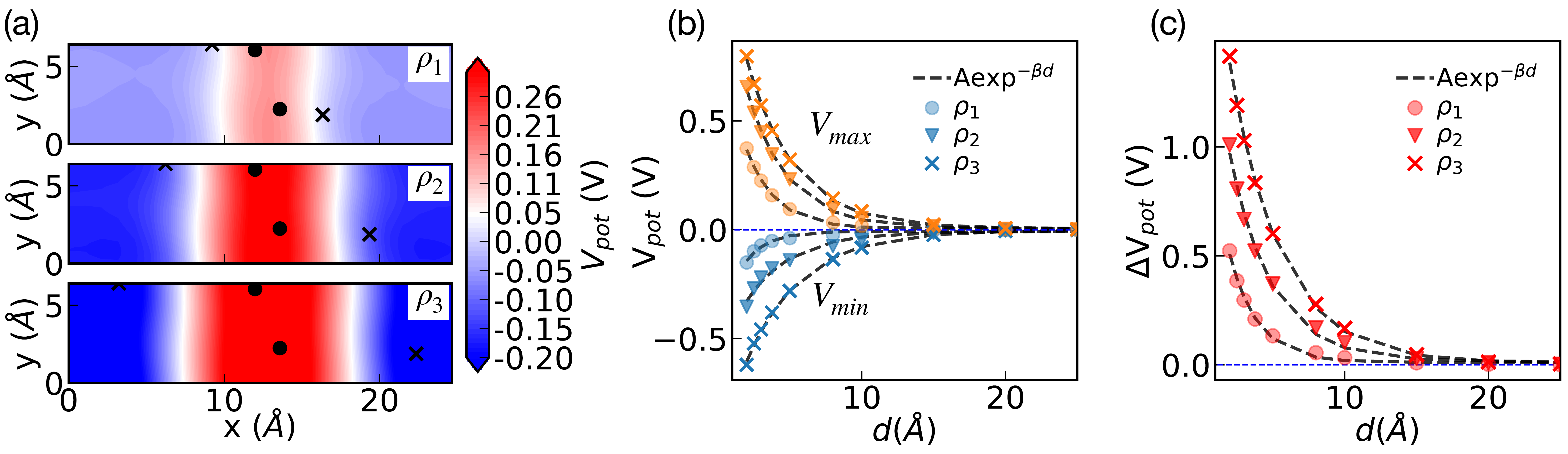}
	\caption{(a) Electrostatic potentials from discrete charge densities $\rho_1, \rho_2, \rho_3$ where the black dots and crosses mark the positions of the point charges in negative and positive values, respectively, as that in Fig.~\ref{figSI:dVbtbt}. (b)-(c) The minimum and maximum potentials, $V_{min}, V_{max}$, and $\Delta V_{pot}=V_{max}-V_{min}$ as a function of distance for planar charges of $\rho_1$ (circles), $\rho_2$ (triangles), $\rho_3$ (crosses). The dashed black lines are fitted using the exponential functions with $\beta/A$ values in Table~\ref{tbl:rho}. Corresponding quadrupole moments in $x$ direction are 7.65, 26.90, 57.54 (D.\AA), respectively.}
	%dV: 0.311, 0.549, 0.737
	\label{figSI:rho}
\end{figure}

\begin{figure}[H]
	\includegraphics[width=0.9\linewidth, center]{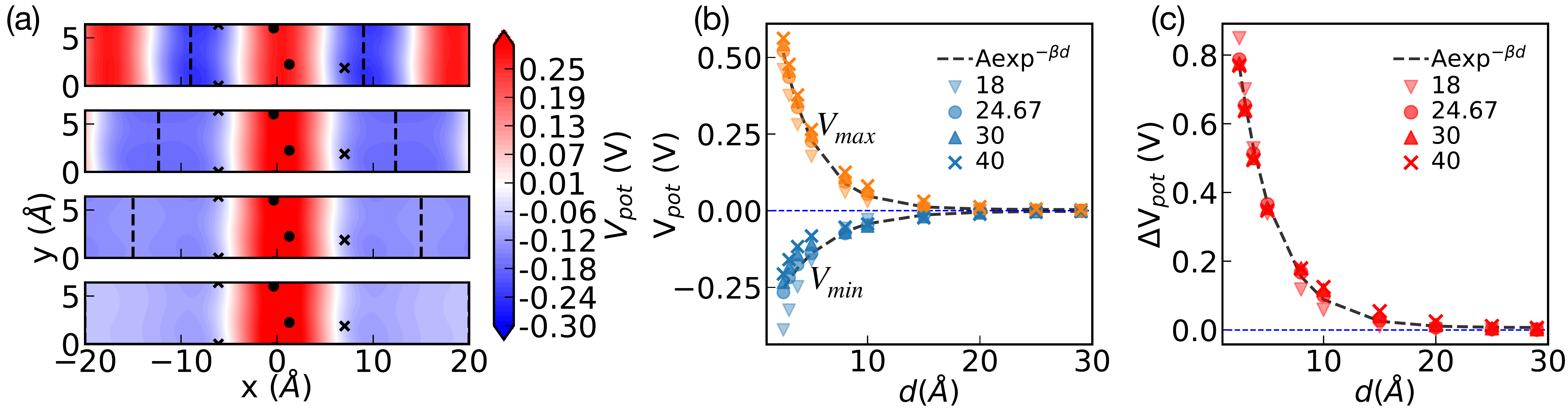}
	\caption{(a) Electrostatic potential from discrete charge densities for C8-BTBT, same as shown in Fig.2 (b) and $\rho_2$ in Fig.~\ref{figSI:rho}, but with different unit cell sizes with $b=6.41$\AA~for all but $a$ of 18 \AA, 24.67 \AA, 30 \AA, and 40 \AA~respectively, as shown by the dashed vertical lines. (b-c) $V_{min}, V_{max}$ and $\Delta V_{pot}$ as a function of $d$ for the four unit cell sizes. Values and parameters of the exponential functions are summarized in Table~\ref{tbl:rho}.}
	%dV: 0.311, 0.549, 0.737
	\label{figSI:cell}
\end{figure}

\begin{table}[H]
	\small
	\captionsetup{margin={1.0cm,1.0cm}}
	\caption{In-plane modulation of the electrostatic potential $\Delta V_{pot}$ from DPCD model as shown in Fig.~\ref{figSI:rho}(a). The parameters in the fitted function $\Delta V=A$exp$(-\beta d)$ for the distance dependence of potential. The unit cell has $b=6.41$ for all and $a$ shown below.} 
	\label{tbl:rho}
	\centering
	\begin{tabular*}{0.8\textwidth}{@{\extracolsep{\fill}}ccccccccc}
		\hline
		\multirow{2}{*}{charges} & \multirow{2}{*}{$a$ (\AA)} & \multirow{2}{*}{$\Delta V$ (V)}  & \multicolumn{3}{c}{$\beta$ (\AA$^{-1}$)} & \multicolumn{3}{c}{$A$ (V)}  \\ \cline{4-6} \cline{7-9}
		&   & & $V_{min}$ & $V_{max}$ & $\Delta V_{pot}$ & $V_{min}$ & $V_{max}$ & $\Delta V_{pot}$  \\ \hline
%		$\rho_1$ & 24.67 & 0.31 & 0.48 & 0.39 & 0.42 & -0.63 & 0.99 & 1.60 \\
%		$\rho_2$ & 24.67 & 0.55 & 0.30 & 0.32 & 0.32 & -0.77 & 1.12 & 1.89 \\
%		$\rho_3$ & 24.67 & 0.737 & 0.32 & 0.30 & 0.31 & -1.31 & 1.18 & 2.48 \\ \hline
		$\rho_1$ & 24.67 & 0.22 & 0.35 & 0.44 & 0.42 & -0.22 & 0.84 & 1.06 \\
		$\rho_2$ & 24.67 & 0.51 & 0.24 & 0.33 & 0.29 & -0.46 & 2.24 & 1.61 \\
		$\rho_3$ & 24.67 & 0.81 & 0.25 & 0.29 & 0.27 & -0.96 & 1.29 & 2.24 \\ \hline
		$\rho_2$ & 18      & 0.53 & 0.39 & 0.41 & 0.40 & -1.11 & 1.36 & 2.47 \\
		$\rho_2$ & 24.67 & 0.51 & 0.24 & 0.33 & 0.29 & -0.46 & 2.24 & 1.61 \\
		$\rho_2$ & 30		& 0.50 & 0.25 & 0.31 & 0.29 & -0.39 & 1.14 & 1.54 \\
		$\rho_2$ & 40		& 0.49 & 0.35 & 0.29 & 0.30 & -0.43 & 1.13 & 1.53 \\
		\hline
	\end{tabular*} 
\end{table}

\begin{figure}[H]
	\includegraphics[width=0.9\linewidth, center]{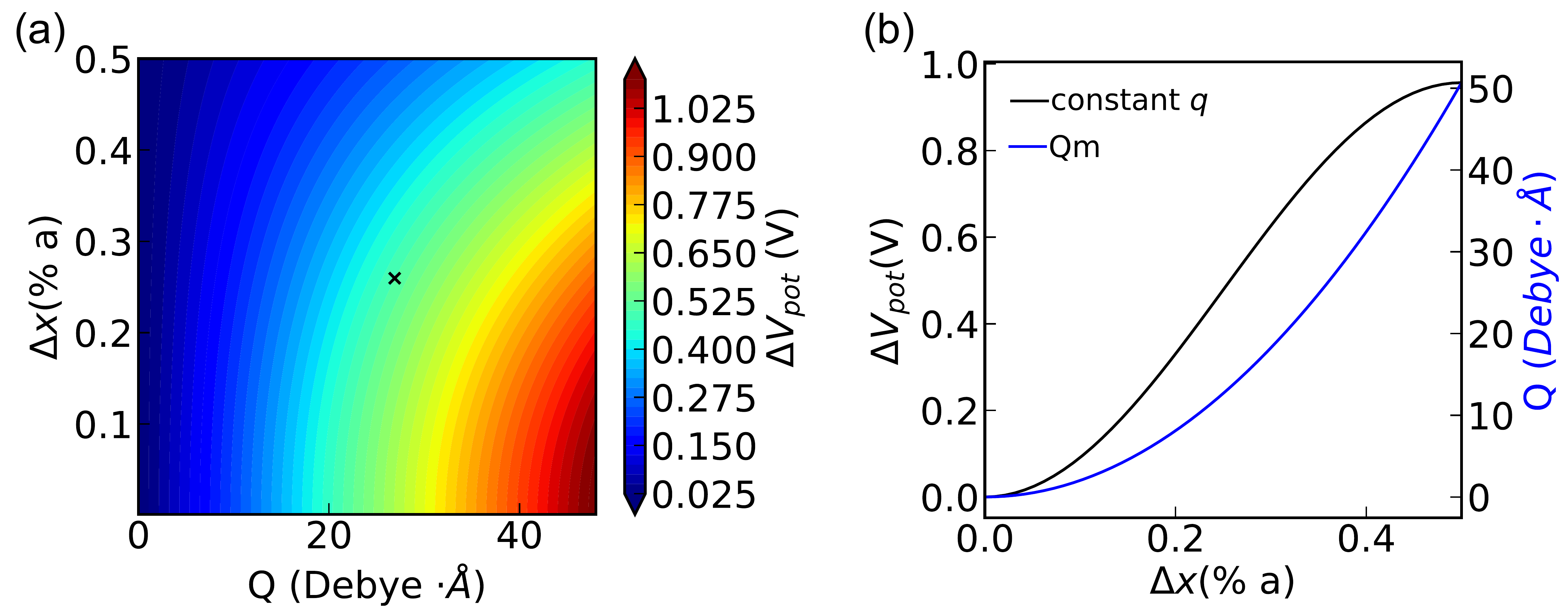}
	\caption{(a) In-plane potential modulation $\Delta V_{pot}$ as a function of quadrupole moment $Q$ and $\Delta x = x_0-x_1$ calculated from Eq.~\ref{EqS:dVsimple} using $x_1 =0$ where $x_1, x_0$ are shown in Fig.~\ref{figSI:4ptcell}, and $q=Q/(2(a\Delta x)^2)$ . The black cross depicts results for the actual C8-BTBT molecular layer. (b) $\Delta V_{pot}$ and $Q$ as a function of $\Delta x$ calculated from Eq.~\ref{EqS:dVsimple} using constant $q=0.07$. }
	\label{figSI:Vqx}
\end{figure}

\begin{figure}[H]
	\includegraphics[width=0.85\linewidth, center]{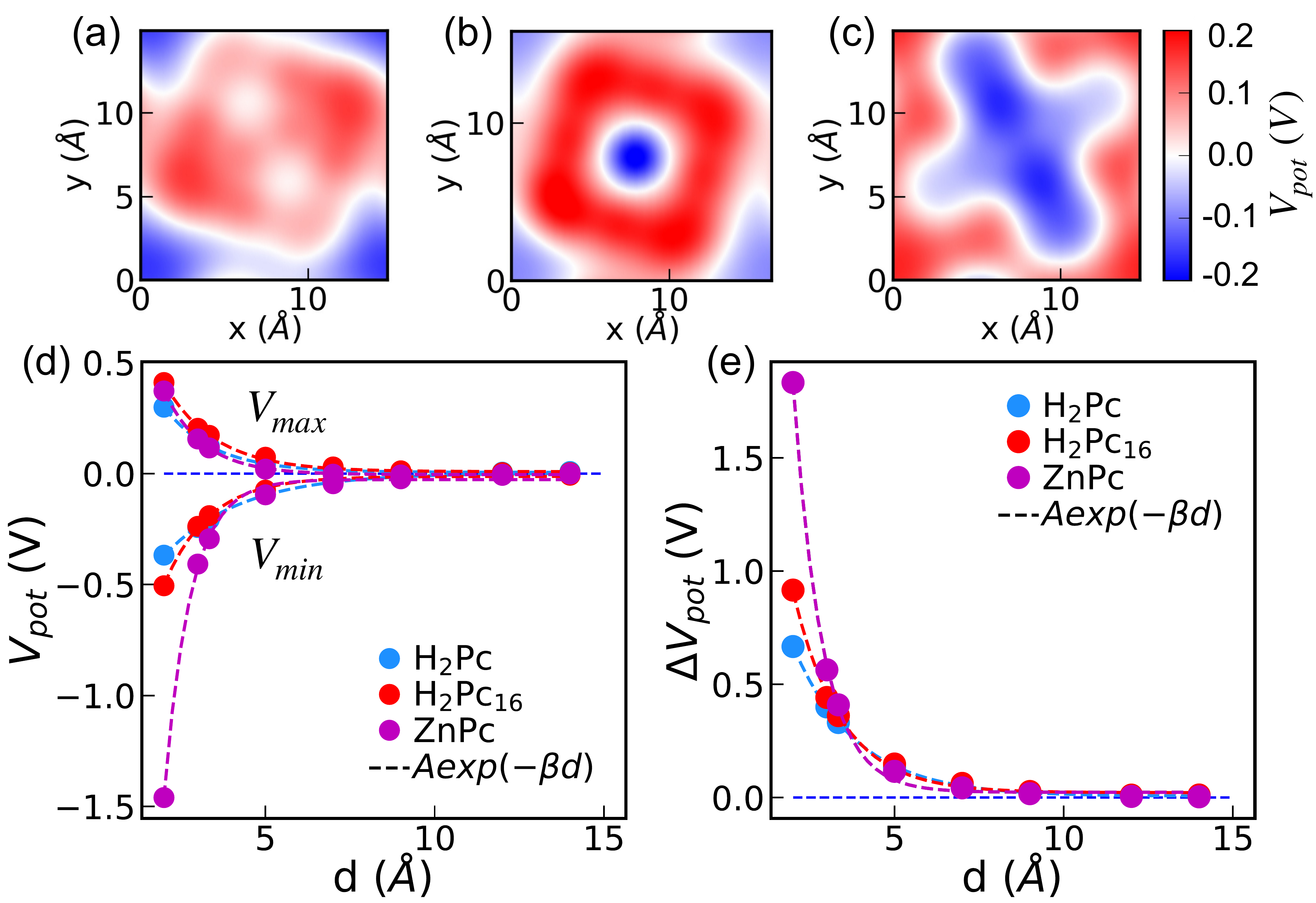}
	\caption{The in-plane potential $V_{pot}$ at the plane of graphene from molecular assembly of (a) H$_2$Pc (b) ZnPc (c) H$_2$PcF$_{16}$, and (d) their $V_{min}, V_{max}$ and (e) $\Delta V_{pot}$ as functions of distance $d$. The $\beta$ values for $\Delta V_{pot}$ ($V_{min}, V_{max}$) are 0.53 (0.44,0.68), 1.17 (1.28, 0.86), 0.71 (0.75, 0.66), respectively, for the three molecules.}
	\label{figSI:Vpcs}
\end{figure}

In Fig.~\ref{figSI:rho}, we show in-plane electrostatic potential from same unit cells but different discrete planar charge densities, and in Fig.~\ref{figSI:cell} the in-plane electrostatic potential from the same charge densities but different unit cell sizes. As summarized in Table~\ref{tbl:rho}, the decay rate, implied by $\beta$ value, is negligibly affected by the unit cell size, but is inversely proportional to the density of point charges, which is equivalent to the size of the molecules. This is reasonable as the smaller the molecule is, the shorter the distance from which the positive and negative charges are indistinguishable is. Therefore the decay length is also proportional to the size of the molecule, see Fig.~\ref{figSI:rho}(c), independent of the unit cell size, as shown in Fig.~\ref{figSI:cell}(c). Meanwhile, from $\rho_1$ to $\rho_3$ with increased charge separation, the quadrupole moments along $x$ direction increases from 7.65, 26.90 to 57.54 in units of D.\AA, leading to increase of in-plane potential modulation, $\Delta V_{pot}$ for same distances.

As a result, molecules with larger quadrupole moment and larger donor-acceptor separation, can lead to both larger in-plane potential modulation, and effect extending to a longer distance.

\begin{figure}[H]
	\includegraphics[width=0.95\linewidth, center]{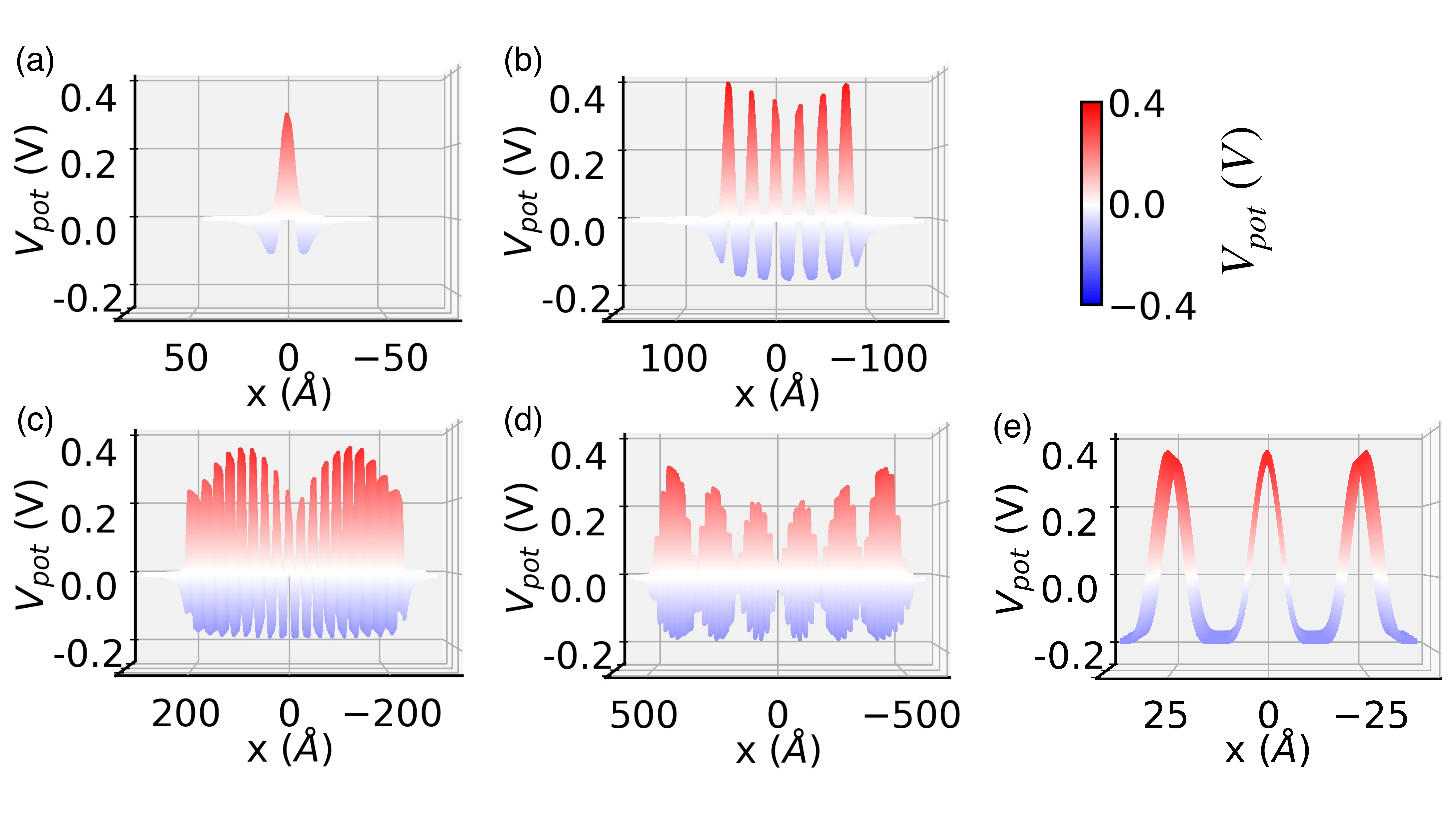}
	\caption{ 3D plot of the electrostatic potential from molecular layers with finite domain size in unit of \AA$\times$\AA\ of (a) 25$\times$26, (b) 147$\times$153, (c) 442$\times$ 460, (d) 984$\times$1022, (e) infinite. }
	%why dV larger: larger values at the edges, same as for dipole effects found in previous work}
	%dV: 0.311, 0.549, 0.737
	\label{figSI:sizemap}
\end{figure}

Using our DPCD model, we can investigate the in-plane electrostatic potential of molecular layers with finite domain size. As shown in Fig.~\ref{figSI:sizemap}(a-c), the electrostatic potential in the plane of graphene with finite domain size of C8-BTBT molecules differ from that of infinite molecular layers as shown in Fig.~\ref{figSI:sizemap}(d). The periodicity of the in-plane potentials are the same as the periodicity of the molecules. However, the maximum and minimum values of the electrostatic potentials are sine functions, as indicated in Eq. 3, for finite domain sizes, leading to an additional periodicity of the potentials. Different edge effects have also shown in previous study~\cite{Natan2007SAMrev} for finite domain of dipole arrays.

\section{Band structures}

As reported in previous work~\cite{Kuo2012GrPh}, the bandgap opening can be achieved by adsorbed molecular layers, an effect dependent on the adsorption sites and normal molecule-graphene distance. 

\begin{figure}[H]
	\centering
	\includegraphics[width=0.5\linewidth, center]{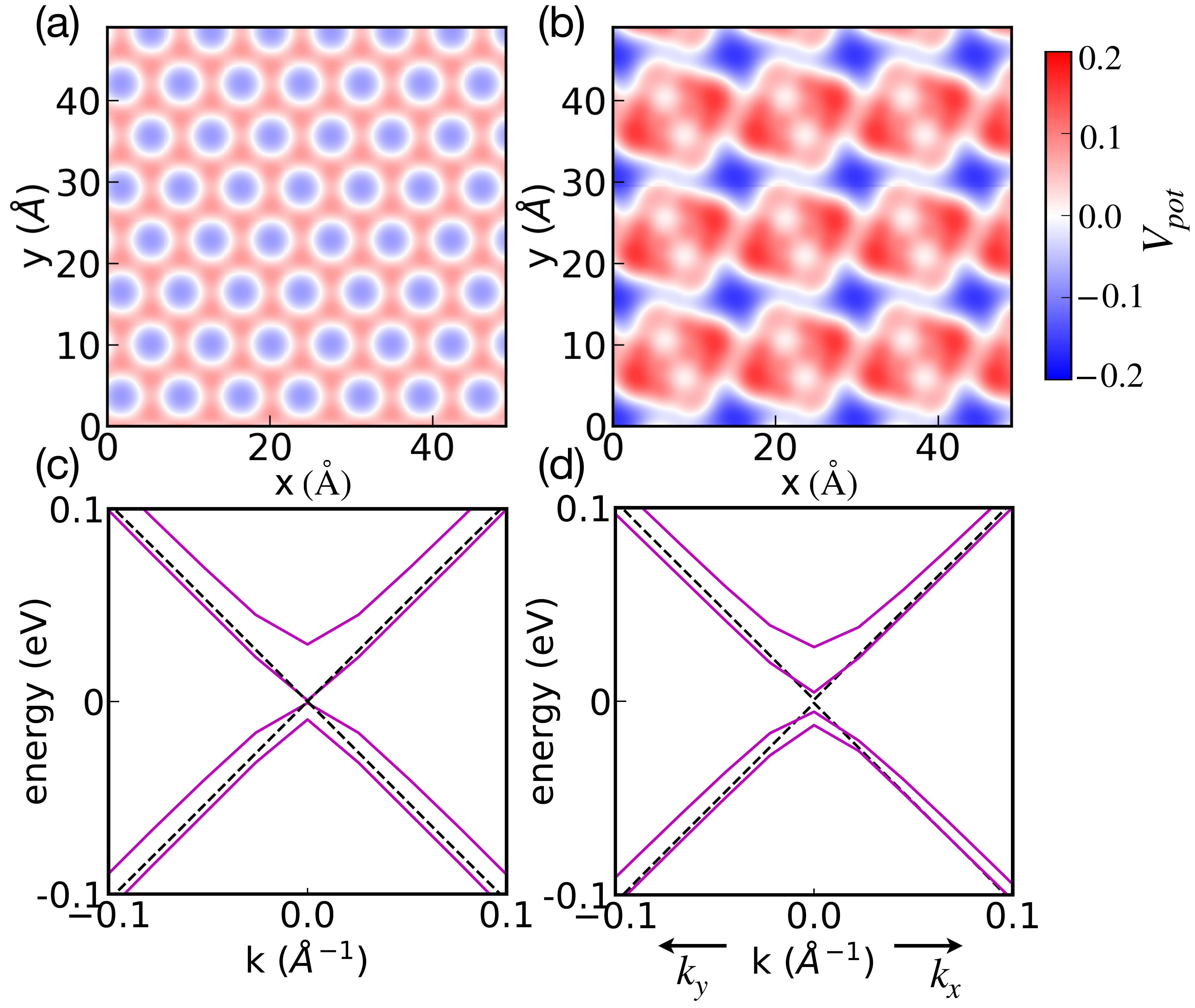}
	\caption{In-plane electrostatic potentials for graphene from (a) C$_6$F$_6$, (b) H$_2$Pc and their energy dispersion (purple) near the Dirac point at $\Gamma$. The dashed black lines are that for pristine graphene.}
	\label{figSI:Vbnd}
\end{figure}

\begin{figure}[H]
	\centering
	\includegraphics[width=0.5\linewidth, center]{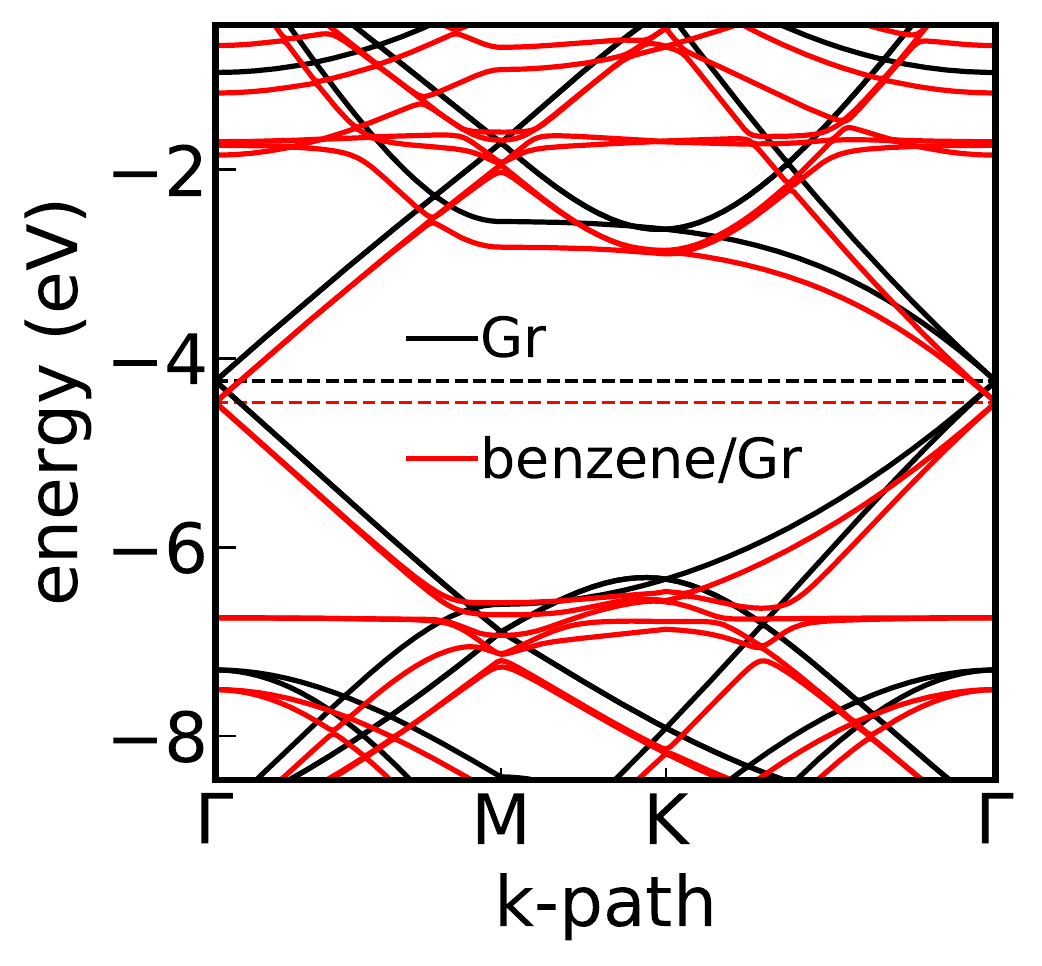}
	\caption{Band structures for pristine graphene (black) and benzene/graphene (red), both in supercell size of 3$\times$3 of graphene primitive cell. The vacuum energy is at zero.}
	\label{figSI:bndPh}
\end{figure}

\begin{figure}[H]
	\centering
	\includegraphics[width=0.85\linewidth, center]{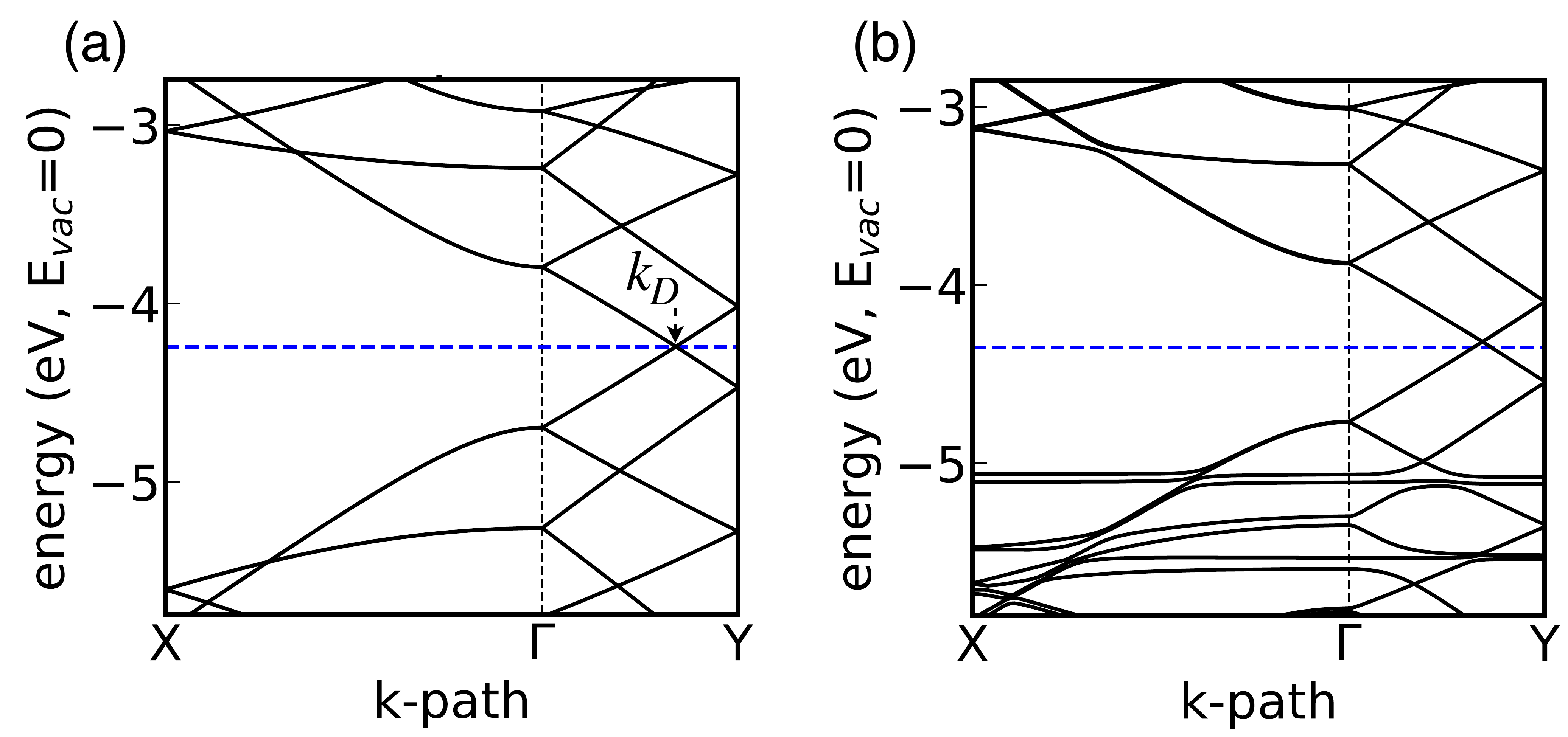}
	\caption{Band structures for (a) pristine graphene (b) C8-BTBT/graphene, both in supercell size of 10 $\times$ 3 $\sqrt{3}$ of graphene primitive cell. The Dirac point $k_D$ is along the path of $\Gamma$-Y.}
	\label{figSI:btall}
\end{figure}

\begin{figure}[H]
		\centering
	\includegraphics[width=0.9\linewidth, center]{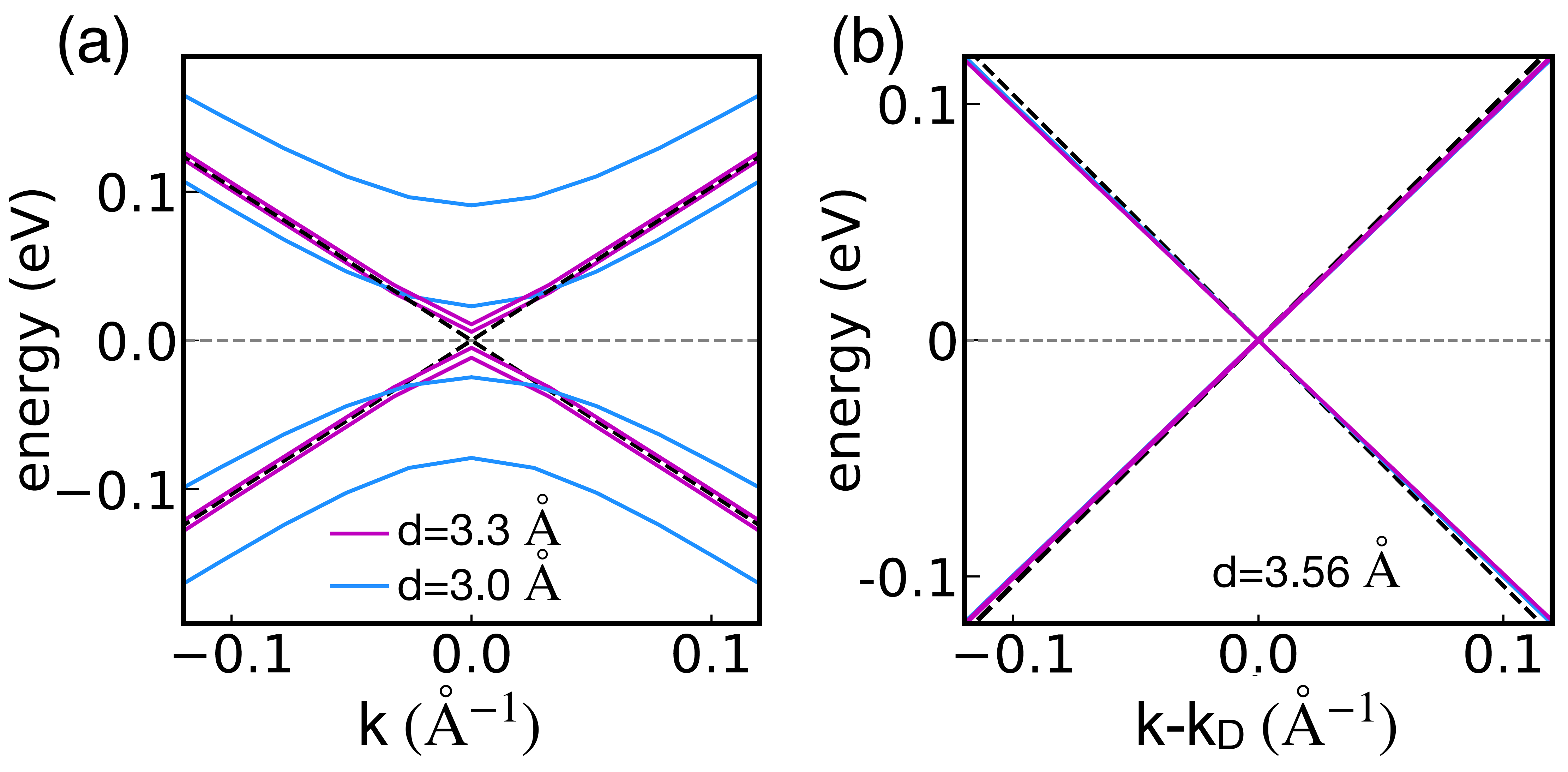}
	\caption{Band dispersion near the Dirac point (k$_D$) for (a) benzene/graphene (bandgaps are 12 meV and 48 meV for $d$ of 3.3 \AA~and 3.0 \AA, respectively) and (b) C8-BTBT/graphene with different molecule-graphene normal distance $d$. For C8-BTBT/graphene, the ground state separation is $d=3.76 \AA$ and the band structure near k$_D$ is shown in Fig. 3. The group velocity for monolayer graphene with C8-BTBT in (b) is reduced by about 3.5\% and 3.9\% along $y$ and $x$ directions, respectively. }
	\label{figSI:bandd}
\end{figure}

\section{Tight-Binding Model Results} \label{Sec:TB}

\begin{figure}[H]
	\centering
	\includegraphics[width=0.8\linewidth, center]{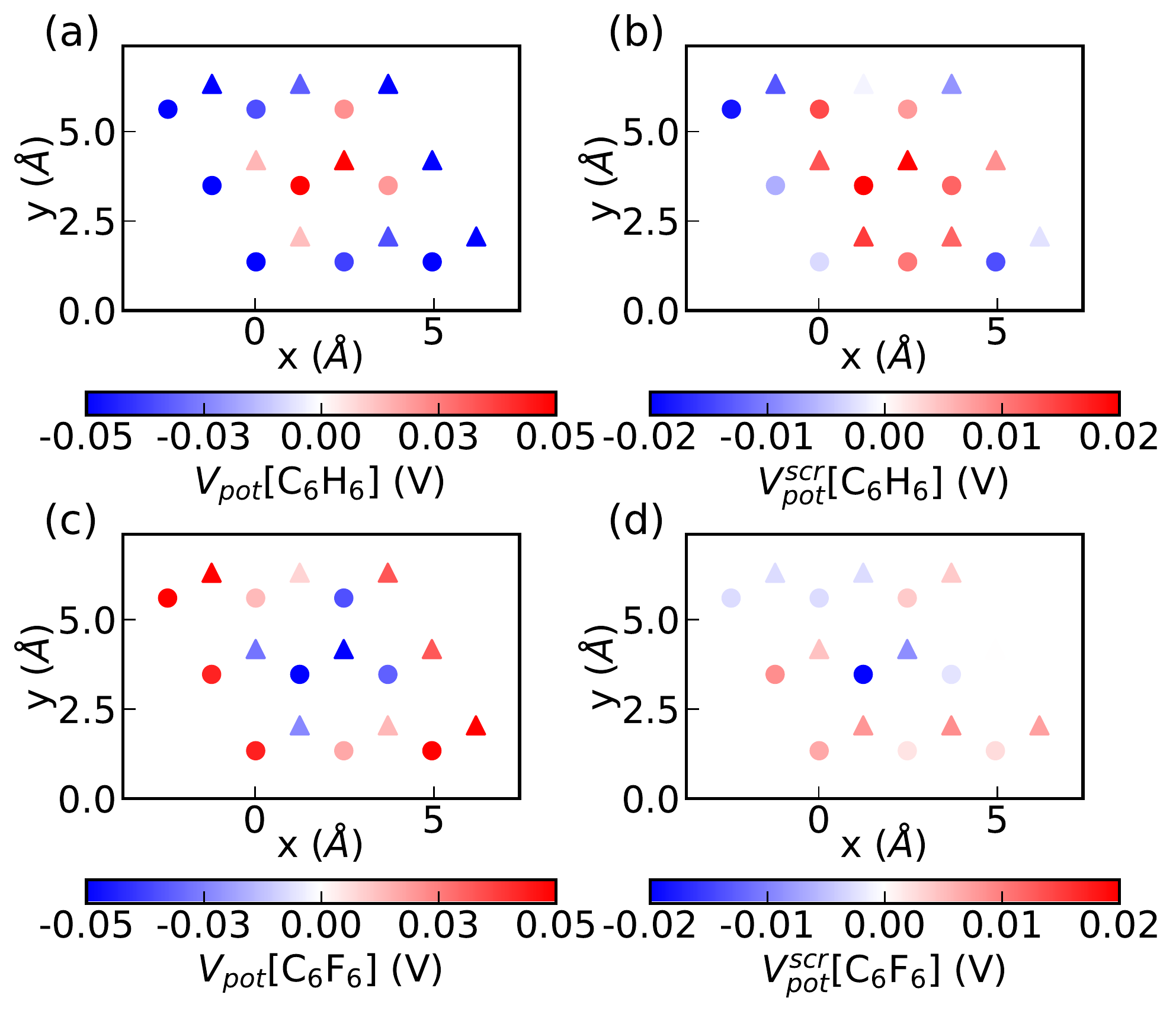}
	\caption{On-site potentials of the two sublattice C atoms represented in dots and triangles, respectively, in a unit cell of benzene/graphene from the DFT-calculated potentials (a) and screened potentials (b). Same for C$_6$F$_6$/graphene in (c) and (d). }
	\label{figSI:VC12Ph}
\end{figure}

\begin{figure}[H]
	\centering
	\includegraphics[width=0.8\linewidth, center]{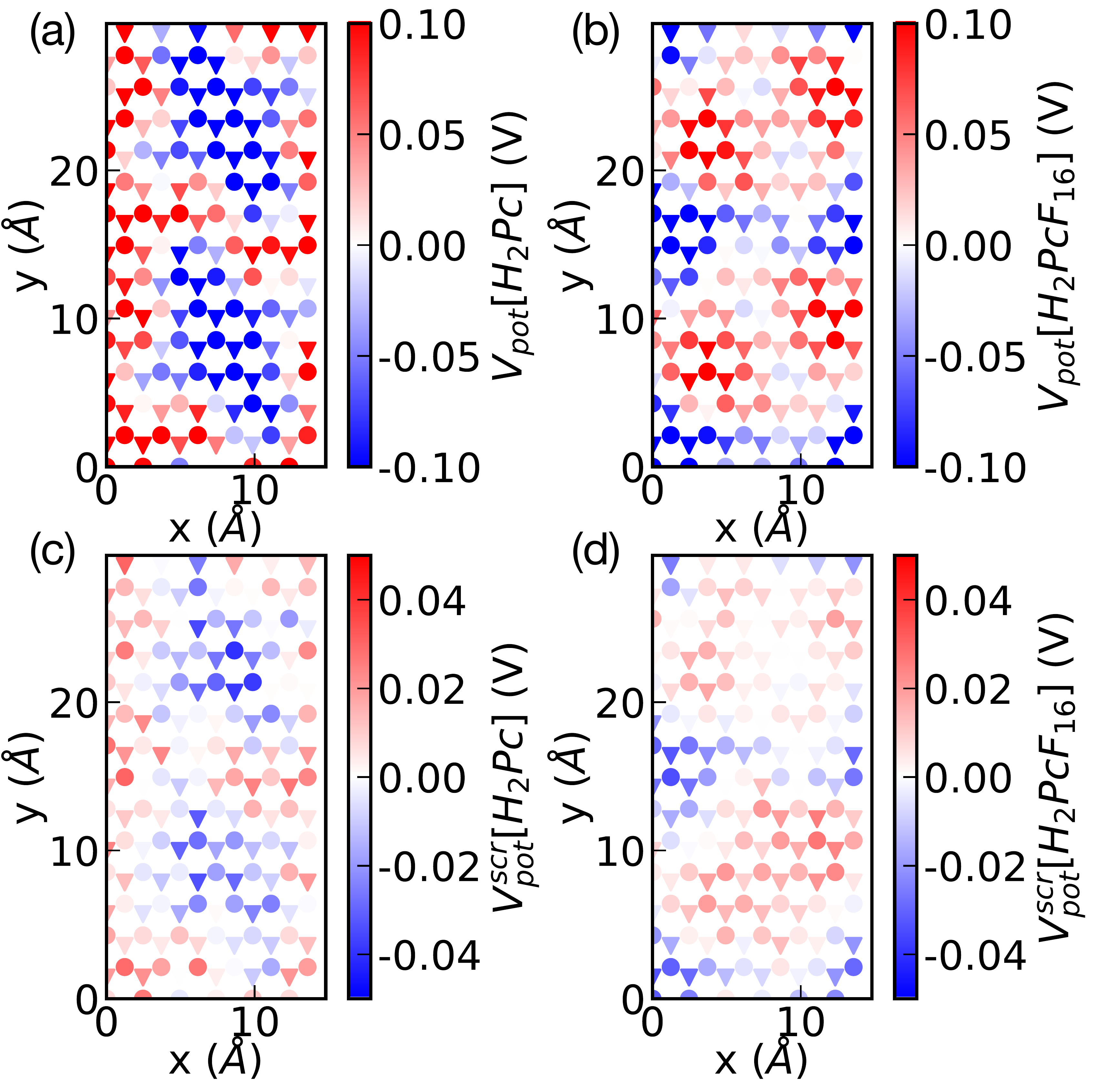}
	\caption{On-site potentials of the two sublattice C atoms represented in dots and triangles, respectively, in a unit cell of H$_2$PcF$_{16}$/graphene from the DFT-calculated potentials (a) and screened potentials (b). Same for H$_2$Pc/graphene in (c) and (d). }
	\label{figSI:VC12Pc}
\end{figure}

Tight-binding model calculations of the band structures for graphene are performed using the PythTB package~\cite{vanderbilt_PythTB2018}. To investigate the effect of the electrostatic potentials from molecular monolayers, we use the simple tight-binding model for monolayer graphene including only nearest-neighbor hopping parameter of 2.7 eV~\cite{Geim2009GrRev} to calculate the band structure. The DFT-computed onsite potentials for each C atom in graphene for benzene/C$_6$F$_6$ and H$_2$Pc(F$_{16}$) molecular monolayers are shown in Fig.~\ref{figSI:VC12Ph} and Fig.~\ref{figSI:VC12Pc}, respectively.

\begin{figure}[H]
	\centering
	\includegraphics[width=0.8\linewidth, center]{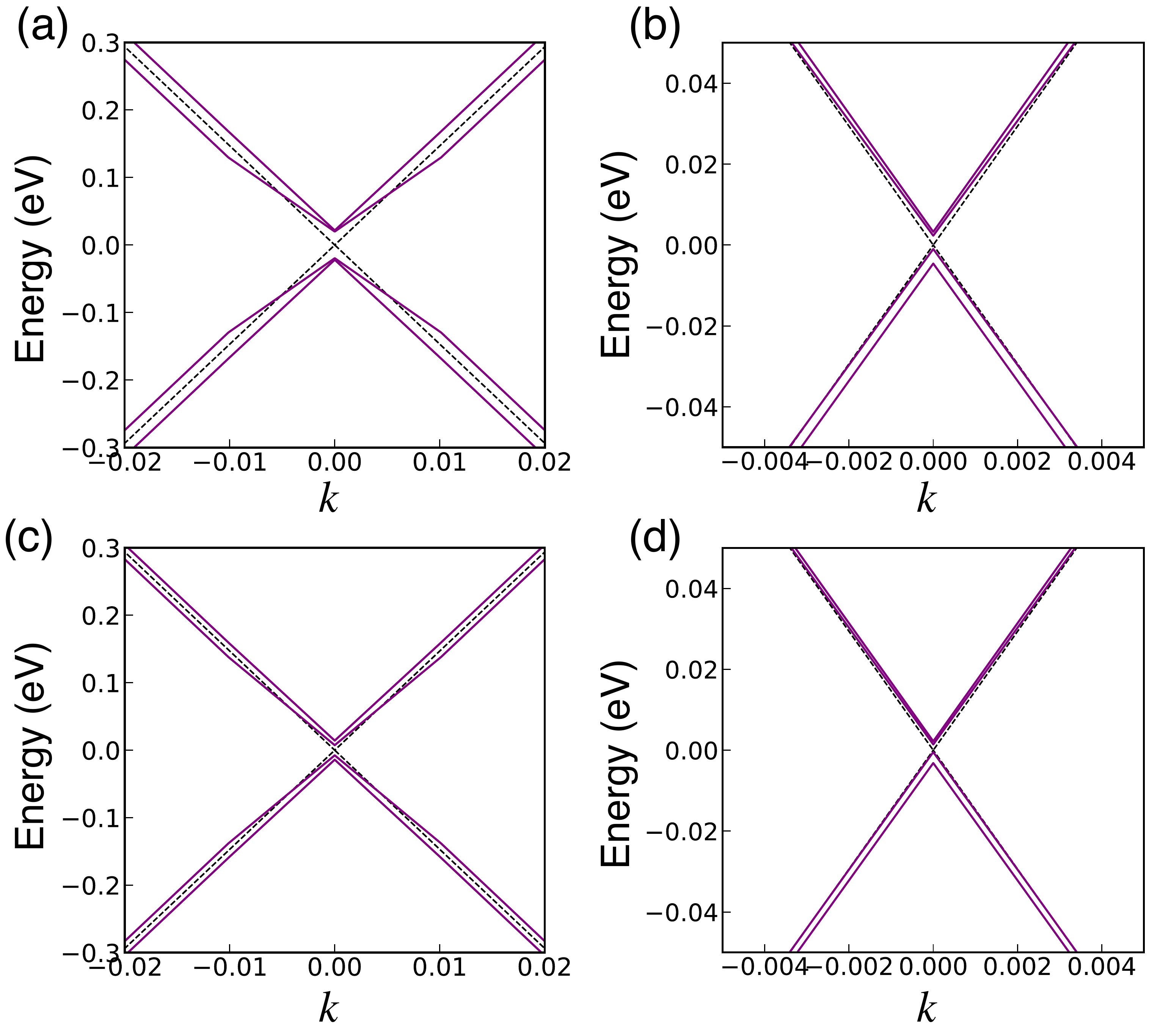}
	\caption{Band structures near the Dirac point at $\Gamma$ for graphene 3$\times$3 superlattice with on-site potentials as shown in Fig.~\ref{figSI:VC12Ph}. (a), (b) band structure from graphene with on-site potentials as shown in Fig.~\ref{figSI:VC12Ph}(a), (b), respectively. (c), (d) with on-site potentials as shown in Fig.~\ref{figSI:VC12Ph}(c), (d), respectively. The dashed lines are for pristine graphene in the same superlattice.}
	\label{figSI:tbPhV}
\end{figure}

\begin{figure}[H]
	\centering
	\includegraphics[width=0.8\linewidth, center]{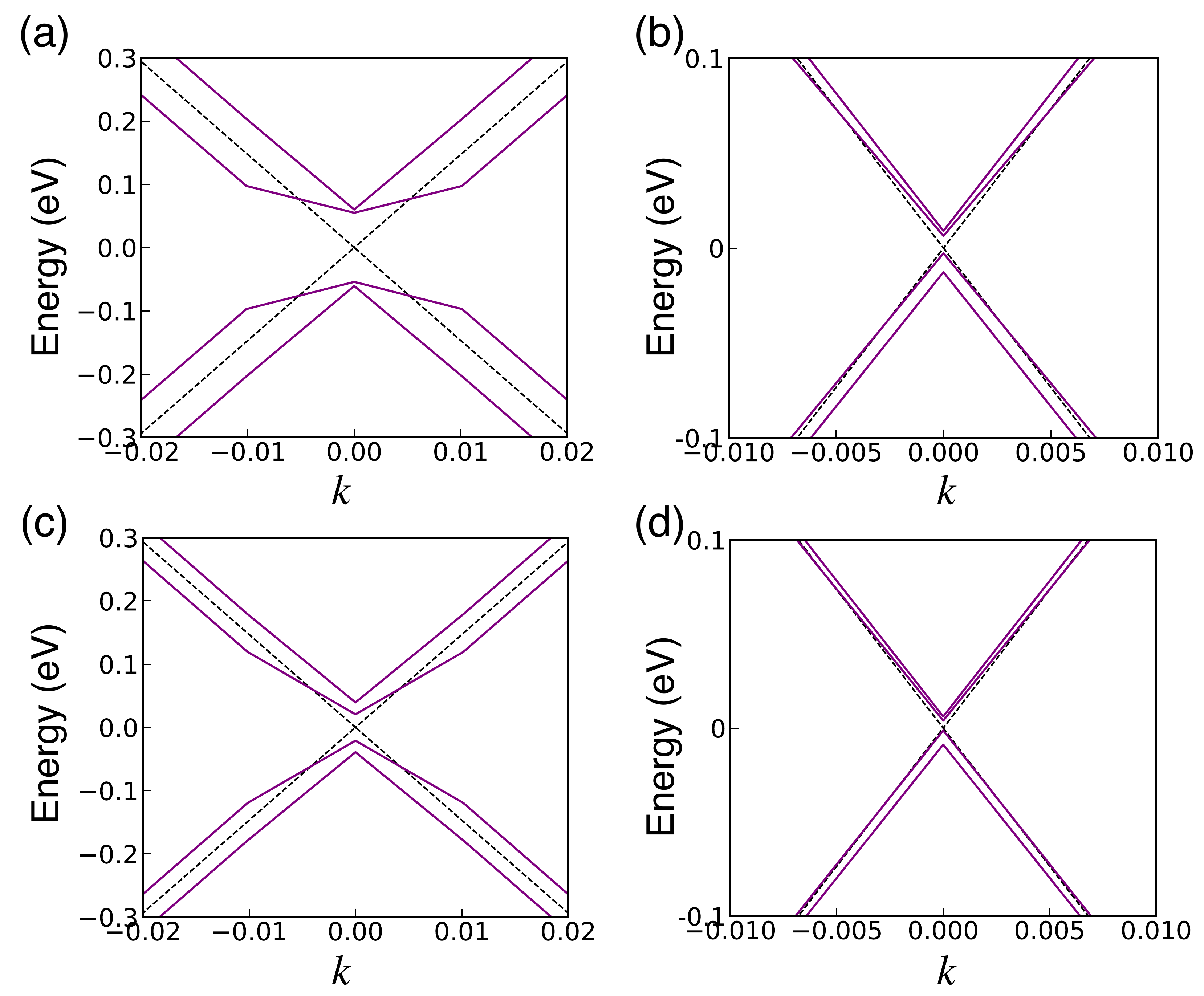}
	\caption{Same as that in Fig.~\ref{figSI:tbPhV} but with on-site potentials from benzene and C$_6$F$_6$ molecular layers of 1\AA\ smaller distance from graphene than the equilibrium structures, i.e. $d=2.31$\AA. The on-site potentials are predicted using the fitted exponential functions with parameters shown in Table~\ref{tbl:dV}, which are about 2.77 times of that shown in Fig.~\ref{figSI:VC12Ph}. }
	\label{figSI:tbPhV1}
\end{figure}

\begin{figure}[H]
	\centering
	\includegraphics[width=0.8\linewidth, center]{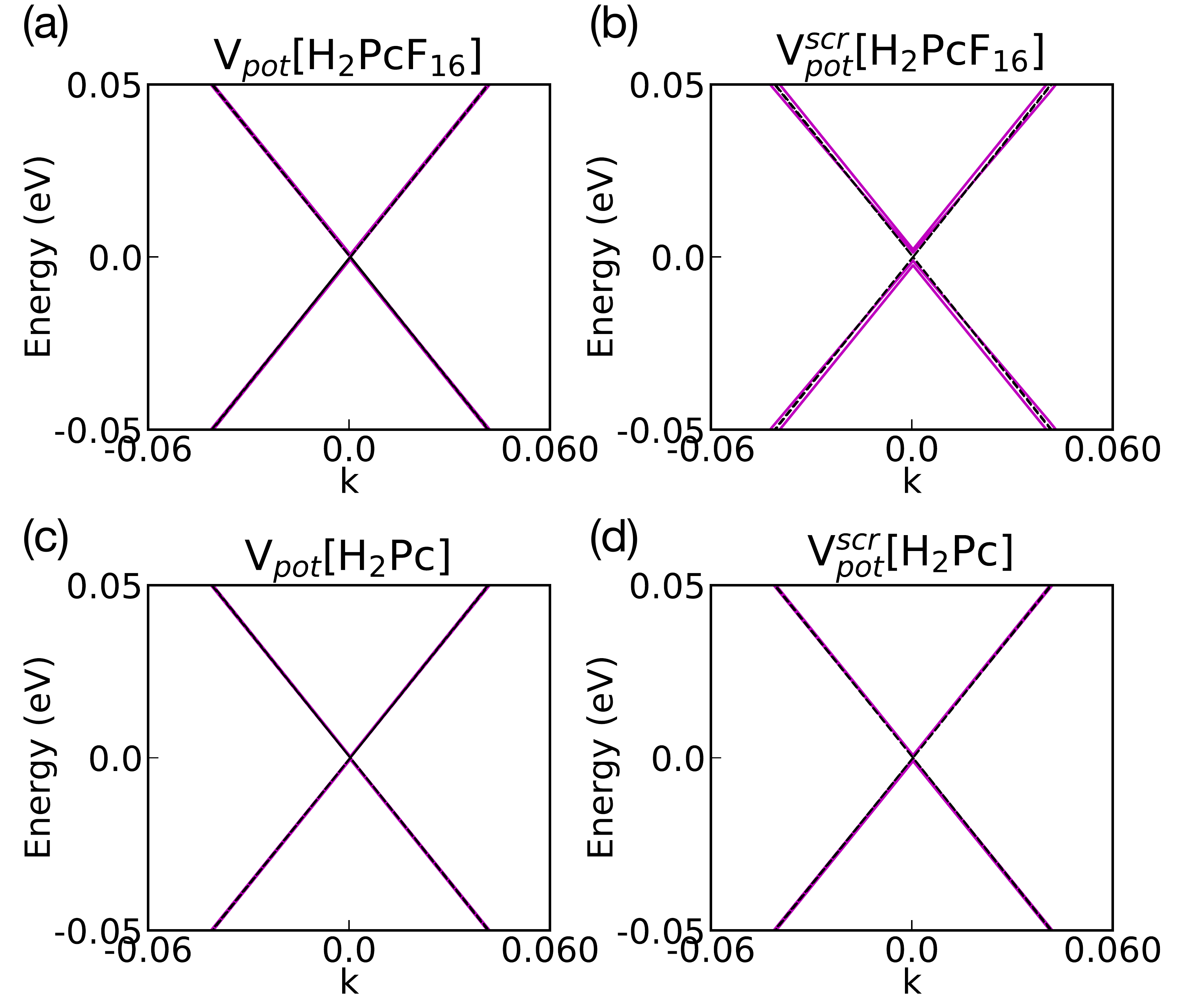}
	\caption{Band structures near the Dirac point at $\Gamma$ using tight-binding calculations for graphene with superlattices and on-site potentials as shown in Fig.~\ref{figSI:VC12Pc} (a-d), respectively.}
	\label{figSI:tbPcV}
\end{figure}

\begin{figure}[H]
	\centering
	\includegraphics[width=0.8\linewidth, center]{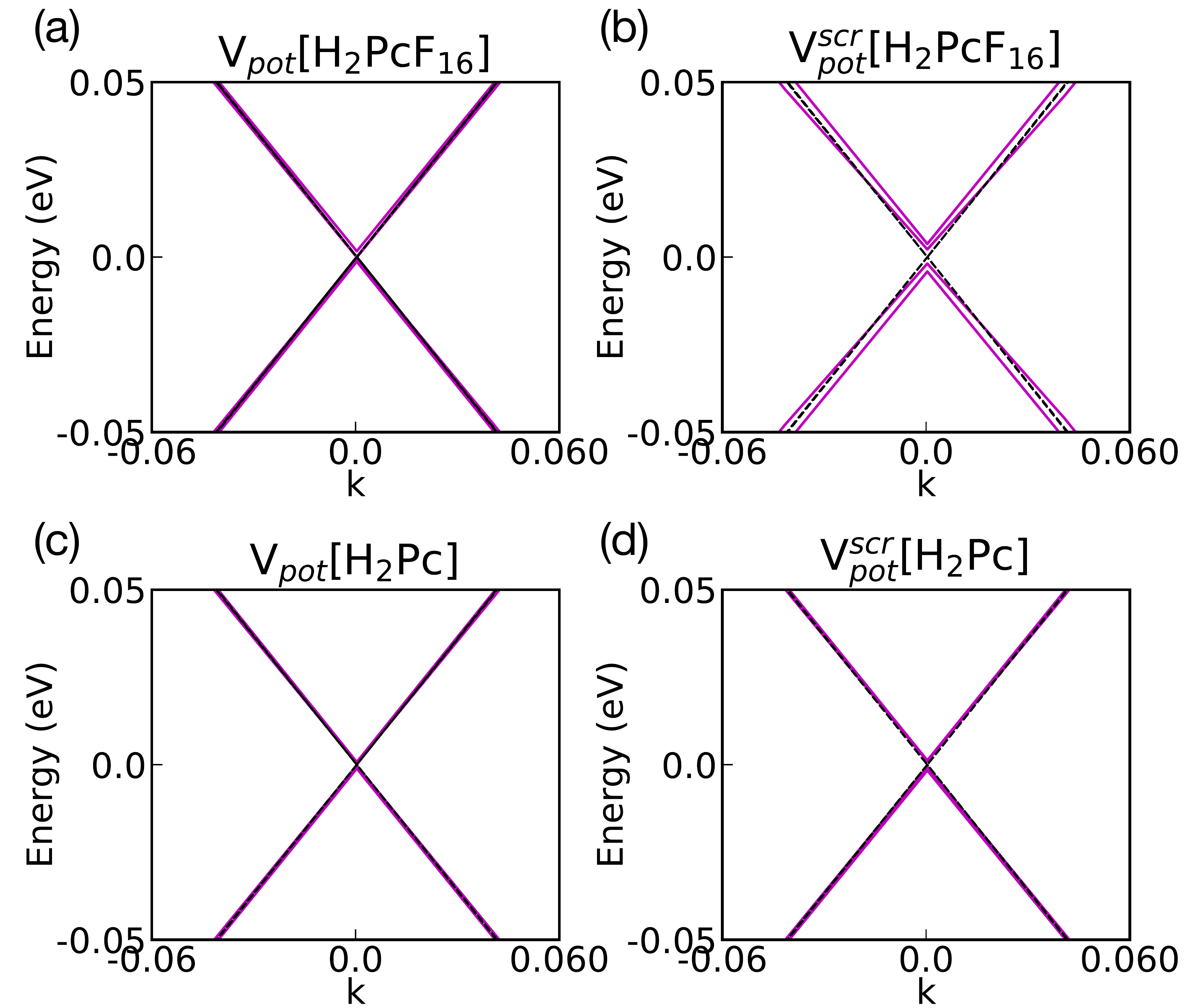}
	\caption{Same as that in Fig.~\ref{figSI:tbPcV} but with on-site potentials from H$_2$Pc(F$_{16}$) molecular layers of 1\AA\ smaller distance from graphene than the equilibrium structures, i.e. $d\simeq 3.35$\AA. The on-site potentials are predicted using the fitted exponential functions with parameters shown in Table~\ref{tbl:dV}, which are about 1.7 times of that shown in Fig.~\ref{figSI:VC12Pc}.}
	\label{figSI:tbPcV1}
\end{figure}

\section{Interface Charge Redistribution}

\begin{figure}[H]
	\centering
	\includegraphics[width=0.8\linewidth, center]{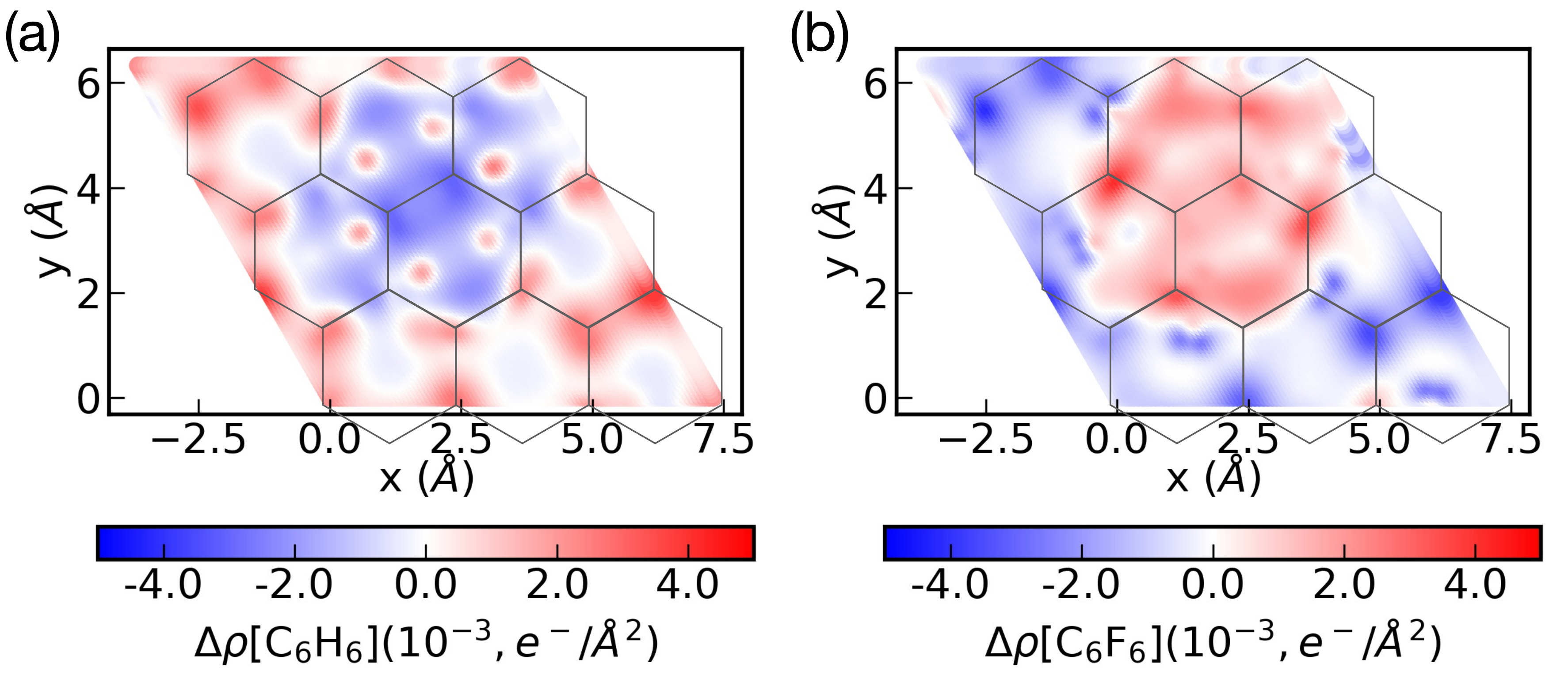}
	\caption{Charge redistribution $\Delta \rho=\rho[SAM/Gr]-\rho[Gr]-\rho[SAM]$ where SAM is the self-assembled molecular layers of benzene (a) and C$_6$F$_6$ (b).}
	\label{figSI:drhoPh}
\end{figure}

\begin{figure}[H]
	\centering
	\includegraphics[width=0.9\linewidth, center]{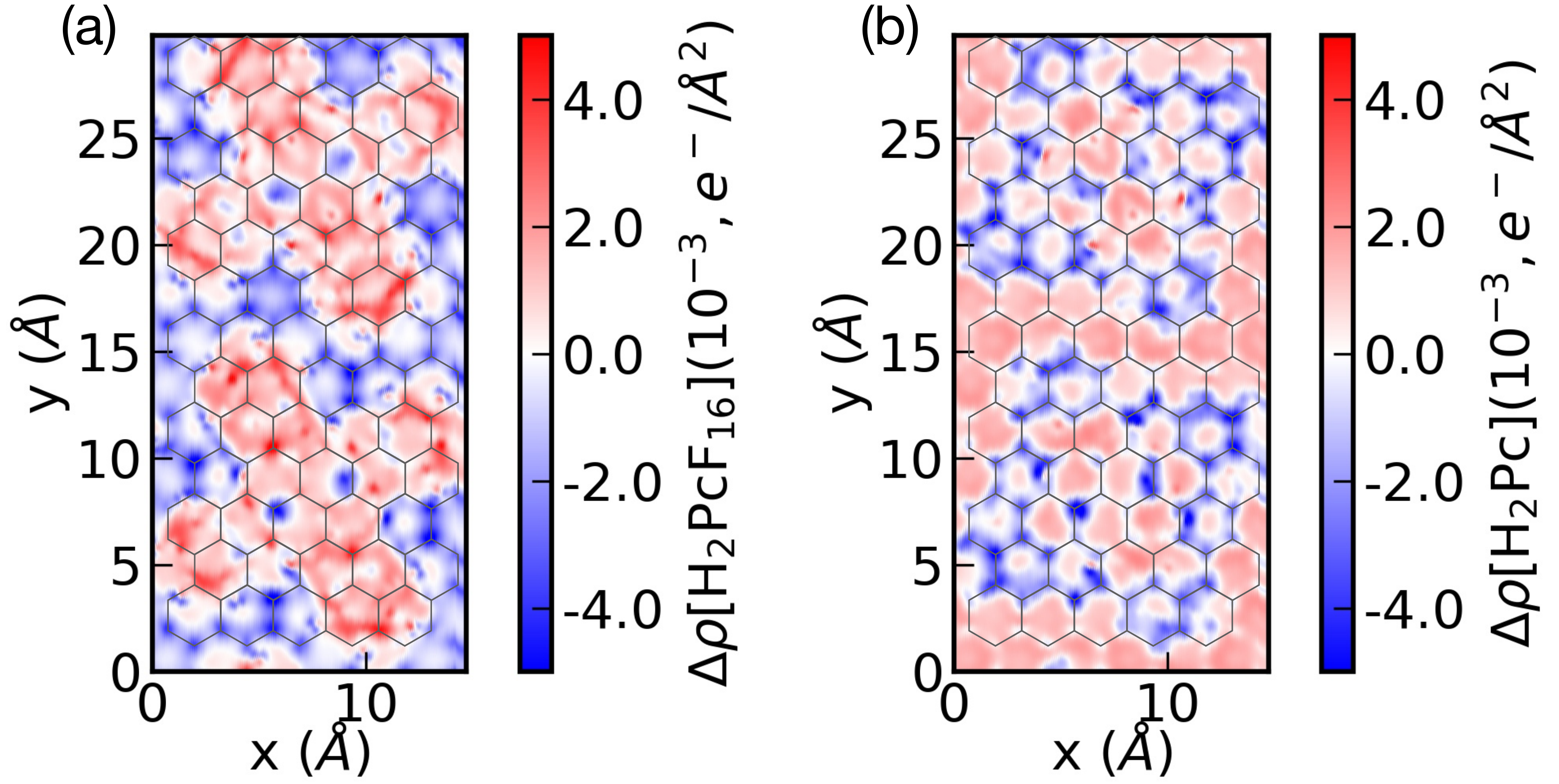}
	\caption{Same as in Fig.~\ref{figSI:drhoPh} for H$_2$PcF$_{16}$ (a) and H$_2$Pc (b).}
	\label{figSI:drhoPc}
\end{figure}

\section{Analytical Solution to the DPCD Model} \label{SecS:formula}
\subsection{Convergence of the Analytical Formula}
\begin{figure}[H]
	\includegraphics[width=0.95\linewidth, center]{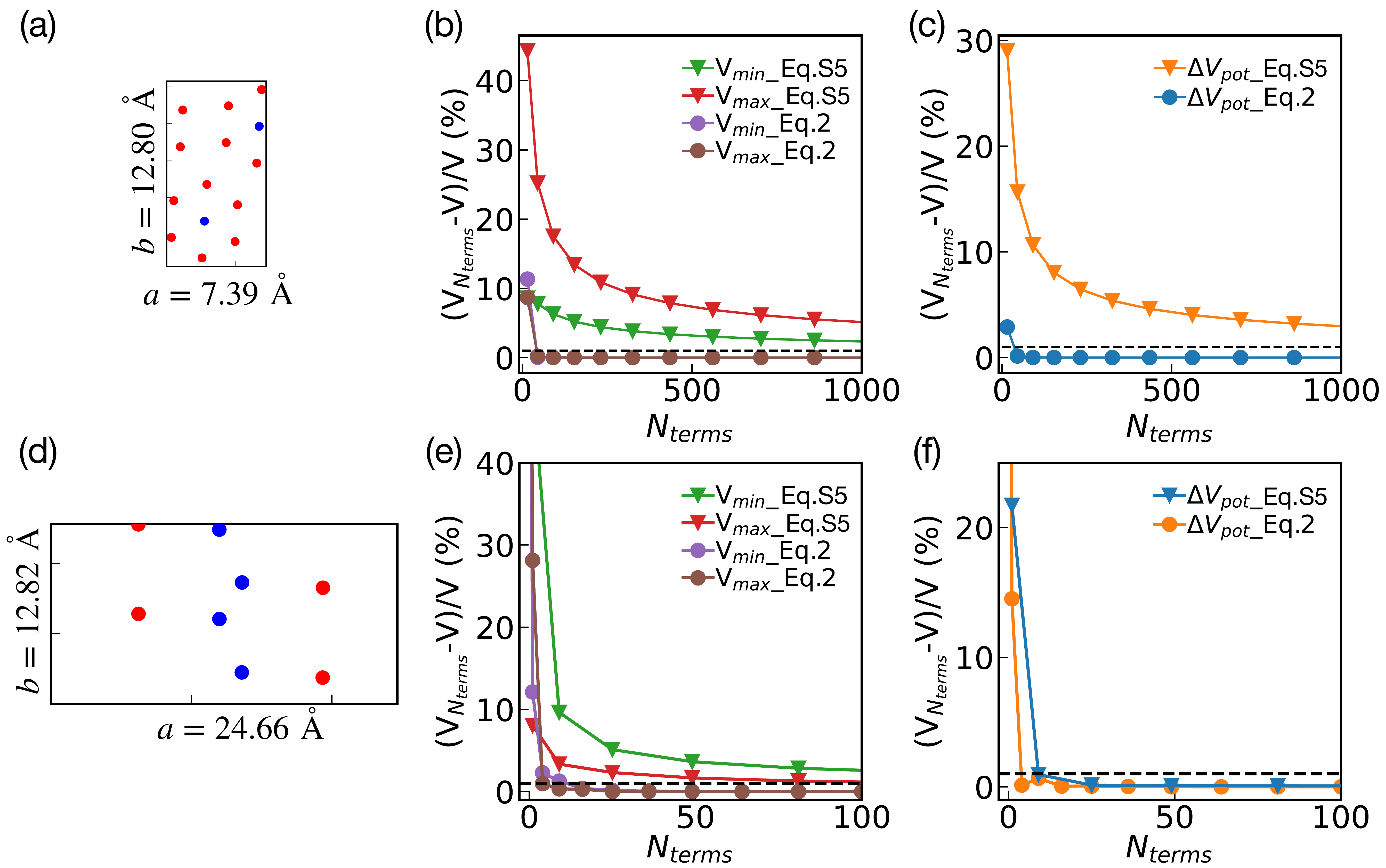}
	\caption{Convergence test for electrostatic potential calculations using Eq.~\ref{Eq:direct} and Eq. 2, for unit cell sizes and charge distributions as shown in (a) and (d), which are that for benzene/graphene as shown in Fig.~\ref{figSI:Vbenzene} (b) and C8-BTBT/graphene (Fig.~\ref{figSI:dVbtbt}), respectively. Corresponding convergence of $V_{min}$, $V_{max}$ and $\Delta V_{pot}=V_{max}-V_{min}$ are shown in (b-c) and (d-e) for charges in (a) and (d), respectively. $N_{terms}$ denotes the number of terms in the summation in Eq.~\ref{Eq:direct} and Eq.2 for each point charge $q_i$. The dashed black lines show error of 1\%, accuracy of 99\%.}
	\label{figSI:Nterms}
\end{figure}

\subsection{Analytical Derivation}

Here we consider the electrostatic potential of a neutral, periodic charge configuration contained by a plane. For simplicity we restrict the charges to be in the plane $z=0$. Furthermore, we consider the unit cell of the charge configuration to be a rectangular lattice with lattice vectors $\vec{a} = a\hat{x}$ and $\vec{b} = b\hat{y}$. Thus, we have a potential of the form
\begin{align}\label{Eq:direct}
V(x,y,z=d) = k\sum_{n_1,n_2,i} \frac{q_i}{\sqrt{(n_1a+x_i-x)^2+(n_2b+y_i-y)^2+d^2}} \quad \& \quad \sum_i q_i = 0
\end{align}
where $i$ is a sum over charges and $n_1$ and $n_2$ are sums over lattice points. We can rearrange this into a dimensionful piece and a dimensionless sum of the form
\begin{align}
V(x,y,z=d) = k\sum_i\frac{q_i}{a}\sum_{\vec{n}} \left((n_1-\alpha)^2+(\gamma n_2-\beta)^2+\delta^2\right)^{-1/2}\label{eq:V},
\end{align}
where $\alpha = (x-x_i)/a$, $\beta = (y-y_i)/a$, $\gamma = b/a$, and $\delta = d/a$. This sum converges very slowly via direct summation. We remark that the dimensionless sum is the $s=1/2$ case of the inhomogeneous Epstein-Hurwitz $\zeta$-function of the form
\begin{align}
\zeta(s) = \sum_{\vec{n}}\left( (n_1-\alpha)^2+(\gamma n_2-\beta)^2+\delta^2 \right)^{-s}.\label{eq:z}
\end{align}
For such two-dimensional $zeta$-functions, the sum converges much faster in Fourier space. Furthermore, by constructing an appropriate $\theta$-function, using Poisson resummation and taking a Mellin transform, we can obtain the Fourier Transform of the above sum.

First we define the Mellin transform. Given a function $\phi(t)$, the {\em Mellin Transform}, $\tilde{\phi}(s)$ is defined as:
\begin{align}
\tilde{\phi}(s) = \int_0^{\infty}t^{s-1}\phi(t) dt
\end{align}
This is well defined if $\phi(t)$ goes to 0 sufficiently fast at both $t=0$ and $\infty$. When this is not the case, the transform can be defined in a piecewise nature for $t\rightarrow 0$ and $t\rightarrow \infty$. We note that for $\phi(t) = e^{-t}$
\begin{align}
\tilde{\phi}(s) = \int_0^{\infty}e^{-t}t^{s-1}dt = \Gamma(s)
\end{align}
\vs
A {\em $\theta$-function} has the following form:
\begin{align}
\theta (t) \equiv \sum_{n=-\infty}^{n=\infty} e^{-\pi n^2 t} = 1+2\sum_{n=1}^{n=\infty}  e^{-\pi n^2 t}
\end{align}
We note that the Mellin Transform of a constant is 0; therefore, $\tilde{\theta}(s) = 2\Gamma(s)\pi^{-s}\zeta(2s)$. Finally, we derive the {\em $\theta$-function identity} in one dimension. We will use Poisson resummation, namely: $\sum_n f(n) = \sum_k\hat{f}(k)$ where $n,k\in \mathbb{Z}$, where $\hat{f}$ denotes the Fourier Transform.

\begin{align*}
f_t(x) &= e^{-\pi t x^2}\\
\hat{f}_t(y) &= \int_{-\infty}^{\infty} e^{-\pi t x^2+2\pi i x y} dx \\
&= e^{-\pi y^2/t}\int_{-\infty}^{\infty} e^{-\pi t (x+iy/t)^2} dx \\
&= f_{1/t}(y) \int_{-\infty}^{\infty} e^{-\pi t x^2}dx\\
&=\frac{1}{\sqrt{t}}f_{1/t}(y)
\end{align*}
Now, applying the Poisson resummation:
\begin{align*}
\sum_n f_t(n) = \theta(t) = \sum_n\hat{f}(n)=\frac{1}{\sqrt{t}}\theta(1/t)
\end{align*}

With this machinery in place, we define a  $\theta$-function whose Mellin transform is related to the $\zeta$-function of eqn.~\ref{eq:z}.
\begin{align}
\theta (t) = \sum_{\vec{n}}  \exp{\left[-\pi \left( (n_1-\alpha)^2+(\gamma n_2-\beta)^2+\delta^2 \right) t\right]}   \label{eq:th}
\end{align}
Taking the Fourier transform of one component of eqn.~\ref{eq:th}. We can use Poisson resummation to write our needed $\theta$-function identity.
\begin{align*}
\int_{-\infty}^{\infty}\int_{-\infty}^{\infty} \exp{\left[-\pi \left( (n_1-\alpha)^2+(\gamma n_2-\beta)^2+\delta^2 \right) t - 2\pi i n_1k_1 - 2\pi i n_2k_2
	\right]} dn_1dn_2
\end{align*}
Separating components
\begin{align*}
e^{-\pi\delta^2 t}\left(\int_{-\infty}^{\infty}\exp{\left[-\pi (n_1-\alpha)^2t  - 2\pi i n_1k_1\right]} dn_1\right)\left(\int_{-\infty}^{\infty}\exp{\left[-\pi (\gamma n_2-\beta)^2t  - 2\pi i n_2k_2\right]} dn_2\right)
\end{align*}
Completing the square of the argument of the exponentials, 
\begin{align*}
e^{-\pi\delta^2 t}\left(\int_{-\infty}^{\infty}\exp{\left[-\pi t(n_1-\left(\alpha-ik_1/t\right))^2-\pi t\left(k_1^2/t^2+2\alpha i k_1/t\right)\right]} dn_1\right)\\
\times \left(\int_{-\infty}^{\infty}\exp{\left[-\pi t(\gamma n_2-\left(\beta-ik_2/t\right))^2-\pi t\left(k_2^2/\gamma^2t^2+2\beta i k_2/\gamma t\right)\right]} dn_2\right),
\end{align*}
we find the product of two Gaussian integrals multiplied by some exponential factors. The first Gaussian integral is simply $1/\sqrt{t}$ as the argument just required a linear shift in the change of variables. The the second Gaussian integral, requires the changes of variables $n' = \gamma n_2-\left(\beta-ik_2/t\right)$, and thus $dn' = \gamma dn_2$ and the integral evaluates to $1/\gamma\sqrt{t}$:
\begin{align*}
\exp{\left[-\pi\left(\delta^2+ \left(k_1^2/t+2\alpha i k_1 \right)+\left(k_2^2/\gamma^2t+2\beta i k_2/\gamma\right)\right)\right]}
\frac{1}{\sqrt{t}}\frac{1}{\gamma\sqrt{t}}
\end{align*}
Finally, using Poisson resummation we have the following identity:
\begin{align}
\theta (t) =& \sum_{\vec{n}}  \exp{\left[-\pi \left( (n_1-\alpha)^2+(\gamma n_2-\beta)^2+\delta^2 \right) t\right]}\nonumber\\
=& \frac{1}{\gamma t}
\sum_{\vec{k}}  \exp{\left[-2\pi i \alpha k_1- 2\pi i \beta k_2/\gamma - \frac{\pi}{t}\left( k_1^2/+k_2^2/\gamma^2\right)-\pi t\delta^2\right]}
\label{eq:pr}
\end{align}

Taking the Mellin transform of eqn.~\ref{eq:th} we find the relation to eqn.~\ref{eq:z}.
\begin{align}
\tilde{\theta}(s) = \int_0^{\infty}t^{s-1}\theta(t)dt &= \int_0^{\infty}t^{s-1} \sum_{\vec{n}}  \exp{\left[-\pi \left( (n_1-\alpha)^2+(\gamma n_2-\beta)^2+\delta^2 \right) t\right]} dt\nonumber\\
&= \sum_{\vec{n}} \int_0^{\infty}t^{s-1}  \exp{\left[-\pi \left( (n_1-\alpha)^2+(\gamma n_2-\beta)^2+\delta^2 \right) t\right]} dt\nonumber\\
&= \pi^{-s}\sum_{\vec{n}} \Gamma(s) \left( (n_1-\alpha)^2+(\gamma n_2-\beta)^2+\delta^2 \right)^{-s}\nonumber\\
\tilde{\theta}(s)
&= \pi^{-s}\Gamma(s)\zeta(s)\label{eq:rs}
\end{align}
This time instead of evaluating the Mellin transform in real-space, we evaluate it in Fourier-space using the identity of eqn.~\ref{eq:pr}
\begin{align*}
\tilde{\theta}(s)&= \int_0^{\infty}t^{s-1} \sum_{\vec{n}}  \exp{\left[-\pi \left( (n_1-\alpha)^2+(\gamma n_2-\beta)^2+\delta^2 \right) t\right]} dt\\
&= \int_0^{\infty}t^{s-1} \frac{1}{\gamma t} \sum_{\vec{k}}  \exp{\left[-2\pi i \alpha k_1- 2\pi i \beta k_2/\gamma - \frac{\pi}{t}\left( k_1^2/+k_2^2/\gamma^2\right)-\pi t\delta^2\right]} dt
\end{align*}.
First we note that the $\vec{k}=0$ term of the above expression evaluates to
\begin{align*}
\int_0^{\infty}t^{s-1-2}\frac{1}{\gamma}e^{-\pi t \delta^2} = \Gamma(s-1)\frac{1}{\gamma}\left(\pi\delta^2\right)^{-(s-1)}
\end{align*}
To evaluate the $\vec{k}\ne 0$ terms, we make use of the integral representation of the modified Bessel function, $K_{\nu}$
\begin{align*}
K_{\nu}(2\sqrt{\alpha \beta}) = \frac{1}{2}\left(\frac{\beta}{\alpha}\right)^{\nu/2}\int_0^{\infty}t^{\nu-1}e^{-\frac{\alpha}{t}-\beta t}dt,
\end{align*}

with $\nu = s-1$, $\alpha  = \pi(k_1^2+k_2^2/\gamma^2)$, $\beta=\pi\delta^2$. Therefore we can write the Mellin transform of eqn.~\ref{eq:th} as

\begin{align}
\tilde{\theta}(s)=&\frac{1}{\gamma}\sum_{\vec{k}}{}' \exp{\left[-2\pi i \alpha k_1- 2\pi i \beta k_2/\gamma\right]} 2\left(\frac{a}{b} \right)^{\nu/2}K_{\nu}\left(2\sqrt{ab}\right)+\frac{1}{\gamma} \frac{ \Gamma(s-1)}{(\pi\delta^2)^{(s-1)}}\nonumber\\
=&\frac{2}{\gamma\delta^{s-1}}\sum_{\vec{k}}{}' \exp{\left[-2\pi i \alpha k_1- 2\pi i \beta k_2/\gamma\right]} \sqrt{k_1^2+\left(\frac{k_2}{\gamma}\right)^2}
K_{s-1}\left(2\pi\delta \sqrt{k_1^2+\left(\frac{k_2}{\gamma}\right)^2}\right)\nonumber\\
&+\frac{ \Gamma(s-1)}{\gamma(\pi\delta^2)^{s-1}}\label{eq:fs}
\end{align}
where the prime on the sum indicates omitting the  $\vec{k}=0$ term. Setting eqns.~\ref{eq:rs} and .~\ref{eq:fs} equal and solving for $\zeta(s)$ we find
\begin{flalign*}
&\zeta(s) = \frac{\pi^s}{\Gamma(s)}\times\\
&\left[ \frac{2}{\gamma\delta^{s-1}}\sum_{\vec{k}}{}' \exp{\left[-2\pi i \alpha k_1- 2\pi i \beta k_2/\gamma\right]} \sqrt{k_1^2+\left(\frac{k_2}{\gamma}\right)^2}
K_{s-1}\left(2\pi\delta \sqrt{k_1^2+\left(\frac{k_2}{\gamma}\right)^2}\right)+\frac{ \Gamma(s-1)}{\gamma(\pi\delta^2)^{s-1}}\right].&&
\end{flalign*}
As the ratio of sequential integer values of the $\Gamma$-function is $\Gamma(n-1)/\Gamma(n) = (n-2)!/(n-1)!  = 1/(n-1)$ and for $s=1/2$, $\Gamma(1/2)=\sqrt{\pi}$ and $K_{1/2}(z) = K_{-1/2}(z) = \sqrt{\pi/2z}e^{-z}$, we find
\begin{align}
\zeta\left(\frac{1}{2}\right) = \frac{1}{\gamma}\sum_{\vec{k}}{}'
\frac{e^{-2\pi\left[ i \alpha k_1+ i \beta k_2/\gamma+\delta \sqrt{k_1^2+\left(\frac{k_2}{\gamma}\right)^2}\right]}}
{\sqrt{k_1^2+\left(\frac{k_2}{\gamma}\right)^2}}-\frac{\pi\delta}{\gamma}
\end{align}
We note that the $\frac{\pi\delta}{\gamma}$ diverges as $\delta$ increases; however, as we have required that the periodic charge configuration is neutral, all the divergent terms cancel and we can ignore this term going forward. Thus the potential, eqn.~\ref{eq:V}, can be written in terms of a sum over Fourier components as
\begin{align}\label{Eq:Vexp}
V(x,y,z=d)  \nonumber
&= \sum_i\frac{q_i}{a} \zeta\left(\frac{1}{2}\right)\\   
&=\sum_i \frac{q_i}{a} \sum_{\vec{k}}{}' 
\frac{e^{-2\pi\left[ i k_1(x-x_i)/a+ i  k_2(y-y_i)/b+d \sqrt{k_1^2/a^2+k_2^2/b^2}\right]}}
{\sqrt{k_1^2b^2/a^2+k_2^2}} 
\end{align}

\subsection{Analytical Formula For the Case of C8-BTBT Monolayer} 

\begin{figure}[H]
	\centering
	\includegraphics[width=0.6\textwidth]{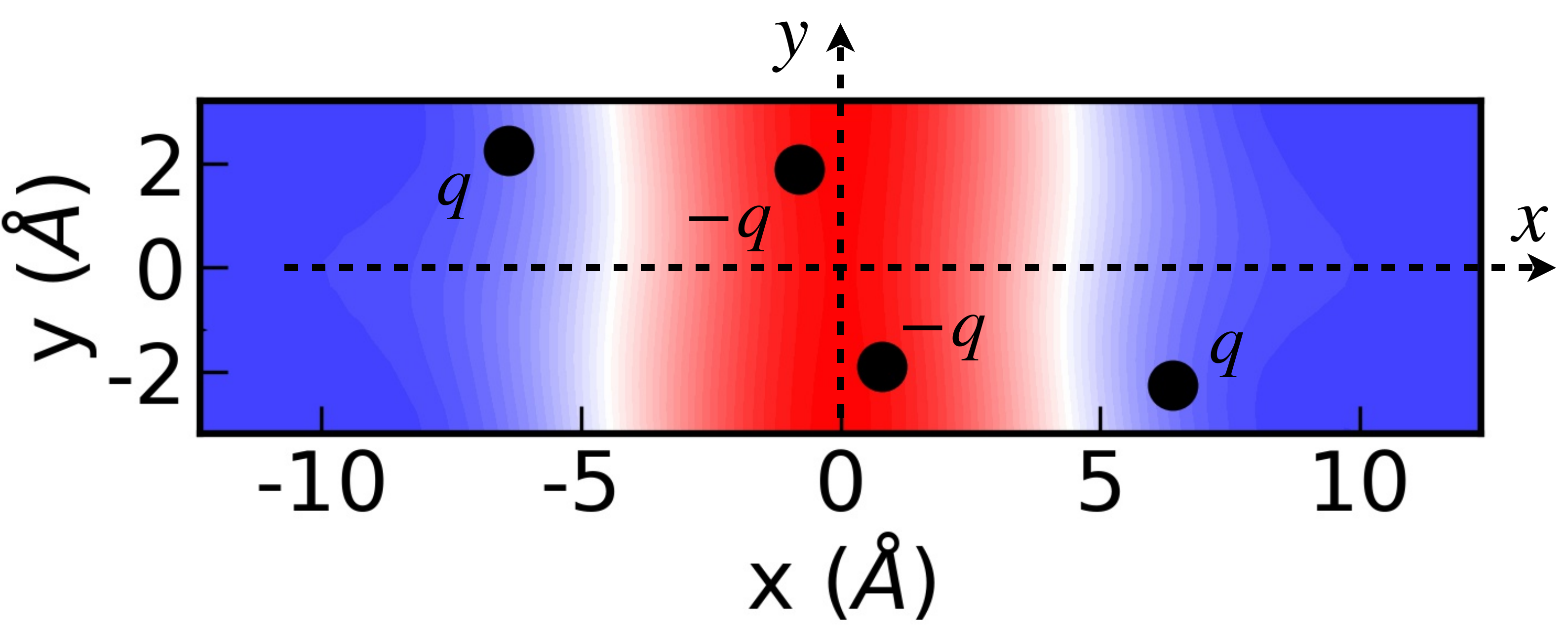}
	\caption{Point charges representing C8-BTBT molecules in one cell. The positions for each point charge from left to right can be defined as $(-x_0a,y_0b,0), (-x_1a, y_1b, 0), (x_1a, -y_1b, 0), (x_0a, -y_0b,0)$ where $a,b$ are the cell sizes along $x, y$ axis, respectively.}
	\label{figSI:4ptcell}
\end{figure}

For C8-BTBT molecular layer, each C8-BTBT molecule is represented by four point charges with values of $q, q, -q, -q$. As shown in Fig.~\ref{figSI:4ptcell}, the positions $\bs r(x,y,z=0)$ of the point charges in one cell can be defined as $ (-x_0a,y_0b,0)$, $(x_0a,-y_0b,0)$, $(-x_1a,y_1b,0)$, $(x_1a, -y_1b,0)$, respectively. The maximum(minimum) potential for electrons lies below the center(edge) of the supercell with lattice parameters $a,b$. The in-plane potential modulation $\Delta V$ therefore is:
\begin{align}
\Delta V & = V_{max}(0,0,d)  - V_{min}(a/2,0,d)
\end{align}

We can expand the summation for each $q_i$ using Eq. 2, %~\ref{Eq:q}, 
and define $M_k=\sqrt{k_1^2/a^2+k_2^2/b^2}$. As $y_0 \approxeq y_1$:
\begin{align}
V_{max}(0,0,d)  & = \frac{q}{4\pi ab}\sum_{\vec{k}}{}'\frac{\cos[2\pi (k_1 x_0+k_2(-y_0)]e^{-2\pi dM_k}}{M_k} \nonumber \\ \nonumber
& + \frac{q}{4\pi ab}\sum_{\vec{k}}{}'\frac{\cos[2\pi (k_1(-x_0)+k_2 y_0]e^{-2\pi dM_k}}{M_k} \\ \nonumber
& + \frac{-q}{4\pi ab}\sum_{\vec{k}}{}'\frac{\cos[2\pi (k_1 x_1+k_2(-y_1))]e^{-2\pi dM_k}}{M_k} \\ \nonumber
& + \frac{-q}{4\pi ab}\sum_{\vec{k}}{}'\frac{\cos[2\pi (k_1 (-x_1)+k_2y_1]e^{-2\pi dM_k}}{M_k} \\ 
& = \frac{2q}{4\pi ab}\sum_{\vec{k}}{}'\frac{e^{-2\pi dM_k}}{M_k}\{\cos[2\pi(k_1x_0-k_2y_0)]-\cos[2\pi(k_1x_1-k_2y_0)]\}
\end{align}
Similarly,
\begin{align}
V_{min}(a/2,0,d)  = &   \\ 
& \frac{2q}{4\pi \varepsilon ab}\sum_{\vec{k}}{}'\frac{e^{-2\pi dM_k}}{M_k}\{\cos[2\pi(k_1(1/2-x_0)-k_2y_0)]-\cos[2\pi(k_1(1/2-x_1)-k_2y_0)]\} \nonumber
\end{align}
Therefore, 
\begin{align}
\Delta V = &V_{max}(0,0,d) - V_{min}(a/2,0,d)  \nonumber \\
& = \frac{2q}{4\pi \varepsilon ab}\sum_{\vec{k}}{}'\frac{N_{k_1,k_2}e^{-2\pi dM_k}}{M_k}
\end{align}
where
\begin{align}
N_{k_1,k_2} & = \cos[2\pi(k_1x_0-k_2y_0)]-\cos[2\pi(k_1x_1-k_2y_0)] \nonumber\\ 
& - \cos[2\pi(k_1(1/2-x_0)-k_2y_0)]+\cos[2\pi(k_1(1/2-x_1)-k_2y_0)]  \nonumber \\ 
& = 
\begin{cases}
2[\sin(2\pi k_1x_0) - \sin(2\pi k_1 x_1)]\sin(2\pi k_2y_0) \quad (k_1=2n, n=0,\pm 1,\pm 2...)\\
2[\cos(2\pi k_1x_0) - \cos(2\pi k_1x_1)]\cos(2\pi k_2y_0) \quad (k_1=2n+1, n=0,\pm 1,\pm 2...)
\end{cases} \\
& = 
\begin{cases}
4 \cos[\pi k_1(x_0+x_1)] \sin[\pi k_1(x_0-x_1)] \sin(2\pi k_2y_0) \quad (k_1=2n, n=0,\pm 1,\pm 2...)\\
-4 \sin[\pi k_1(x_0+x_1)] \sin[\pi k_1(x_0-x_1)] \cos(2\pi k_2y_0) \quad (k_1=2n+1, n=0,\pm 1,\pm 2...)
\end{cases}
\end{align}

For the superlattices of C8-BTBT monolayers on graphene, the in-plane potential modulation $\Delta V$ can be obtained within 2.5\% accuracy by using $k_1=[-1,1], k_2=0$. Therefore
\begin{align}\label{EqS:dVk}
\Delta V %& = \frac{2q}{4\pi ab}\sum_{\vec{k}}\frac{N_{k_1,k_2} e^{-2\pi dM_k}}{M_k}  \nonumber \\ 
& = \frac{-8q}{4\pi \varepsilon}\sum_{\vec{k}}{}'\frac{\sin[\pi k_1(x_0+x_1)] \sin[\pi k_1(x_0-x_1)]\cos(2\pi k_2y_0)}{\sqrt{k_1^2b^2+k_2^2 a^2}\ e^{2\pi d \sqrt{\frac{k_1^2}{a^2}+\frac{k_2^2}{b^2}}}} \nonumber \\
& = \frac{-16q}{4\pi \varepsilon} \frac{\sin[\pi (x_0+x_1)] \sin[\pi (x_0-x_1)]}{b e^{2\pi d/a}}  \qquad (k_1=0,\pm 1, k_2=0)  
%& = \frac{-16q}{4\pi \varepsilon} \frac{sin^2(\pi x_0) }{b e^{2\pi d/a}}  \qquad (x_1 = 0) \\
%& = \frac{-16}{4\pi \varepsilon} \frac{Q_x sin^2(\pi x_0) }{ 2(ax_0)^2 b e^{2\pi d/a}}  \qquad (Q_x=2q(ax_0)^2)
%\begin{cases}
% \frac{-4q sin[\pi \varepsilon (x_0+x_1)] sin[\pi (x_0-x_1)]}{\pi b e^{2\pi d/a}} \\% \qquad (else) 
% \noalign{\vskip9pt}
% \frac{-4q(x_0^2-x_1^2)}{\pi  b e^{2\pi d/a}} = \frac{-4Q_x}{\pi b e^{2\pi d/a}} \qquad (when \ x_0\approx x_1 \rightarrow 0) 
%\end{cases}
\end{align}

We can further simplify Eq.~\ref{EqS:dVk} by taking $x_1=0$, analogous to the representation of each C8-BTBT molecule using only three point charges (see Fig.~\ref{figSI:dVbtbt3pt}), in which case the quadrupole moment is $Q_x=2q(ax_0)^2$. This results in:
\begin{align}\label{EqS:dVsimple}
 \Delta V %& = \frac{-16q}{4\pi \varepsilon} \frac{sin^2(\pi x_0) }{b e^{2\pi d/a}}  \qquad (x_1 = 0) \\
 = \frac{-16}{4\pi \varepsilon} \frac{Q_x \sin^2(\pi x_0) }{ 2(ax_0)^2 b e^{2\pi d/a}}  
\end{align}

% \newpage

%\singlespacing
\bibliography{MultipoleDoping}